\documentclass[10pt,aps,prd,superscriptaddress,nofootinbib,nobibnotes,longbibliography,floatfix,twocolumn]{revtex4-2}

\usepackage{bm}
\usepackage{mathtools,
amsmath,
amssymb,
amsfonts,
mathrsfs,
chngcntr,
multirow}

\let\cc\corresponds
\let\corresponds\relax
\usepackage{mathabx}
\let\corresponds\cc

\usepackage[utf8]{inputenc}
\usepackage[T1]{fontenc}
\usepackage[dvipsnames]{xcolor}
\usepackage[unicode]{hyperref}
\hypersetup{colorlinks=true, citecolor=MidnightBlue,
            linkcolor=MidnightBlue, urlcolor=MidnightBlue, linktocpage=true}
\usepackage[normalem]{ulem}
\usepackage{orcidlink}
\usepackage[capitalize]{cleveref}
\usepackage{float}
\usepackage{comment}

\usepackage{orcidlink}

\newcommand{\orcid}[1]{\href{https://orcid.org/#1}{\includegraphics[width=10pt]{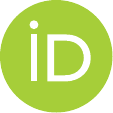}}}

\newcommand{\qm}[1]{``#1''}

\usepackage{placeins}

\begin{document}

\title{Constraining the geometry of rotating black holes with eikonal QNMs}

\author{Ciro De Simone \orcid{0009-0004-0610-1686}}
\email{ciro.desimone@unina.it}
\affiliation{Dipartimento di Fisica \qm{E. Pancini}, Università di Napoli \qm{Federico II}, Complesso Universitario di Monte S. Angelo, Via Cinthia Edificio 6, I-80126 Napoli, Italy}
\affiliation{Istituto Nazionale di Fisica Nucleare, Sezione di Napoli, Complesso Universitario di Monte S. Angelo, Via Cinthia Edificio 6, 80126 Napoli, Italy}

\author{Sebastian H.\,V\"olkel \orcid{0000-0002-9432-7690}}
\email{sebastian.voelkel@uni-tuebingen.de}
\affiliation{Theoretical Astrophysics, IAAT, University of T\"ubingen, D-72076 T\"ubingen, Germany}

\author{Kostas D.\,Kokkotas \orcid{0000-0001-6048-2919}}
\email{kostas.kokkotas@uni-tuebingen.de}
\affiliation{Theoretical Astrophysics, IAAT, University of T\"ubingen, D-72076 T\"ubingen, Germany}

\begin{abstract}
Quasinormal modes of black holes provide a central avenue for confronting general relativity with gravitational-wave observations from the ringdown stage of compact binary coalescence. However, inferring black-hole properties or the background spacetime requires theory-dependent input. In this work, we connect a calibrated eikonal method, which relates quasinormal mode shifts to local metric properties near the light ring, with the underlying rotating black hole metric. Using Bayesian inference, we compare a metric-specific model based on the global Konoplya-Rezzolla-Zhidenko parametrization with a metric-agnostic approach based on local values and radial derivatives of the metric functions at the light ring. For each model, we perform both one-parameter and simultaneous multi-parameter analyses and quantify how their priors map onto the same local quantities. We find that uniform priors on the metric-specific coefficients can induce strongly nonuniform and skewed priors on these local metric quantities. For an injection compatible with general relativity, both models remain consistent with Kerr and yield nearly identical posteriors for the orbital frequency and Lyapunov exponent despite their different induced priors. For injections representing deviations from general relativity, one-parameter analyses can fail to recover the injected local deviations and, in the metric-specific case, can bias the global metric reconstruction. The multi-parameter analyses of both models yield consistent constraints, suggesting sufficient robustness. However, they introduce substantial degeneracies and differences in marginalized parameters, demonstrating practical limitations when considering more flexible models. Finally, we apply the framework to approximate posterior information from the ringdown analysis of GW250114 and find that both models are consistent with the Kerr metric in general relativity.
\end{abstract}

\maketitle

\section{Introduction}

The ringdown stage of binary black hole (BH) mergers provides a direct observational window into the strong-field dynamics of BH spacetimes~\cite{LIGOScientific:2016aoc}. In general relativity (GR), the response of a perturbed Kerr BH~\cite{Kerr:1963ud} at intermediate times is described by a discrete spectrum of quasinormal modes (QNMs)~\cite{Regge:1957td,Zerilli:1970se,Teukolsky1972,Vishveshwara:1970zz,Chandrasekhar:1975zza,Leaver1985,Chandrasekhar:1985kt,Kokkotas:1999bd,Berti:2009kk,Konoplya:2011qq}, whose complex frequencies are determined only by the remnant mass and spin. Measurements of these modes therefore enable tests of the Kerr hypothesis and of the no-hair conjecture~\cite{Israel:1967wq,Carter:1971zc,Hawking:1971vc,Robinson:1975bv} through BH spectroscopy~\cite{Detweiler:1980gk,Dreyer:2003bv,Berti:2005ys,Cardoso:2016ryw,Franchini:2023eda,Berti:2025hly}. 

QNM frequencies can often be measured or parametrized in a theory-agnostic way, but inferring constraints on the underlying spacetime requires additional theoretical input. One common approach circumventing this limitation is to relate QNMs to properties of the null geodesic structure in the eikonal limit, where the QNM frequency is connected to the orbital frequency and instability timescale of the light ring~\cite{1972ApJ...172L..95G,Ferrari:1984zz,Cardoso:2008bp,Dolan:2010wr,Yang:2012he}. 
Although the eikonal correspondence provides a useful connection between QNMs and properties of unstable photon orbits, it is not universal. In particular, the correspondence can fail for gravitational perturbations in certain beyond-GR theories~\cite{Konoplya:2017wot,Konoplya:2022gjp,Chen:2022nlw}, and the physical association between BH ringing and the light ring should be interpreted with care~\cite{Khanna:2016yow}.

To represent possible deviations from the Kerr metric, a wide range of theory-agnostic parametrizations of BH spacetimes have been proposed in the literature, e.g., the Johannsen~\cite{Johannsen:2013szh}, Johannsen-Psaltis~\cite{Johannsen:2011dh,Cardoso:2014rha}, the Effective Metric Description~\cite{DelPiano:2023fiw,DelPiano:2024gvw}, bumpy Kerr~\cite{Collins:2004ex,Vigeland:2009pr}, Rezzolla-Zhidenko (RZ)~\cite{Rezzolla:2014mua}, Konoplya-Rezzolla-Zhidenko (KRZ)~\cite{Konoplya:2016jvv, Konoplya:2018arm}, and many others. 
Each metric follows a different strategy and has its own advantages and disadvantages. 
In this work, we adopt the KRZ metric, whose continued fraction representation provides an efficient expansion of stationary and axisymmetric BH metrics whose coefficients can either be mapped to theory-specific solutions or treated phenomenologically~\cite{Mukazhanov:2024rka}. Although the complete expansion contains infinitely many coefficients, a relatively small subset can reproduce many physically relevant BH observables with good accuracy~\cite{Konoplya:2020hyk,Konoplya:2022tvv}. 

Several approaches connecting QNMs to deviations from the Kerr geometry or from the GR perturbation equations have been developed. 
In the non-rotating case, the RZ metric~\cite{Rezzolla:2014mua} has been used to study constraints from synthetic QNM data with varying numbers of free metric parameters in Ref.~\cite{Volkel:2020daa}. Explicit mappings among the Johannsen–Psaltis, RZ, and Effective Metric Description parametrizations have also been developed for spherically symmetric spacetimes, with applications to the eikonal QNMs of regular BHs~\cite{DelPiano:2025ykr}. 
Constraints on the KRZ spacetime have been obtained from the inspiral part of actual gravitational-wave data, focusing on a subset of KRZ deformation parameters varied individually~\cite{Shashank:2021giy}. Moreover, the eikonal method has also been used to compute bounds on beyond-Kerr metrics from GW data in Ref.~\cite{Dey:2022pmv}. 

Complementary to parametrized BH metrics, parametrized QNM frameworks~\cite{Cardoso:2019mqo,McManus:2019ulj,Kimura:2020mrh,Volkel:2022aca,Volkel:2022khh,Franchini:2022axs,Thomopoulos:2025nuf} at the level of perturbation equations of non-rotating black holes have been extended to modified Teukolsky equations describing rotating BHs~\cite{Cano:2024jkd,DeSimone:2026waz}, following earlier works~\cite{Li:2022pcy,Hussain:2022ins,Cano:2023tmv}. 
Related to the eikonal method, the framework has been studied within the WKB approximation in Ref.~\cite{Tang:2025qaq}. Using the gravitational wave event GW250114~\cite{LIGOScientific:2025rid,LIGOScientific:2025wao}, first bounds on the beyond-Teukolsky parametrization~\cite{Cano:2024jkd} have recently been reported in Ref.~\cite{Volkel:2026qqz}. 

In theoretical setups, constraints on the parametrized modifications of the effective potentials governing non-rotating BH perturbations were studied for single fields in Ref.~\cite{Volkel:2022khh}, and extended to couplings to new fundamental degrees of freedom in Ref.~\cite{Volkel:2022aca}. 
These studies showed that individual expansion coefficients are often poorly constrained from QNM data because of degeneracies, while local quantities near the light ring, such as the potential and its derivatives, can be more robustly inferred. 
RZ-based scattering studies have explored Bayesian reconstruction of nonrotating analogue geometries from simulated time-domain signals in Ref.~\cite{Albuquerque:2025eny}.
A related setup identified an optimal incident-pulse width for black-hole metric inference, set approximately by the inverse square root of the potential-barrier height, and demonstrated that most information is typically contained in the excited fundamental mode~\cite{Albuquerque:2026bgc}.

Taken together, these results motivate a systematic comparison, for rotating BHs, between constraints on global metric deformations and on the local geometry probed by the eikonal correspondence. Such a comparison should account for the priors induced on the local quantities and for the differences between restricted one-parameter and simultaneous multi-parameter analyses. 

In this work, we apply the eikonal method to rotating BH spacetimes in order to determine the information that can be extracted from the fundamental QNM frequency on the underlying geometry. We compare two models: a \textit{metric-specific model}, in which deviations from Kerr are described by selected coefficients of the global KRZ parametrization, and a \textit{metric-agnostic model}, in which the free parameters are the local values and derivatives of the metric functions evaluated at the light ring. In both cases, the eikonal approximation is calibrated to the Kerr QNM spectrum and used to compute frequency shifts away from the GR prediction. 

We study both models at two levels of complexity: restricted analyses in which one deformation parameter is varied at a time and simultaneous analyses in which all selected parameters are allowed to vary. This enables us to distinguish effects associated with the choice of metric description from those caused by the dimensionality of the inference problem. We also quantify the role of Bayesian priors by expressing the prior distributions of both models in terms of the same local metric quantities. In particular, uniform priors on the KRZ coefficients can induce strongly nonuniform and skewed distributions for the metric quantities at the light ring.

We apply this framework to three cases: a GR injection corresponding to a Kerr BH, a non-GR injection defined by a nontrivial realization of the KRZ metric, and observational constraints from GW250114. For the GR injection, as expected, all analyses remain consistent with Kerr, but already highlight the importance of priors between the two models, i.e., uniform in the KRZ metric represents non-trivial priors in the local light ring properties of the metric. While single parameter constraints differ, simultaneously varying parameters of both models gives robust constraints. 
For the non-GR injection, however, restricting either model to a single deformation parameter can prevent the injected local metric deviations from being recovered. In the metric-specific case, this restriction can also bias the inferred KRZ coefficients and the resulting global metric reconstruction. Simultaneous multi-parameter analyses provide again consistent constraints, but exhibit substantial degeneracies. Applied to the damped sinusoid analysis of GW250114, both models yield results consistent with Kerr. 

Although our analysis relies on the calibrated eikonal approximation and is therefore only approximate away from GR, its qualitative lessons concerning metric-specific versus metric-agnostic inference, prior dependence, and single- versus multi-parameter Bayesian analyses may apply more broadly, and thus extend beyond gravitational-wave tests of general relativity.

The rest of this work is organized as follows. In Sec.~\ref{method}, we introduce the theoretical framework and Bayesian inference methodology. In Sec.~\ref{applications}, we apply the framework to synthetic GR and non-GR injections, and to publicly available posterior samples from GW250114. We summarize our results and discuss their implications in Sec.~\ref{conclusions}. 
Additional material is provided in  Appendix~\ref{appendix:retro}.

\section{Methods}\label{method}

In this work, we combine the following methods to quantify how QNM information can be used to constrain BH geometries. 
First, in Sec.~\ref{method_eikonal} we use the eikonal approximation to relate QNMs to the local spacetime geometry near the photon ring. 
Second, in Sec.~\ref{method_krz} we describe the KRZ parametrization as a global BH metric to capture deviations from the Kerr BH. 
Third, we formulate a Bayesian inference framework that maps measured ringdown frequencies to spacetime parameters in Sec.~\ref{method_bayesian}. 
Finally, we introduce a metric-agnostic strategy that reconstructs the local geometry directly from the quantities probed by the observations in Sec.~\ref{method_constraints}.

\subsection{Eikonal method}\label{method_eikonal}

The eikonal approximation establishes a direct connection between QNM frequencies and the local spacetime geometry near the unstable photon orbit~\cite{1972ApJ...172L..95G,Ferrari:1984zz,Cardoso:2008bp,Dolan:2010wr,Yang:2012he}. In the large-angular-momentum limit, $\ell\gg 1$, the QNM spectrum can be expressed in terms of the photon-orbit frequency $\Omega$ and the Lyapunov exponent $\gamma$, which characterizes the orbit's instability timescale~\cite{Cardoso:2008bp,Konoplya:2017wot,Pedrotti:2024znu}. This correspondence considerably simplifies the inverse problem, reducing it to the inference of a small set of local geometric quantities from the QNMs observed in the ringdown signal.

In this work, we focus on the fundamental modes, $n=0$, with $\ell=|m|$. These modes are of particular interest for two reasons. First, they are typically among the most strongly excited modes in gravitational perturbations. Second, they are associated with prograde ($m>0$) and retrograde ($m<0$) equatorial photon orbits. 

Circular orbits around a rotating BH at radius $r_{\rm ph}$ satisfy the equations~\cite{Glampedakis:2017dvb}
    \begin{align}
        g_{tt}(r_{\rm ph})+2g_{t\phi}(r_{\rm ph})\,\Omega+g_{\phi\phi}(r_{\rm ph})\,\Omega^2&=0,\label{eq:eq_orbital_1}\\
        g'_{tt}(r_{\rm ph})+2g'_{t\phi}(r_{\rm ph})\,\Omega+g'_{\phi\phi}(r_{\rm ph})\,\Omega^2&=0,\label{eq:eq_orbital_2}
    \end{align}
where $\Omega$ is the orbital frequency of the photon orbit and primes denote derivatives with respect to the radial coordinate. These equations can be used to obtain an explicit expression of $\Omega$ in terms of the metric functions (see Eq.~\eqref{eq:eik_frequency} below) or combined to give an equation for the photon orbit radius
    \begin{align}\label{eq:photon_orbit_radius}
        g_{\phi\phi}(g'_{tt})^2+2g_{tt}(g'_{t\phi})^2-g'_{tt}(g_{tt}g'_{\phi\phi}+2g_{t\phi}g'_{t\phi})\nonumber\\\mp2W^{1/2}(g_{t\phi}g'_{tt}-g_{tt}g'_{t\phi})=0,
    \end{align}
where $W=(g'_{t\phi})^2-g'_{tt}g'_{\phi\phi}$. Given a BH metric, this equation provides the radius of the prograde and retrograde circular orbits.

In the eikonal approximation, the corresponding QNM frequency can be written as
\begin{equation}\label{eikonal_QNMs}
    \omega=m\,\Omega-\frac{i}{2}|\gamma|\,,
\end{equation}
where
\begin{subequations}\label{eikonal_quantites}
\begin{align}
    \Omega&=\frac{g'_{tt}}{-g'_{t\phi}\mp W^{1/2}}\,, \label{eq:eik_frequency}\\
    \gamma&=\Omega \,\sqrt{\frac{(g_{t\phi}^2-g_{tt}g_{\phi\phi})(g''_{tt} b^2+2g''_{t\phi} b+g''_{\phi\phi})}{2g_{rr}(g_{tt} b+g_{t\phi})^2}} \label{eq:eik_Lyapunov}\,.
\end{align}
\end{subequations}
In the above expressions $b=\Omega^{-1}$ denotes the impact parameter of the photon orbits and all quantities are evaluated either at the prograde or retrograde photon orbit~\cite{Glampedakis:2017dvb}.

The appeal of Eq.~\eqref{eikonal_QNMs} lies in its broad applicability, since it only requires knowledge of the black-hole metric. At the same time, it should be regarded as an approximation, and it is not guaranteed to provide accurate QNM frequencies for a generic black-hole spacetime~\cite{Konoplya:2017wot}. This limitation is expected, as the eikonal construction does not account for the full perturbation dynamics determined by the field equations of the underlying gravitational theory.

Even in GR, the eikonal approximation is not highly accurate for the dominant $\ell=2$ modes, while modes with much larger $\ell$ may only become accessible with future detectors such as LISA~\cite{Berti:2016lat,Barausse:2020rsu, LISA:2022yao,LISA:2024hlh}. For example, for a Kerr BH with $a=0.7$ the relative error compared to Leaver's method~\cite{Leaver1985} is $\sim 5-7\%$ for the real and imaginary parts at $\ell=m=2$, with significant improvement for larger $\ell$. To avoid interpreting this intrinsic systematic error as a spurious deviation from GR, we follow the common approach of calibrating the eikonal prediction to the exact GR result~\cite{Dey:2022pmv,Pani:2026yzi,DeSimone:2026mkz}. In practice, we use the eikonal approximation only to estimate the shift away from GR, and write
\begin{align}
\omega^\text{model}(\boldsymbol{\theta})
=
\omega_\text{GR}
+
\Delta \omega_\text{Eik}(\boldsymbol{\theta})\,,
\end{align}
where
\begin{align}
\Delta \omega_\text{Eik}(\boldsymbol{\theta})
=
\omega_\text{Eik}(\boldsymbol{\theta})
-
\omega_\text{Eik}(\boldsymbol{\theta}_\text{GR})\,.
\end{align}
Here $\omega_\text{GR}$ denotes the exact GR QNM frequency for the Kerr BH with the same mass $M$ and spin $a$, while $\boldsymbol{\theta}_\text{GR}$ denotes the corresponding Kerr values of the metric parameters. This prescription ensures that the model reproduces the correct GR limit, while retaining the eikonal estimate for beyond-GR frequency shifts. The accuracy of this approximation depends on the specific model under consideration~\cite{DeSimone:2026mkz,Pani:2026yzi} and, crucially, on the validity of the eikonal correspondence~\cite{Cardoso:2008bp}. Under those assumptions, it is also clear that the approximation improves for larger values of the multipole $\ell$.

The dependence on the mass $M$ and spin $a$ is not introduced as an additional explicit correction, since these quantities already determine the metric functions entering $\boldsymbol{\theta}$, which we treat as the fundamental independent parameters.

\subsection{Konoplya-Rezzolla-Zhidenko metric}\label{method_krz}

The KRZ metric~\cite{Konoplya:2016jvv} generalizes the RZ parametrization~\cite{Rezzolla:2014mua} to axisymmetric spacetimes while preserving the two main features: a compactified radial coordinate and a continued-fraction expansion for the near horizon behavior. 
Its line element is given by
\begin{align}
    ds^2 = &-\frac{N^2(r,\theta)-W^2(r,\theta)\sin^2\theta}{K^2(r,\theta)} dt^2+ \nonumber\\&- 2W(r,\theta)\,r\sin^2\theta dtd\phi+ \nonumber\\&+ K^2(r,\theta)\,r^2\sin^2\theta d\phi^2+ \nonumber\\&+\Sigma(r,\theta)\left(\frac{B^2(r,\theta)}{N^2(r,\theta)}dr^2+r^2 d\theta^2\right),
\end{align}
in terms of the functions: $N^2(r,\theta)$, $W(r,\theta)$, $K^2(r,\theta)$, $B(r,\theta)$ and $\Sigma(r,\theta)$. The key idea of the KRZ parametrization is to express all the metric functions in terms of the coordinates $y=\cos\theta$ and the compactified radial coordinate 
\begin{equation}
    x = 1-\frac{r_0}{r},
\end{equation}
where the event horizon $r_0$ can be identified with the largest root of $N^2(r,\pi/2)$. 
In terms of the coordinate $x \in [0,1)$, where $x=0$ corresponds to the event horizon position and $x=1$ to infinity, the parametrized metric functions are
\begin{align}
    N^2(x) &= xA_0(x) + \sum_{i=1}^\infty A_i(x)\,y^i\,,\\
    B(x) &= 1 + \sum_{i=0}^\infty B_i(x)\,y^i\,,\\
    W(x) &= \sum_{i=0}^\infty \frac{W_i(x)\,y^i}{\Sigma}\,,\\
    K^2(x) &= 1+\frac{aW(x)}{r}+\sum_{i=0}^\infty \frac{K_i(x)\,y^i}{\Sigma}\,,
\end{align}
where
\begin{equation}
    \Sigma(r,\theta) = 1+\frac{a^2}{r^2}\cos^2\theta\,.
\end{equation}
It is important to notice, however, that the $m=\pm \ell$ QNMs are associated to photon orbits that lie on the equatorial plane, thus $\theta=\pi/2$ and $y=0$. As a consequence, only the $i=0$ terms survive in the expressions above. Moreover, $g_{\theta\theta}$ in the KRZ parametrization has the same form as in the Kerr BH.  Following Ref.~\cite{Konoplya:2016jvv}, the KRZ metric functions on the equatorial plane can be expressed as
\begin{align}
    B_0(x) = & b_{00}\,(1-x)+\tilde{B_0}(x)(1-x)^2\,,\\
    W_0(x) = & w_{00}\,(1-x)^2+\tilde{W_0}(x)(1-x)^3\,,\\
    K_0(x) = & k_{00}\,(1-x)^2+\tilde{K}_0(x)(1-x)^3\,,\\
    A_0(x) = &1-\epsilon_0\,(1-x)+(a_{00}-\epsilon_0+k_{00})(1-x)^2 \nonumber\\
           & +\tilde{A}_0(x)(1-x)^3\,, 
\end{align}
and all the quantities labeled by a tilde are written in terms of continued fractions, such as
\begin{equation}
    {\tilde A_0}(x)=\frac{a_{01}}{\displaystyle 1+\frac{\displaystyle
    a_{02}\,x}{\displaystyle 1+\ldots}}\,,
\end{equation}
and analogous expressions for the other functions can be found in Ref.~\cite{Konoplya:2016jvv}.

Constraints on the KRZ parameters can be obtained by requiring asymptotic flatness~\cite{Konoplya:2016jvv}
\begin{align}
    \epsilon_0 &= \frac{2M-r_0}{r_0}\,, \label{eps0}\\
    w_{00} &= \frac{2a}{M}\frac{(1+\epsilon_0)^2}{2}\,,
\end{align}
while the coefficients that appear in the continued fractions specify the near-horizon metric. Moreover, we also require that
\begin{equation}
    k_{00} = \frac{a^2}{r_0^2}\,,
\end{equation}
in order to recover the Kerr equatorial metric at lowest order in the expansion, while the higher-order coefficients $k_{0j}$ remain free. Constraints on the $a_{00}$ and $b_{00}$ parameters can also be obtained from PPN expansion~\cite{Will:2014kxa} based on the asymptotic behavior of the metric. In the following, we do not impose the PPN constraints in order to allow for more general metric profiles, even though they can be naturally introduced in the statistical analysis.

Additional constraints~\cite{Abdikamalov:2021zwv} can be obtained by requiring that, as for the GR BHs, the metric determinant has to be negative and the signature of the metric has to be preserved. We further require the absence of closed time-like curves by imposing the condition $g_{\phi\phi}>0$. Notice also that the $g_{tt}$ function changes sign at the ergosphere and can thus have positive or negative values. In order to recover the Kerr metric when the deformations are set to zero, we redefine the $\epsilon_0$ parameter as
    \begin{equation}
        \epsilon_0 = \epsilon_0^{GR}+\delta \epsilon,
    \end{equation}
where $\epsilon_0^{GR}$ corresponds to the expression given in Eq.~\eqref{eps0} where the radius $r_0$ coincides with the outer horizon of the Kerr BH $r_+=M+\sqrt{M^2-a^2}$. With this choice, the free KRZ parameters that we consider in the rest of this paper are $\{\delta\epsilon, a_{00},b_{00},a_{01},b_{01},w_{01},k_{01}\}$ and the Kerr BH is recovered when they are all zero. Notice also that the position of the equatorial photon orbit for a given combination of KRZ parameters can be obtained from Eq.~\eqref{eq:photon_orbit_radius}.

\subsection{Bayesian methods}\label{method_bayesian}

Bayesian analysis provides a framework for inferring model parameters by combining prior information with the information contained in the data. For a model described by parameters $\boldsymbol{\theta}$ and data $d$, Bayes' theorem gives the posterior probability distribution
\begin{equation}
p(\boldsymbol{\theta}\mid d)\,\propto\,\mathcal{L}(d\mid \boldsymbol{\theta})\,\pi(\boldsymbol{\theta})\,,
\end{equation}
where $\mathcal{L}(d\mid \boldsymbol{\theta})$ is the likelihood and $\pi(\boldsymbol{\theta})$ is the prior. The posterior distribution contains the inferred parameter values, their uncertainties, and correlations. Since this distribution is generally high-dimensional and cannot be sampled directly, we explore it numerically using Markov-Chain-Monte-Carlo (MCMC) methods. In particular, we use the Python package \texttt{emcee}~\cite{Foreman-Mackey:2012any}, based on the affine-invariant ensemble sampler~\cite{2010CAMCS...5...65G}, to generate samples $\{\boldsymbol{\theta}_i\}$ from the posterior distribution, which are then used to estimate marginalized constraints, credible intervals, and parameter correlations. For each case considered in the analysis, the Markov chains contain $\sim10^5$ steps.

For the QNM analysis, we define the likelihood directly in terms of the complex QNM frequencies. For each fundamental mode labeled by $(\ell,m,n=0)$, the data vector is taken to be
\begin{equation}\label{d_theta}
\mathbf{d}_{\ell m}
=
\begin{pmatrix}
\omega_{\ell m,{\rm Re}} \\
\omega_{\ell m,{\rm Im}}
\end{pmatrix}^{\rm obs}\,, \qquad 
\end{equation}
which contains the observed real and imaginary parts of the mode frequency. This is compared with the corresponding theoretical prediction
\begin{equation}\label{omega_theta}
\boldsymbol{\omega}_{\ell m}(\boldsymbol{\theta})
=
\begin{pmatrix}
\omega_{\ell m,{\rm Re}}(\boldsymbol{\theta}) \\
\omega_{\ell m,{\rm Im}}(\boldsymbol{\theta})
\end{pmatrix}\,.
\end{equation}
Assuming Gaussian uncertainties, with covariance matrix $\mathbf{C}_{\ell m}$ for the real and imaginary components, the likelihood for a single mode is
\begin{equation}
\mathcal{L}_{\ell m}(d\mid \boldsymbol{\theta})
\propto\exp\left[-\frac{1}{2}\Delta\boldsymbol{\omega}_{\ell m}^{T}(\boldsymbol{\theta})
\mathbf{C}_{\ell m}^{-1}\Delta\boldsymbol{\omega}_{\ell m}(\boldsymbol{\theta})\right]\,,
\end{equation}
where
\begin{equation}
\Delta\boldsymbol{\omega}_{\ell m}(\boldsymbol{\theta})=\mathbf{d}_{\ell m}-\boldsymbol{\omega}_{\ell m}(\boldsymbol{\theta})\,.
\end{equation}
Equivalently, up to an additive constant, the log-likelihood is
\begin{equation}
\ln \mathcal{L}_{\ell m}=-\frac{1}{2}\Delta\boldsymbol{\omega}_{\ell m}^{T}(\boldsymbol{\theta})
\mathbf{C}_{\ell m}^{-1}\Delta\boldsymbol{\omega}_{\ell m}(\boldsymbol{\theta})\,.
\end{equation}
Those equations can be naturally generalized to the case where more than one mode is present by including the real and imaginary parts of the new mode in Eq.~\eqref{d_theta} and Eq.~\eqref{omega_theta}.

We further include external constraints on the remnant mass and spin by defining
\begin{equation}
    M=M_0+\delta M \quad \textrm{and} \quad a = a_0+\delta a\,,
\end{equation}
where $M_0,a_0$ are the reference mass and spin, and $\delta M,\delta a$ the corresponding shifts to those values. We adopt a multivariate Gaussian prior on $\delta M$ and $\delta a$ as
\begin{equation}
\mathbf{x}
=
\begin{pmatrix}
M \\
a
\end{pmatrix},
\qquad
\mathbf{x}_0
=
\begin{pmatrix}
M_0 \\
a_0
\end{pmatrix},
\end{equation}
in the general form
\begin{equation}
\pi(M,a)\,\propto\,\exp\left[-\frac{1}{2}\left(\mathbf{x}-\mathbf{x}_0\right)^T\mathbf{\Sigma}_{Ma}^{-1}\left(\mathbf{x}-\mathbf{x}_0\right)
\right].
\end{equation}
Here $\mathbf{\Sigma}_{Ma}$ is the covariance matrix for $\delta M$ and $\delta a$, including their possible correlation. Equivalently, up to an additive constant,
\begin{equation}
\ln \pi(M,a)=-\frac{1}{2}\Delta \mathbf{x}^{T}\mathbf{\Sigma}_{Ma}^{-1}\Delta \mathbf{x},\qquad\Delta \mathbf{x}=\mathbf{x}-\mathbf{x}_0 .
\end{equation}
The posterior sampled in the analysis is therefore
\begin{equation}
p(\boldsymbol{\theta}\mid d)\,\propto\,\mathcal{L}(d\mid \boldsymbol{\theta})\,\pi(M,a)\,\pi_{\rm flat}(\boldsymbol{\theta}),
\end{equation}
where $\pi_{\rm flat}(\boldsymbol{\theta})$ denotes uniform priors on the remaining parameters over the ranges specified in the relevant application sections. To have a clear comparison between prior and posterior knowledge of the metric-agnostic constraints introduced in Sec.~\ref{method_constraints}, we will also map the uniform priors in the KRZ parameters to the corresponding, in general, non-uniform priors for the metric-agnostic parameters. 

It is important to notice that mass and spin parameters enter the eikonal estimates in both the Kerr and KRZ cases. As a consequence, for all values of those parameters the eikonal shifts at the prograde photon sphere are zero and only non-zero values of the KRZ parameters will give rise to non-zero eikonal shifts. In the following, the shifts in mass and spin are treated as free parameters in both the theory-agnostic and theory-specific applications.

We adopt these priors on the mass and spin following the proposal of Ref.~\cite{Volkel:2026qqz} and earlier discussions in Refs.~\cite{Volkel:2022khh,Volkel:2022aca}. If deviations from GR are small, estimates of the final mass and spin obtained from the inspiral-merger signal under the assumption of GR are expected to remain approximately valid, up to small theory-dependent corrections. The remnant mass and spin are therefore not treated as completely unknown in the ringdown analysis. Instead, we use the inspiral-merger information as external input, while allowing for possible systematic offsets by choosing priors that are sufficiently broad.

\subsection{Metric-specific and metric-agnostic constraints}\label{method_constraints}

We now turn to the central methodological development of this work.
To facilitate the discussion of our results, we first introduce the two types of constraints that we aim to obtain with our framework, namely ``metric-specific'' and ``metric-agnostic'' constraints. 
We refer to constraints as metric-specific when they are directly tied to an explicit parametrization of a black-hole metric, i.e., to a particular realization of the KRZ parameters $\boldsymbol{\theta}_{\rm KRZ}$. 
We refer to constraints as metric-agnostic when they concern only the metric functions themselves, i.e., evaluated at the corresponding light ring, without reference to a specific metric parametrization $\boldsymbol{\theta}_{g}$. 

In practice, these metric-agnostic quantities are evaluated at the light ring associated with the QNM mode under consideration, so that prograde and retrograde modes may probe different radii. Based on the eikonal formulae~\eqref{eq:eik_frequency}-\eqref{eq:eik_Lyapunov}, the total number of parameters in the agnostic case is the ten values of the metric functions and their radial derivatives evaluated at the photon sphere: 
$\{\delta g_{tt},\delta g'_{tt},\delta g''_{tt},\delta g_{t\phi},\delta g'_{t\phi},\delta g''_{t\phi},\delta g_{rr},\delta g_{\phi\phi},\delta g'_{\phi\phi},\delta g''_{\phi\phi}\}$. Three metric functions appear together with their radial derivatives up to second order, while the derivatives of $g_{rr}$ and the function $g_{\theta\theta}$ do not enter the eikonal formulae. Moreover, the value of the metric functions and their derivatives at any point in the spacetime is not invariant under a coordinate transformation. This implies that in the agnostic approach we are implicitly assuming Boyer-Lindquist-like coordinates. 

Another important feature of the metric agnostic approach concerns the equations for the photon orbit properties. While Eq.~\eqref{eq:eq_orbital_2} is used to determine the expression of the photon orbital frequency, Eq.~\eqref{eq:eq_orbital_1} provides an additional constraint on the metric shifts. As a consequence, not all the metric shifts are independent and Eq.~\eqref{eq:eq_orbital_1} will be used to express $\delta g_{tt}$ in terms of the other nine metric shifts, which induces a non-trivial prior on $\delta g_{tt}$. According to Eq.~\eqref{eq:eq_orbital_1}, it depends on the other metric shifts and the orbital frequency, which itself depends on the other metric shifts. This implies that in the agnostic approach, we cannot impose flat priors on $\delta g_{tt}$, but only the inherited prior and the regularity conditions discussed in Sec.~\ref{method_krz}. 

In the metric-specific case, the number of possible KRZ parameters is, in principle, infinite, whereas only a finite number of metric functions enter the eikonal formula. Moreover, since the eikonal formula relates the metric information to a single complex quantity, namely the fundamental QNM frequency, the inference problem is generically highly degenerate. One therefore expects simple, well-behaved constraints only when the number of free parameters is sufficiently small. For the purposes of this paper, we select the KRZ parameters up to first order: $\{\delta\epsilon,a_{00},a_{01},b_{00},b_{01},k_{01},w_{01}\}$, while all the others are set to zero. The metric-specific QNM estimates are thus obtained by applying Eqs.~\eqref{eq:eik_frequency}-\eqref{eq:eik_Lyapunov} to the corresponding KRZ metric functions. Moreover, in the theory-specific approach the two constraints in Eqs.~\eqref{eq:eq_orbital_1}-~\eqref{eq:eq_orbital_2} are directly enforced to determine the orbital frequency and the position of the photon orbit via Eq.~\eqref{eq:photon_orbit_radius}.

To compare the metric-specific constraints with the agnostic ones, we map a given metric-specific constraint on the KRZ parameters, $\boldsymbol{\theta}_{\rm KRZ}$, to the corresponding shift on the metric-function parameters compared to the Kerr case, $\boldsymbol{\theta}_{g}$. This is done by numerically determining the location of the light ring and then evaluating the relevant metric functions at that radius. Applying this map directly to posterior samples of $\boldsymbol{\theta}_{\rm KRZ}$ gives samples from the induced posterior on $\boldsymbol{\theta}_{g}$ for the metric-specific analysis. Since the metric-agnostic parameters can also be sampled directly, this allows us to compare the two approaches and assess how robustly the metric functions are constrained in each case. Note also that in the agnostic approach the position of the prograde photon sphere is not known. This is an important difference compared to the metric-specific case that gives access to the entire spacetime outside the event horizon.

It is crucial to discuss the choice of the priors on the parameters for the metric-specific and metric-agnostic case. In the KRZ metric, all the parameters are dimensionless and we adopt flat priors on the selected KRZ coefficients according to Table~\ref{tab:priors_KRZ}. The ranges have been chosen in such a way that the maximum shift in the real or imaginary part of the QNMs is up to $10\%$ when only one KRZ parameter at a time is nonzero. As for the agnostic case, the priors on the metric shifts have been chosen as uniform distributions over the range $\mathcal{U}(-1.5,1.5)$. 

{\renewcommand{\tabcolsep}{5mm}
\renewcommand{\arraystretch}{1.3}

\begin{table*}[ht]
\centering
\huge
\resizebox{\textwidth}{!}{
\begin{tabular}{c|c|c|c|c|c|c|c}
\hline
Parameters & $\delta\epsilon$ & $a_{00}$ & $b_{00}$ & $a_{01}$ & $b_{01}$ & $w_{01}$ & $k_{01}$ \\
\hline
Priors &
$\mathcal{U}(-0.2,\,0.2)$ &
$\mathcal{U}(-0.7,\,1)$ &
$\mathcal{U}(-0.5,\,1.5)$ &
$\mathcal{U}(-1,\,1)$ &
$\mathcal{U}(-0.8,\,1)$ &
$\mathcal{U}(-0.3,\,0.2)$ &
$\mathcal{U}(-0.3,\,0.5)$ \\
\hline
\end{tabular}
}
\caption{Uniform prior distributions used throughout our statistical analysis for the KRZ parameters.
}
\label{tab:priors_KRZ}
\end{table*}
}

As for the metric-agnostic case, one has to take into account that some of the parameters are dimensionful. For instance, $\delta g'_{tt}$ has dimensions of $[M]^{-1}$ since it corresponds to the shifts in the radial derivative of the dimensionless metric function $g_{tt}$. Similar considerations apply to all the other metric shifts. As a consequence, the priors have to be chosen appropriately to the parameter under consideration, by suitably rescaling the flat priors with the correct units of mass via the total mass $M_0+\delta M$. 

In both the metric-agnostic and specific case, we adopt Gaussian priors on the shifts in the mass $\delta M$ and angular momentum shift $\delta a$. Moreover, we require that the metric functions preserve the signature as well as the condition for a non-extremal BH: $a_0+\delta a < M_0+\delta M$. This is imposed because we are investigating small deviations from GR and the $\epsilon$ parameter is calibrated to the GR value $(\epsilon^{GR})$, which depends crucially on the position of the outer Kerr event horizon. This quantity is not well-defined for BHs beyond the extremal limit and can give rise to anomalous behaviors. Our approach is thus limited to non-extremal BHs in the GR sense even for non-Kerr BHs. Since also mass and spin are treated as free parameters, the application of this methodology to extremal BHs needs to be handled with care. 

When analyzing GR injections, the posterior distributions should be statistically consistent with the GR values, although sizable deviations may still be allowed because of the degeneracies inherent in the problem. If only one deviation parameter is varied at a time, as is commonly done in similar tests, the proposed framework effectively acts as a null test of GR within that restricted parameter subspace. For QNMs injected from Kerr, such an analysis should recover constraints consistent with the corresponding GR values. For non-Kerr injections, however, one-parameter analyses should be interpreted with caution: since the true metric may not lie within the restricted one-parameter family being tested, the inferred deviation parameter need not include the correct metric value.

If, instead, all relevant parameters are varied simultaneously, the correct values should in principle lie within the full posterior, provided that the parametrization is sufficiently flexible and that the priors include the true point. The corresponding marginalized one-dimensional posteriors may nevertheless be broad and individually uninformative due to the strong degeneracies of the problem. In such cases, the relevant information may reside primarily in correlations within the higher-dimensional posterior rather than in the marginalized constraints on individual parameters.

\section{Applications}\label{applications}

In the following, we apply our framework to three scenarios. To demonstrate the framework in a simple setup, we first discuss a GR injection in Sec.~\ref{app_GR}, followed by a non-GR injection in Sec.~\ref{app_nonGR}. Finally, we apply the framework to publicly available posterior data from GW250114 in Sec.~\ref{app_GW}.

\subsection{GR injection}\label{app_GR}

To isolate the effects of model dependence and prior choice, we begin with a controlled GR injection based on the fundamental $\ell=m=2$ mode of a Kerr BH with $M_0=1$ and $a_0=0.7$. 
Since consistency with Kerr is expected by construction, this analysis serves primarily as a validation of the inference framework and as a baseline for identifying any spurious deviations introduced by the parametrization or its induced priors. 
It also provides a reference for the subsequent non-GR injection and GW250114 analyses.

As priors on $\delta M$ and $\delta a$, we adopt uncorrelated Gaussian distributions centered at zero with standard deviation $\sigma=0.01$. 
The same Gaussian uncertainty model has been used for the uncertainty on the real and imaginary parts of the eikonal shifts in the log-likelihood. 
Even though the priors for the KRZ and the local metric-agnostic parameters are both uniform in their respective parameters, mapping the KRZ prior to local metric quantities produces substantially different distributions. 

\begin{figure}[ht]
\centering
\includegraphics[width=1.0\linewidth]{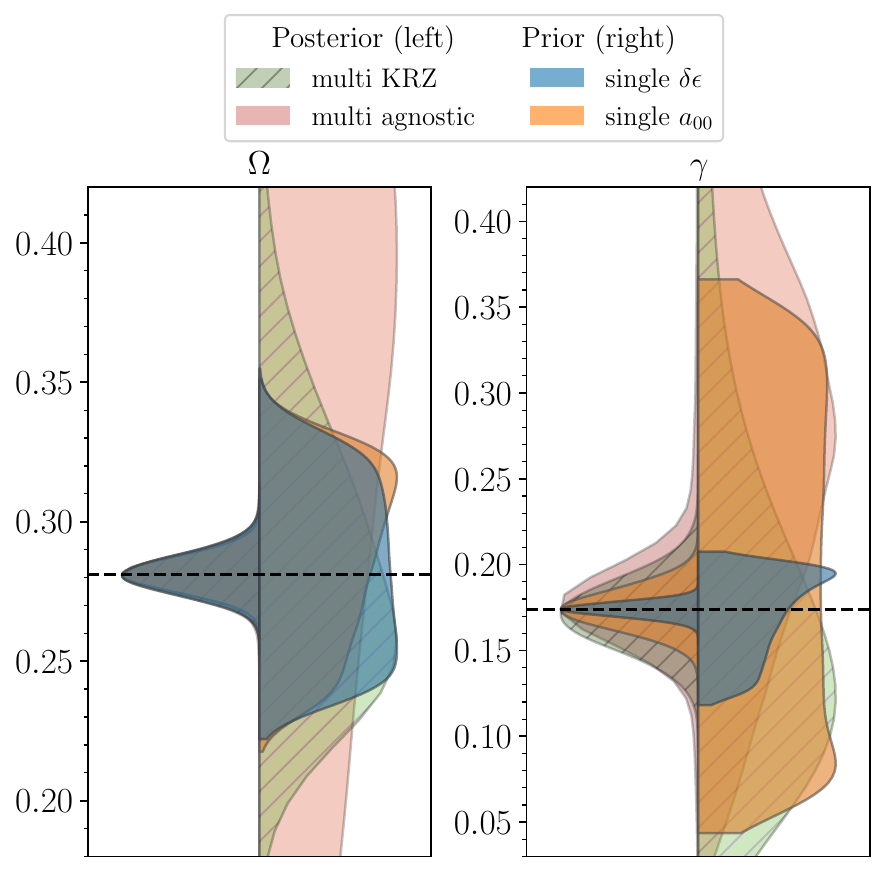}
\caption{Each panel shows Gaussian kernel density estimations (KDEs) for the posteriors (left) and priors (right) of the orbital frequency $\Omega$ and Lyapunov exponent $\gamma$ in the case of a GR injection with $M_0=1$ and $a_0=0.7$ (black dashed). 
Different cases correspond to the multi-parameter metric agnostic, and single- and multi-parameter KRZ analyses (probabilities not normalized), respectively. 
}
\label{fig:freq_lyap}
\end{figure}

\begin{figure*}[!t]
\centering
\includegraphics[width=1.0\linewidth]{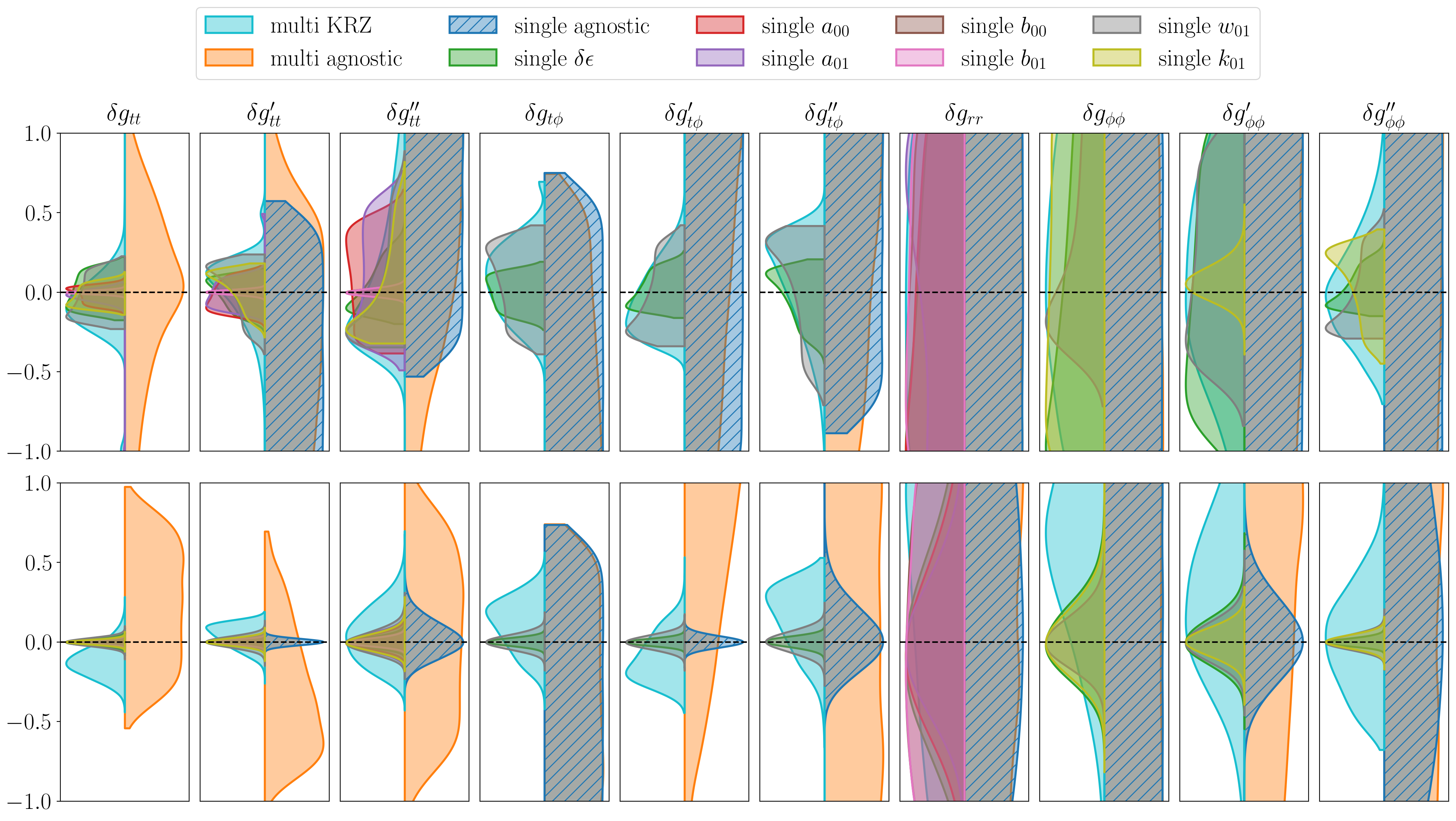}
\caption{Each panel shows metric shifts corresponding to the prior (top row) and posterior (bottom row) distributions for the single- and multi-parameter KRZ and metric agnostic analyses for the GR injection (probabilities not normalized). We plot the priors and posteriors over the smaller range $[-1,1]$ for clarity. The black dashed lines represent the corresponding GR values. Within each panel, the KRZ-derived distributions are shown on the left and the metric-agnostic distributions on the right; colors distinguish the single- and multi-parameter analyses.
}
\label{fig_gr}
\end{figure*}

\begin{figure*}[!t]
\centering
\includegraphics[width=1.0\linewidth]{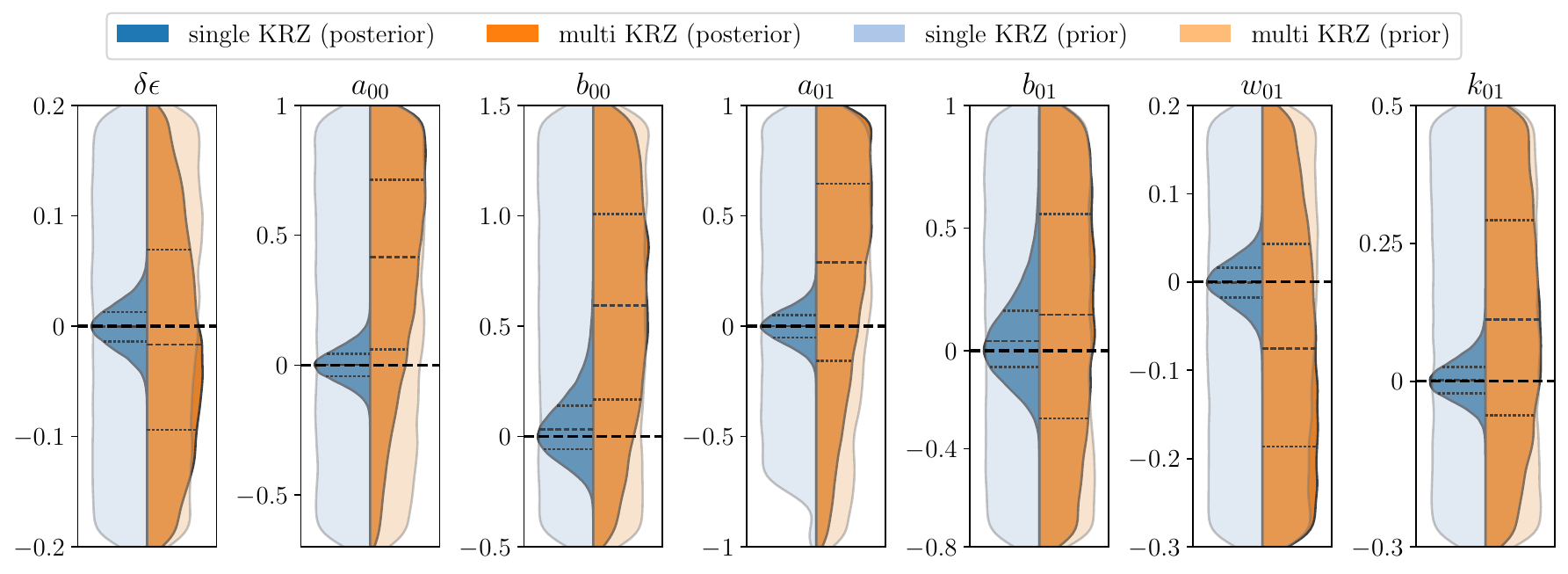}
\caption{Each panel shows the prior and posterior distributions of a KRZ parameter for the single-parameter (left violins) and multi-parameter (right violins) KRZ analyses of the GR injection (probabilities not normalized). The posterior distributions include the QNM information, whereas the prior distributions rely only on the adopted priors. The black dashed lines represent the corresponding GR values, while the black dotted lines identify the \(1/4\), \(1/2\), and \(3/4\) quantiles.
}
\label{GR_prior_KRZ}
\end{figure*}

Since the eikonal formulae depend on the orbital frequency $\Omega$ and Lyapunov exponent $\gamma$ of the prograde photon orbit, it is interesting to see how they are reconstructed from the MCMC samples of the KRZ parameters and the local metric-agnostic parameters.

The prior and posterior distributions of $\Omega$  and $\gamma$ are shown in Fig.~\ref{fig:freq_lyap} for four cases: all the considered KRZ parameters are varied simultaneously, all metric functions are varied simultaneously, and only the single deformation parameters $\delta \epsilon$ and $a_{00}$ are considered (as examples). 

First, it is evident that the priors in the multi-parameter cases are different from each other, and clearly less informative than the ones for the single-parameter cases. Moreover, the latter admit non-trivial shapes, whose maxima are not close to the GR value. This demonstrates the non-trivial mapping of uniform priors on KRZ coefficients to non-uniform prior on other observables. 

Second, comparing the posteriors for $\Omega$, it is remarkable that all four cases are quite similar. As should be expected from informative data in a noise-less injection, all the posterior distributions have their peak near the Kerr values. Minor differences are likely due to the drastically different priors. 
Considering the $\gamma$ posteriors, it is evident that the two multi-parameter posteriors are again very similar, while the single-parameter ones are narrower, especially the one for $\delta \epsilon$. Again, as should be expected, the respective peak locations are close to the injected value. 

In summary, one should expect that single-parameter tests may drastically underestimate the uncertainty in $\Omega$ or $\gamma$, while the two different multi-parameter cases are flexible enough to represent similar posterior distributions. 

Next, in Fig.~\ref{fig_gr}, we show the results for the shifts in the metric functions at the prograde photon orbit in the metric-specific and agnostic case. For $\delta g_{tt}$, we do not report the single-parameter agnostic analysis since this parameter has been expressed via Eq.~\eqref{eq:eq_orbital_1} in terms of the other metric shifts. The posterior distributions are shown in the bottom row. Even though the same range of values is used for all the metric functions, it is worth stressing that they have different units of $M$.

From the top panel of Fig.~\ref{fig_gr}, which shows the different prior distributions, one can clearly notice that the physical requirements on the metric functions discussed at the end of Sec.~\ref{method_krz} lead to nontrivial behaviors in the priors for $\delta g_{tt}$ and $\delta g_{t\phi}$. Other metric shifts such as $\delta g_{rr}$ and $\delta g_{\phi\phi}$ are not affected considerably since $g_{rr}$ and $ g_{\phi\phi}$ are significantly different from zero already in GR. Moreover, allowing all selected KRZ parameters to vary simultaneously (multi KRZ) leads to posteriors that are compatible with GR in both the theory-agnostic and theory-specific case. 

The comparison of top and bottom rows in Fig.~\ref{fig_gr} shows that the QNM information leads to drastic changes in the posteriors associated to the single KRZ parameters. Even though the priors often peak away from GR, they always include the GR values. Moreover, once the QNM information is added the posteriors are closely centered on GR. Varying multiple KRZ parameters simultaneously typically yields marginalized posteriors for individual KRZ parameters that are often closer to the GR value, but do not always have their maximum located at it. Similar considerations apply to the agnostic case for the one at a time metric shift (single agnostic) and when all the metric shifts are treated as free parameters (multi agnostic).

\begin{figure}[!t]
\centering
\includegraphics[width=1.0\linewidth]{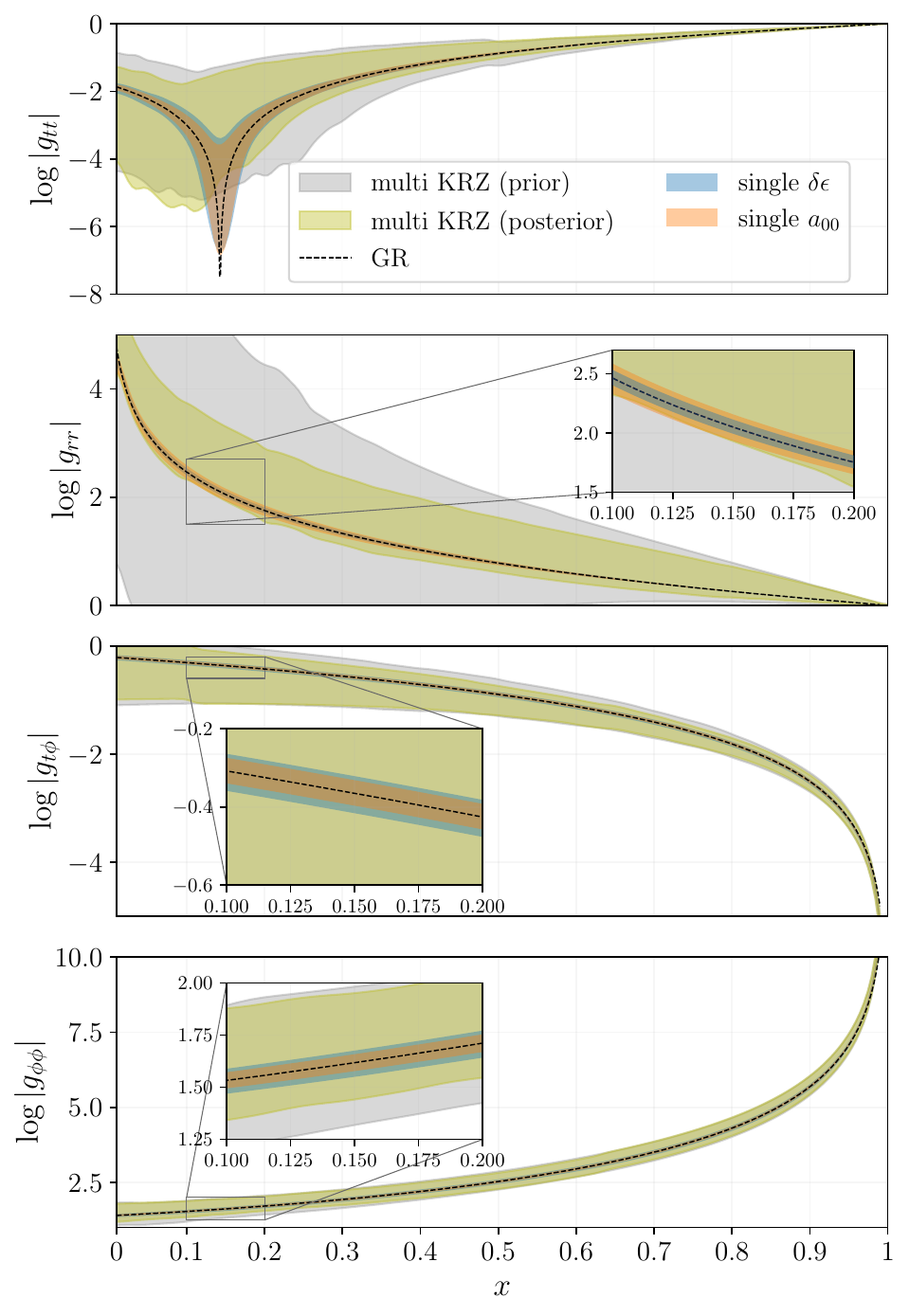}
\caption{$68\%$ HDIs of the equatorial KRZ metric functions reconstructed for the GR injection. The multi-parameter bands are obtained either from posterior samples including the 
QNM information (multi KRZ posterior) or from the KRZ priors alone (multi KRZ prior). The single-parameter bands show the posterior reconstructions obtained by varying only $\delta\epsilon$ or $a_{00}$. The black dashed lines represent the corresponding GR metric functions. The insets in the lower three panels magnify the region around the prograde photon orbit at $x_{\mathrm{ph}}=0.148$.
}
\label{fig_gr_hdi}
\end{figure}

Figure~\ref{GR_prior_KRZ} shows how much we learn on the KRZ parameters by including the information on the QNM shifts compared to the case where only priors are used. We show the single KRZ parameter cases (priors and posteriors) in the left violins, and the marginalized (priors and posteriors) when multiple KRZ parameters are being varied simultaneously in the right ones. The most striking difference is in the left violins for the one-dimensional case, where the QNM information leads to much more informative posteriors which are fully compatible with GR. As for the right violins, the changes are less pronounced since the number of free parameters is much larger than the number of QNM shifts available. Similarly, adding the QNM information associated to the retrograde modes can also affect the posterior distributions, especially in the multi KRZ case (see Appendix~\ref{appendix:retro}).

Finally, it is interesting to look at the reconstructed KRZ metric functions obtained by either changing all the selected KRZ parameters at the same time or only one at a time. Figure~\ref{fig_gr_hdi} shows the $68\%$ highest density intervals (HDIs) for the relevant metric functions in the single-parameter and multi-parameter KRZ cases. The HDIs are computed from the KRZ metric functions sampled from the respective KRZ posterior and evaluated over a grid of $200$ points for the radial compactified coordinate $x \in [0,1)$. For the one-at-a-time cases, we only report the HDIs for $\delta\epsilon$ and $a_{00}$ since the other cases are quite similar and do not add significant information.

Figure~\ref{fig_gr_hdi} also includes the results obtained only from the priors on the KRZ parameters in order to gauge the effect of the QNM information. This naturally leads to wider HDIs as can be clearly seen for the $g_{rr}$ function. Notice, however, that in the other cases the HDIs corresponding to the priors and all the KRZ parameters appear rather similar, thus the QNM information is not providing significantly stronger constraints on the metric functions. The results are consistent with the Kerr metric as the GR profile is always contained in the HDIs. In order to highlight the reconstructed metric near the photon sphere $x_{ph}\approx0.148$, three insets are added in the bottom row.

\subsection{Non-GR injection}\label{app_nonGR}

To isolate the effects of model and prior assumptions in a non-GR setting, we analyze the fundamental $\ell=m=2$ mode of a non-Kerr BH with $M_0=1$ and $a_0=0.7$, described by a nontrivial realization of the KRZ metric. 
Because the injected geometry is known, differences between the two inference strategies can be traced to the chosen parametrization, prior, and number of free deformation parameters.

\begin{figure}[!]
\centering
\includegraphics[width=1.0\linewidth]{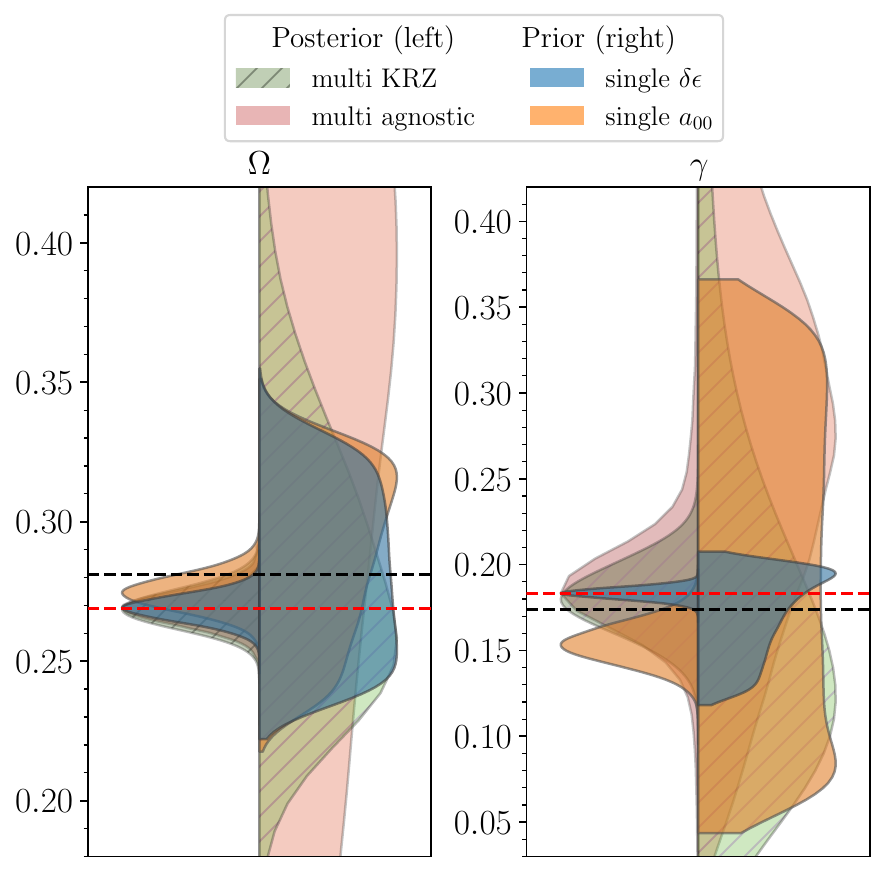}
\caption{Each panel shows Gaussian KDEs of the posterior (left) and prior (right) distributions of the orbital frequency $\Omega$ and Lyapunov exponent $\gamma$ for the non-GR injection. The curves correspond to the multi-parameter KRZ and metric-agnostic analyses and to the single-parameter KRZ analyses varying either $\delta\epsilon$ or $a_{00}$. The probability densities are not normalized. The horizontal dashed lines mark the non-GR injection (red) and the GR values (black).}
\label{fig:freq_lyap_INJ}
\end{figure}

\begin{figure*}[!]
\centering
\includegraphics[width=1.0\linewidth]{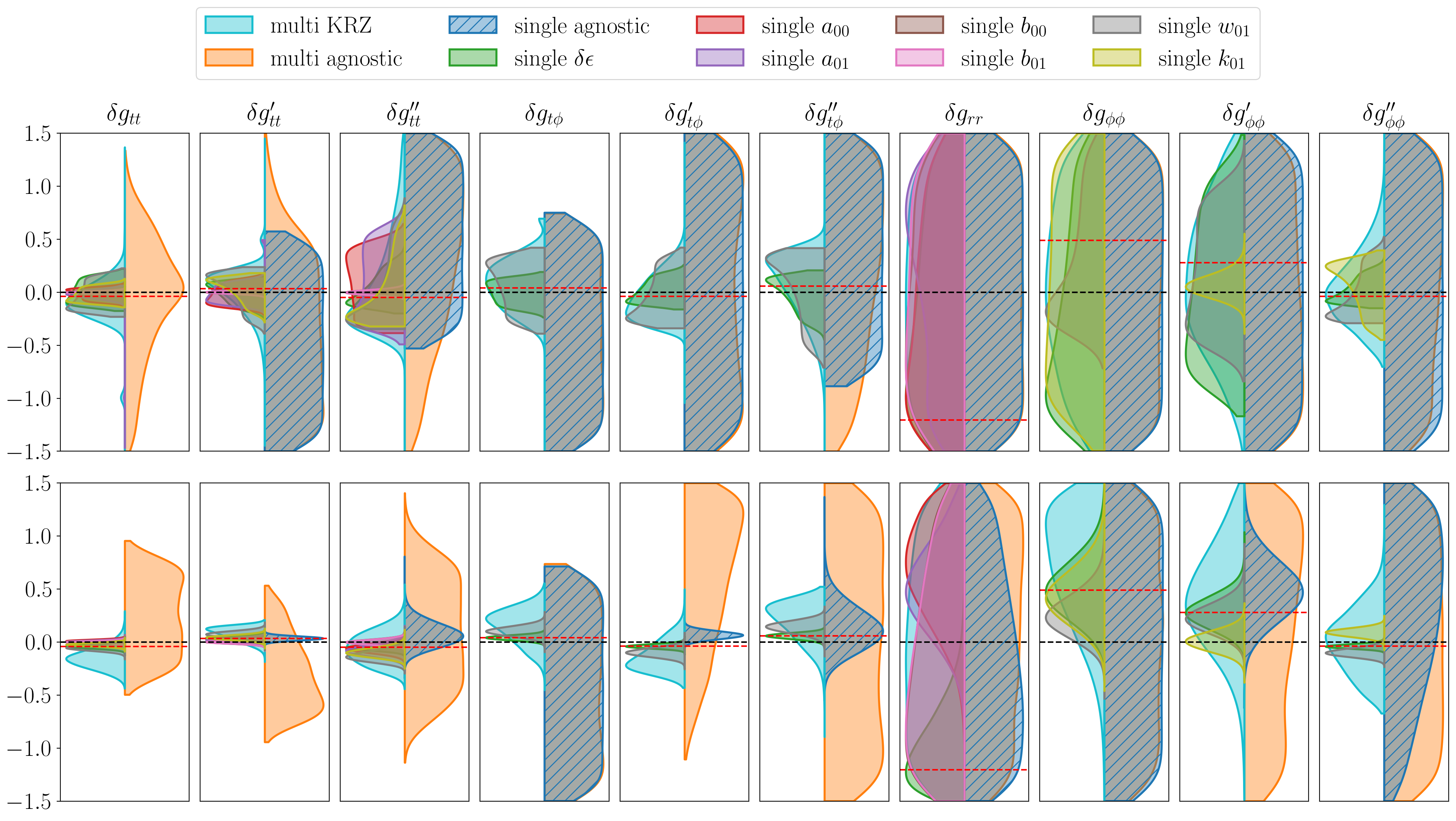}
\caption{Each panel shows metric shifts corresponding to the prior (top row) and posterior (bottom row) distributions for the single- and multi-parameter KRZ and metric-agnostic analyses for the non-GR injection (probabilities not normalized). The red dashed lines represent the injected metric shifts at the prograde photon orbit, while the black dashed lines represent the corresponding GR values. Within each panel, the KRZ-derived distributions are shown on the left and the metric-agnostic distributions on the right; colors distinguish the single- and multi-parameter analyses.
}
\label{fig_nonGR}
\end{figure*}

\begin{figure*}[!]
\centering
\includegraphics[width=1.0\linewidth]{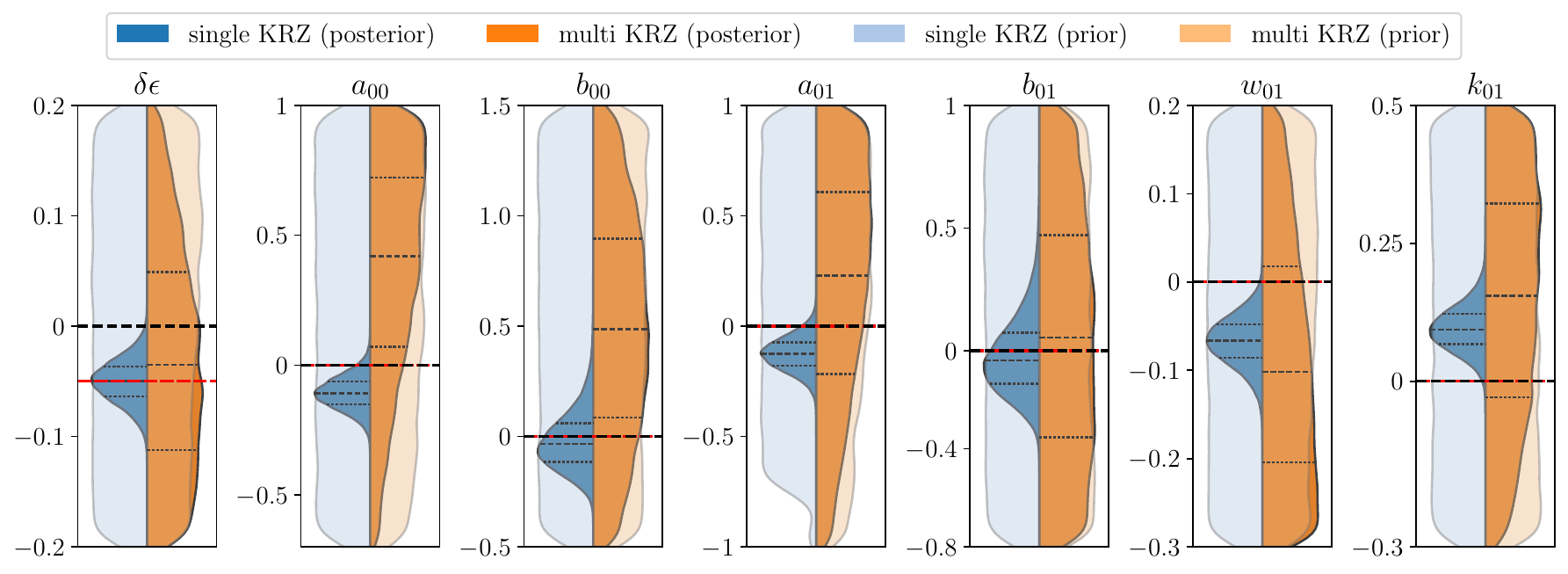}
\caption{Each panel shows the prior and posterior distributions of a KRZ parameter for the single-parameter (left violins) and multi-parameter (right violins) KRZ analyses of the non-GR injection (probabilities not normalized). The posterior distributions include the QNM information, whereas the prior distributions rely only on the adopted priors. The red dashed lines represent the injected non-GR values (only non-zero for $\delta \epsilon$), the black dashed lines represent the corresponding GR values, and the black dotted lines identify the $1/4$, $1/2$, and $3/4$ quantiles.}
\label{inj_prior_KRZ}
\end{figure*}

\begin{figure}[!t]
\centering
\includegraphics[width=1.0\linewidth]{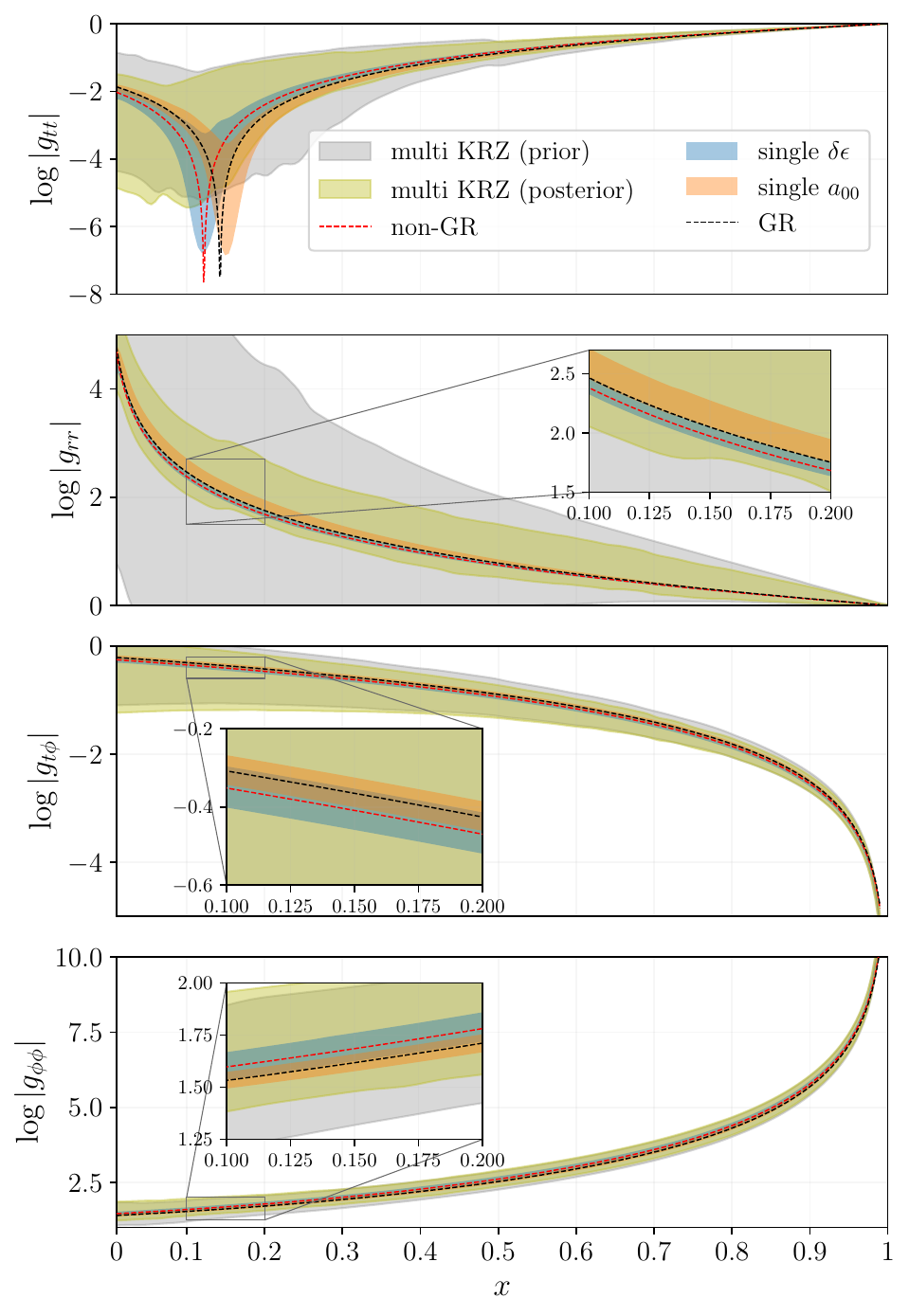}
\caption{$68\%$ HDIs of the equatorial KRZ metric functions reconstructed for the non-GR injection. The multi-parameter bands are obtained either from posterior samples including the QNM information (multi KRZ posterior) or from the KRZ priors alone (multi KRZ prior). The single-parameter bands show the posterior reconstructions obtained by varying only $\delta\epsilon$ or $a_{00}$. The black and red dashed lines represent the corresponding GR and injected non-GR metric functions, respectively. The insets in the lower three panels magnify the region around the prograde photon orbit at $x_{\mathrm{ph}}=0.162$.}
\label{fig_inj_hdi}
\end{figure}

The deformed Kerr BH is obtained by changing only the value of $\delta\epsilon=-0.05$, which more sensitively affects the properties of the metric functions near the horizon, while all the other KRZ parameters are set to zero. Moreover, $\delta\epsilon=-0.05$ affects all the metric functions, unlike other coefficients. This choice induces shifts of approximately $4\%$ in the orbital frequency $\Omega$ and $5\%$ in the Lyapunov exponent $\gamma$, relative to their GR values. As in the GR case, we adopt uncorrelated Gaussian priors on $\delta M$ and $\delta a$, centered on zero with standard deviation $\sigma=0.01$. The same conditions have been implemented for the uncertainty on the real and imaginary parts of the eikonal shifts in the log-likelihood.
Even in the non-GR case, the priors on the KRZ parameters include the conditions on the metric functions such that they preserve the metric signature, the sign of the determinant and the absence of closed time-like curves. Notice, however, that those criteria do not have to be satisfied in any theory of gravity beyond GR. 

As in the GR case, we show the prior and posterior distributions of the orbital frequency and Lyapunov exponent in Fig.~\ref{fig:freq_lyap_INJ} for the same type of cases. The posteriors obtained from varying multiple KRZ parameters or metric functions simultaneously appear very similar. This underlines that both models are flexible enough to represent similar posterior information. 

The posteriors obtained by varying only $\delta\epsilon$ are fully compatible with the injection, since the non-GR BH is obtained by varying that parameter. In contrast, the posterior associated to only varying $a_{00}$ is clearly biased, i.e., its posteriors for $\Omega$ and $\gamma$ do not peak at the injection (neither at its GR value). This is, in general, expected due to the choice of the non-GR injection and is compatible also with the posteriors on the single KRZ parameters shown in Fig.~\ref{inj_prior_KRZ}. Similar considerations apply to the other KRZ parameters (which are not shown in the figure to keep it concise). Furthermore, we found similar results when considering more general non-GR injections where multiple KRZ parameters are varied.

Importantly, the non-GR case highlights a crucial feature of single-parameter analyses. When the injection is generated from an arbitrary set of KRZ coefficients (or potentially a different BH metric), but recovered by varying a single coefficient, the induced prior may not support the injected values of $\Omega$ and $\gamma$, thus no value of the coefficient within its allowed range can recover them. This leads to biased posteriors and reconstructed observables that cannot be compatible with the injected values. A possible solution to this problem consists of enlarging the prior on the KRZ parameter, but it depends crucially on the parameter under consideration. Varying a single KRZ parameter, with strong priors on mass and spin, may not be able to explain the observed QNM. Moreover, the additional conditions imposed on the KRZ parameters to avoid changes in the metric signature, closed time-like curves, etc., have a different impact on each KRZ parameter. In particular, they can result in effective bounds on the single KRZ parameters that are independent of the flat prior range; see e.g., Ref.~\cite{Kocherlakota:2022mro}. Those considerations are especially important in the non-GR case, because the injected values are unknown and thus single-parameter studies can lead to biased results.

Violin plots for the metric-specific and agnostic shifts in the metric functions at the prograde photon sphere are shown in Fig.~\ref{fig_nonGR}. As in the GR case, there is a remarkable difference between the top and bottom rows in Fig.~\ref{fig_nonGR}, due to the addition of the QNM information. However, since the KRZ parameters deviate from GR, the one-at-a-time reconstruction is usually not able to recover the correct injected values, because a single deviation parameter can modify two observables simultaneously, as is the case here.

The violin plots for the KRZ parameters, comparing the posteriors obtained from the priors and the QNM shifts are shown in Fig.~\ref{inj_prior_KRZ}. As in the GR case, the most relevant differences are on the one-at-a-time violin plots (single KRZ (posterior), single KRZ (prior)). Even though the QNM information leads to much more informative posteriors on the KRZ parameters, they never reach the same level of agreement with the injected values as in Fig.~\ref{GR_prior_KRZ}. The remarkable property of Fig.~\ref{inj_prior_KRZ} is that, even though the non-GR injection is not recovered at the level of the KRZ parameter, some of the posteriors still show clear disagreement with the GR values. This is true also for the $w_{01}$ and $k_{01}$ parameters which are zero in the non-GR injection. In contrast, the posteriors of other parameters such as $b_{00}$ and $b_{01}$ are compatible with GR. The impact of retrograde QNMs is quite similar to the GR case, as discussed in Appendix~\ref{appendix:retro}.

As for the metric HDIs, in the non-GR case they appear very different from GR as shown in Fig.~\ref{fig_inj_hdi}. We report only the one-at-a-time reconstructions associated to the parameters $\delta\epsilon$ and $a_{00}$, since the other cases appear quite similar to $a_{00}$ and do not provide new information. The most relevant departure from GR is observed on the $g_{tt}$ function, which is not always recovered in the one-at-a-time reconstructions. In particular, the reconstructed $g_{tt}$ obtained by only varying $a_{00}$ (single $a_{00}$) is not in good agreement with the non-GR injection. The non-GR metric functions are instead contained in the HDIs when all the KRZ parameters are varied at the same time and when only the priors are taken into account. As in the GR case, three insets are added to highlight the spacetime near the prograde photon orbit located at $x_{ph}\approx0.162$. 


\subsection{GW250114}\label{app_GW}

Finally, we apply our framework to the existing gravitational-wave observation GW250114, one of the most suitable events for black hole spectroscopy. Assuming GR, the remnant BH can be well described by a Kerr BH with maximum-likelihood values for the dimensionless spin $\chi=a/M=0.68$ and redshifted mass $M=68.1 M_\odot$, and more than one QNM has been confidently extracted from the ringdown~\cite{LIGOScientific:2025wao,LIGOScientific:2025rid}.

\begin{figure}[ht]
\centering
\includegraphics[width=1.0\linewidth]{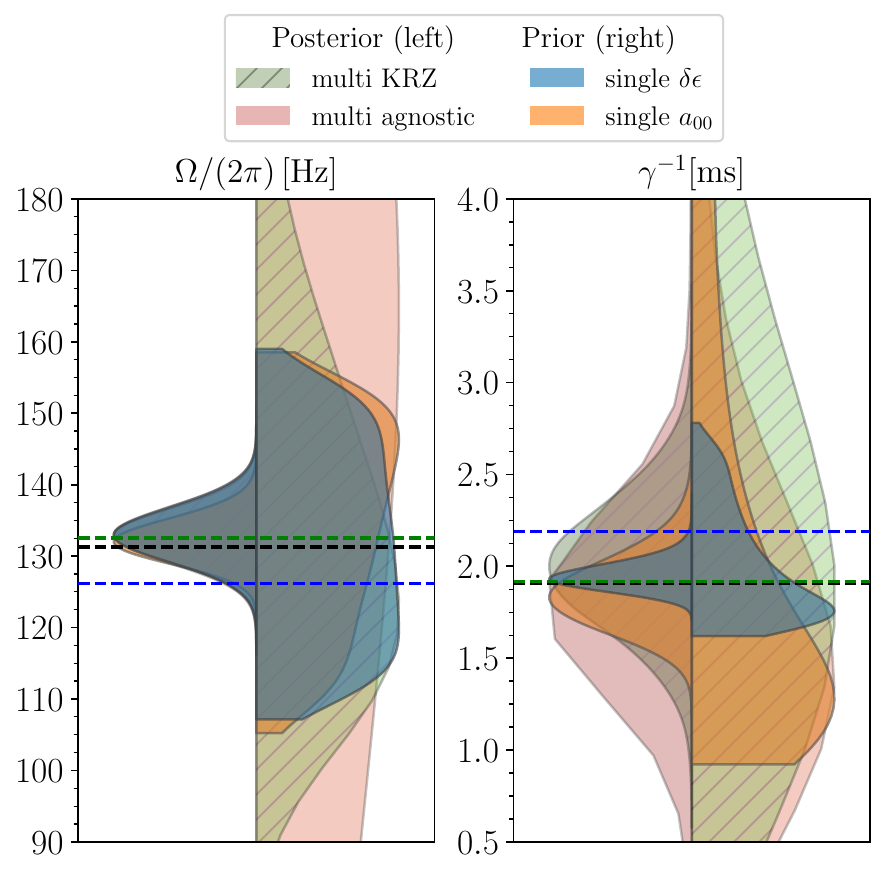}
\caption{In each panel, we report Gaussian KDEs for the posteriors (left) and priors (right) of the orbital frequency $\Omega/2 \pi$ and the time scale $1/\gamma$ computed for the KRZ and metric-agnostic analyses of GW250114. The horizontal black lines represent the corresponding maximum-likelihood GR values (from mass and spin) and the values of $\Omega$ and $1/\gamma$ in SI units. The blue line is obtained by applying the eikonal formulae to the mean values from the agnostic QNM analysis, and the green one from the eikonal formula using the mass and spin obtained by inverting the exact prediction for Kerr QNMs; see text for details.}
\label{fig:freq_lyap_GW}
\end{figure}

\begin{figure*}[ht]
\centering
\includegraphics[width=1.0\linewidth]{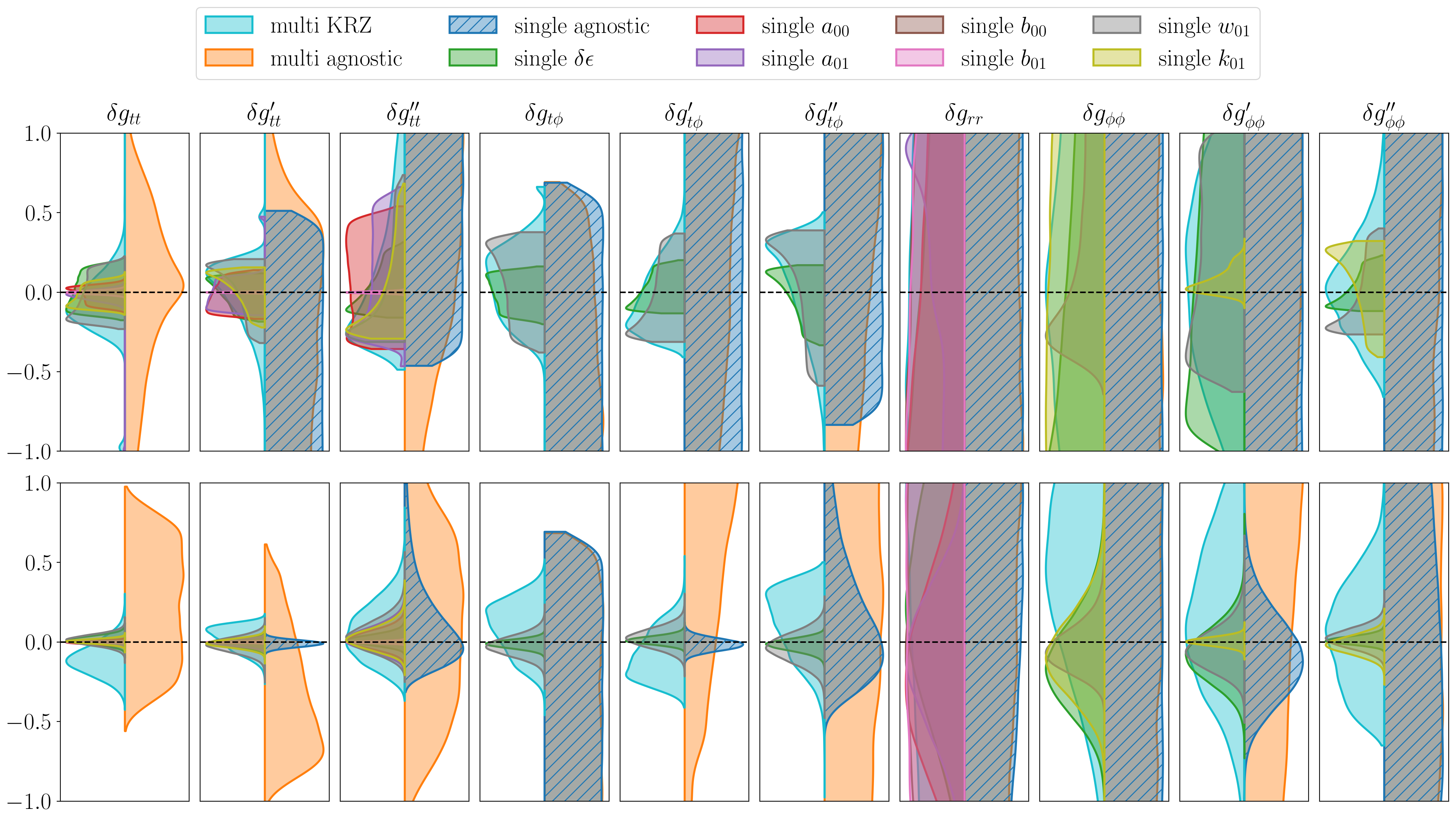}
\caption{Each panel shows metric shifts corresponding to the prior (top row) and posterior (bottom row) distributions for the single- and multi-parameter KRZ and metric-agnostic analyses of GW250114 (probabilities not normalized). We plot the priors and posteriors over the smaller range $[-1,1]$ for clarity. The black dashed lines represent the corresponding maximum-likelihood GR values (for mass and spin) at the prograde photon orbit. Within each panel, the KRZ-derived distributions are shown on the left and the metric-agnostic distributions on the right; colors distinguish the single- and multi-parameter analyses.}
\label{fig_GW}
\end{figure*}

\begin{figure*}[ht]
\centering
\includegraphics[width=1.0\linewidth]{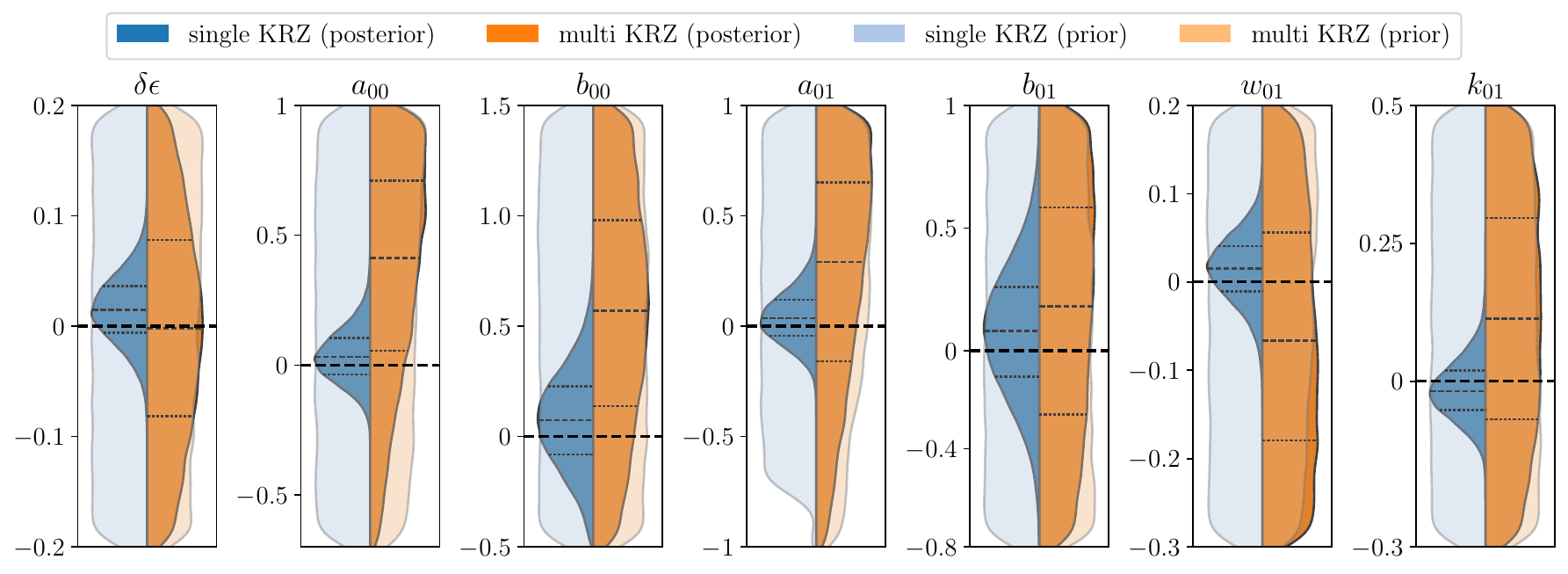}
\caption{Each panel shows the prior and posterior distributions of a KRZ parameter for the single-parameter (left violins) and multi-parameter (right violins) KRZ analyses of GW250114 (probabilities not normalized). The posterior distributions include the QNM information, whereas the prior distributions rely only on the adopted priors. The black dashed lines represent the corresponding maximum-likelihood GR values (for mass and spin), while the black dotted lines identify the $1/4$, $1/2$, and $3/4$ quantiles.}
\label{GW_prior_KRZ}
\end{figure*}

Utilizing this information in our framework, we provide the first eikonal test of the local light-ring geometry in the KRZ spacetime allowing multiple parameters to vary simultaneously. In particular, we use the publicly available posteriors of the fundamental $\ell=m=2$ QNM from the theory-agnostic analysis of two damped sinusoids starting at $t=10\,M$, and use the NRSur7dq4 samples from the full inspiral-merger-ringdown analysis to approximate the final mass and spin~\cite{LIGOScientific:2025wao} with multivariate Gaussians. 
We note that this approach is approximate, using the inspiral-merger-ringdown posterior information on mass and spin, and the agnostic ringdown analysis for the fundamental mode is not equivalent to a full Bayesian analysis of the entire event, which is beyond the scope of this application and left for future work. 
The data-analysis procedure adopted here follows closely Ref.~\cite{Volkel:2026qqz}, which we refer to for further details. 

The covariance matrix $\mathbf{\Sigma}_{Ma}$ for the priors on mass and spin is different from the GR/non-GR cases, and moreover, there is a nonzero correlation between the two parameters 
\begin{align}\label{sigma_Ma}
    \mathbf{\Sigma}_{Ma} = 
    \begin{pmatrix}
        2.57\cdot10^{-1} & 3.08\cdot10^{-3} \\
        3.08\cdot10^{-3}  & 4.27\cdot10^{-5}
    \end{pmatrix}\,,
\end{align}
expressed in units of solar masses and dimensionless spin. Accordingly, using the posteriors for the QNMs, the covariance matrix used in the log-likelihood for the QNM shifts is also different from the previous two cases, as it is obtained from GW250114. In terms of the frequency $f$ and the damping time $\tau$, which is the inverse of the magnitude of the QNM imaginary part, the covariance matrix
\begin{align}\label{sigma_QNM}
    \mathbf{\Sigma}_{f\tau} = 
    \begin{pmatrix}
        52.63 & 0.55 \\
        0.55  & 0.56
    \end{pmatrix}\,,
\end{align}
is in units of Hz and ms. The corresponding maximum-likelihood values obtained from the QNM chains are $f_{220}=252.3$ Hz and $\tau_{220}=4.37$ ms for the $\ell=m=2$ fundamental mode ($n=0$).

The prior and posterior distributions of the orbital frequency $\Omega$ and Lyapunov exponent $\gamma$ for GW250114 are shown in Fig.~\ref{fig:freq_lyap_GW}. Because of statistical fluctuations in the observed data, the true values are not expected to coincide exactly with the posterior maxima, but rather to lie within the corresponding credible regions. This is qualitatively different from our previous examples, which adopted noise-less injections. Besides this difference, the inferred posteriors are qualitatively similar to the GR case shown in Fig.~\ref{fig:freq_lyap} also when only the parameter $\delta\epsilon$ or $a_{00}$ is varied, consistent with the good agreement of GW250114 with GR. Even though the Kerr value of the Lyapunov exponent is consistent with the posterior distribution, the correlations and larger measurement errors clearly affect the shape of the posterior compared to the GR case. The effect is less pronounced for the orbital frequency, with broader priors in the GW250114 case. 

\begin{figure}
\centering
\includegraphics[width=1.0\linewidth]{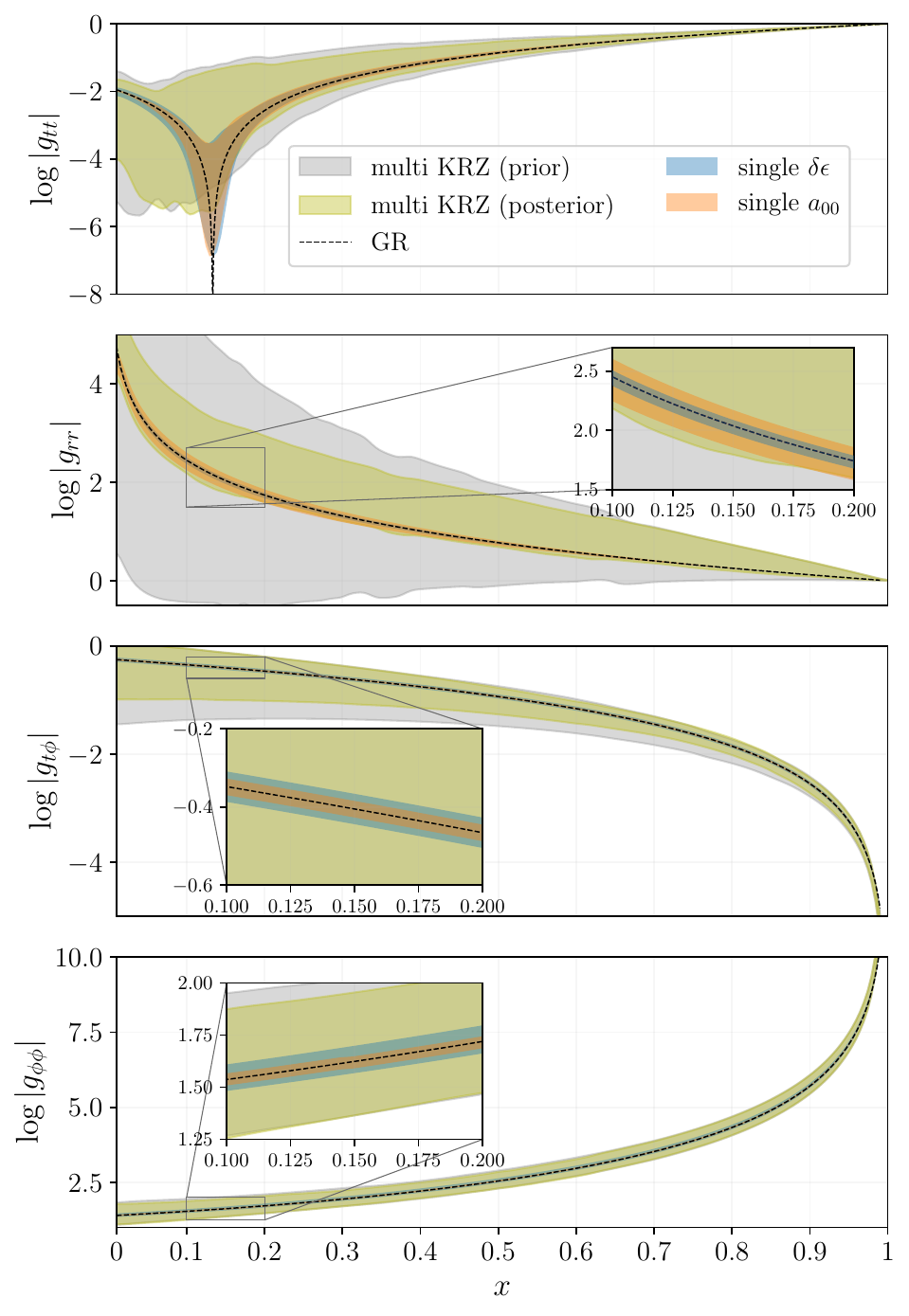}
\caption{$68\%$ HDIs of the equatorial KRZ metric functions reconstructed for GW250114. The multi-parameter bands are obtained either from posterior samples including the prograde QNM information (multi KRZ posterior) or from the KRZ priors alone (multi KRZ prior). The single-parameter bands show the posterior reconstructions obtained by varying only $\delta\epsilon$ or $a_{00}$. The black dashed lines represent the maximum-likelihood Kerr metric functions (for mass and spin).}
\label{fig_GW_hdi}
\end{figure}

We also show the values of $\Omega$ and $\gamma$ obtained by directly inverting the uncalibrated eikonal relation using the mean of the theory-agnostic posterior for the fundamental QNM frequency $\omega_{220}$. For the $\ell=m=2$ mode, this procedure produces a substantial bias in $\Omega$. The inferred value of $\gamma$ lies within the multi-parameter posteriors, although this agreement is largely due to their broader statistical uncertainties. This underlines the importance of using the GR calibration in the eikonal method. Moreover, the agreement with the GR value is expected to improve for larger $\ell$ since the eikonal approximation is more accurate~\cite{DeSimone:2026mkz}. Assuming GR, one can also map the mean of the theory-agnostic fundamental QNM posterior to the corresponding remnant mass and spin. Although these estimates differ slightly from those obtained from the full inspiral-merger-ringdown analysis, they are consistent within the statistical uncertainties. Using the resulting mass and spin to predict $\Omega$ and $\gamma$ yields slightly different values that remain well within our posterior credible intervals.

In Fig.~\ref{fig_GW}, we show the metric shifts in the metric-agnostic and metric-specific case. As expected, the top row looks qualitatively similar to Fig.~\ref{fig_gr}, since they only depend on the prior information on the KRZ parameters. As for the bottom rows of Fig.~\ref{fig_GW}, the GR case differs significantly from GW250114. This is mainly due to the choice of covariance matrix on the QNM shifts Eq.~\eqref{sigma_QNM}, which differs from the uncorrelated Gaussian priors used for the GR and non-GR case. 

Figure~\ref{GW_prior_KRZ} illustrates how adding QNM information constrains the KRZ parameters by comparing their prior and posterior distributions. All posterior constraints are consistent with GR. Since only the information on prograde modes is available for GW250114, we have not considered the effects of retrograde modes for this case. The excitation of those modes is relevant in binary systems where the spins of the progenitors are anti-aligned with the angular momentum~\cite{Berti:2005ys,Lim:2019xrb}.

To further compare the GW250114 case to GR, we also report in Fig.~\ref{fig_GW_hdi} the $68\%$ HDIs on the KRZ metric functions for the one at a time $\delta\epsilon$ and $a_{00}$, and all the KRZ parameters at the same time. Here, the BH mass is fixed to one and the spin to $\chi=0.68$ in such a way that it can be easily compared to Fig.~\ref{fig_gr_hdi}. The metric functions are always contained within the HDIs, even though the intervals are typically larger.

The present analysis demonstrates the feasibility of the framework under current observational conditions. However, its full potential will be realized with next-generation detectors~\cite{LISA:2022yao,ET:2025xjr}, whose substantially higher ringdown signal-to-noise ratios will enable significantly tighter constraints on both local geometric quantities and parametrized deviations from Kerr, as well as additional QNMs for the metric-specific analysis.

\section{Conclusions}\label{conclusions}

In this paper, we have developed a Bayesian framework based on the calibrated eikonal approximation for QNMs to infer the geometry of rotating BHs from ringdown observations in two types of approaches: global metric-specific and local metric-agnostic. The former is essential for testing gravitational theories and reconstructing known spacetime models; whereas the latter provides a more direct and less parametrization-dependent characterization of the local geometry probed by QNM observations. 

In our metric-specific analysis, the eikonal QNMs are computed using the KRZ parametrization, while in the agnostic approach, the eikonal QNMs have been expressed in terms of local metric functions and their radial derivatives evaluated at the equatorial photon orbit. The former provides a global representation of the inferred geometry, while the latter isolates the local information entering the eikonal correspondence without imposing a
particular metric away from the photon orbit. 

Furthermore, we have distinguished between two analyses: one where KRZ parameters or the metric functions are varied one at a time (single parameter), fixing all the others to zero (GR); one where all the considered parameters are allowed to vary at the same time (multi parameter). Within the calibrated eikonal framework, the orbital frequency $\Omega$ and Lyapunov exponent $\gamma$ are the two combinations entering directly into the complex QNM frequency. We therefore use them as likelihood-level quantities for comparing the metric-specific and metric-agnostic approaches. 

In Sec.~\ref{applications}, we applied our framework to three scenarios: a Kerr BH modeled via the KRZ parametrization assuming no shift in the QNMs (GR), a non-GR BH assuming a nontrivial realization of the KRZ parameters (non-GR), and an application to approximate posterior information from the gravitational wave event GW250114. Despite the different parametrizations and induced priors of the KRZ and metric-agnostic models, the multi-parameter analyses yield similar posteriors for these combinations. By contrast, the individual KRZ coefficients, local metric shifts, and reconstructed metric functions remain strongly degenerate and prior-dependent. Single-parameter analyses generally produce tighter constraints on $\Omega$ and $\gamma$, but these constraints are reliable only when the restricted model contains the underlying deformation. A similar hierarchy has previously been found at the level of parametrized perturbation potentials, where individual expansion coefficients were poorly identifiable while suitable combinations of the potential and its derivatives near the local maximum of the potential can be constrained more robustly~\cite{Volkel:2022aca,Volkel:2022khh}.

The GR case presented in Sec.~\ref{app_GR} mainly validated our setup, and importantly, served as a clean test bed for quantifying the importance of non-trivial priors, and the different behavior when varying only single parameters or multiple simultaneously. We show that uniform KRZ parameters correspond to non-trivial priors for the metric functions at the equatorial light ring. Moreover, requiring physical BH metrics implies non-trivial conditions that can be incorporated as rejections during Bayesian analyses, especially when multiple KRZ parameters are varied simultaneously. 
 
The non-GR example in Sec.~\ref{app_nonGR} is obtained by taking a nonzero value for the KRZ parameter $\epsilon$. Varying only single parameters at a time in the metric-specific case allows to recover the correct non-GR injection only for $\epsilon$. Otherwise, posteriors are in general biased. This shows the typical limitation of single-parameter analyses beyond null tests. When varying multiple parameters simultaneously, the marginalized posterior distributions are not guaranteed to show the correct injection, although they are usually wide enough to include it. 

For GW250114~\cite{LIGOScientific:2025wao,LIGOScientific:2025rid}, in Sec.~\ref{app_GW}, we follow the procedure introduced in Ref.~\cite{Volkel:2026qqz}, approximating the publicly released ringdown posterior for the QNM frequency and the inspiral-merger-ringdown posterior for the remnant mass and spin by multivariate Gaussian distributions. While Ref.~\cite{Volkel:2026qqz} used this approximation to constrain parametrized modifications of the Teukolsky potential~\cite{Cano:2024jkd}, here we use it to infer the corresponding KRZ coefficients, local metric shifts, and reconstructed metric functions. Within this approximation and the calibrated eikonal framework, all inferred quantities are consistent with their Kerr reference values, and we find no evidence for a GR deviation.

Across our applications, we have stressed the relation between the priors on the KRZ parameters and the corresponding metric shifts. Because of the nonlinear mapping between them, uniform priors on the KRZ parameters can induce metric-shift distributions that are strongly skewed away from GR. Assessing these induced priors is therefore essential for distinguishing prior-driven structures from QNM information. Aside from the metric shifts at the prograde photon orbit, we have also studied the KRZ parameters themselves by comparing posteriors including the QNM information with the corresponding prior distributions. We have further investigated the effect of adding a retrograde mode with the same assumed uncertainty as the prograde mode in Appendix~\ref{appendix:retro}. Finally, we reconstructed the equatorial metric functions using pointwise $68\%$ HDIs. The multi-parameter intervals contain the corresponding GR or non-GR profiles but remain broad, while single-parameter reconstructions of the non-GR injection can be unreliable. For GW250114, the reconstructed metric functions are consistent with the maximum-likelihood Kerr reference.

Our methodology can be extended in several ways. First, it could be used to test GR with higher multipoles $(\ell,m)$, that are expected to be available with the next-generation of GW detectors such as LISA~\cite{Berti:2016lat,Barausse:2020rsu, LISA:2022yao,LISA:2024hlh}. Since the eikonal approximation improves significantly for larger $\ell$, this will pave the way to more precise tests of GR. It can also be used to quantify the intrinsic inaccuracy of the eikonal approximation and include it in the statistical analysis. Notice, however, that since all equatorial eikonal QNMs are connected to the same geometric properties, additional $\ell=m$ modes do not provide new geometric information. Nonetheless, including multiple equatorial modes will in general lead to better constraints on the orbital frequency and Lyapunov exponent. 

Higher multipoles could still be used to test the consistency of the eikonal constraints and, alternatively, non-equatorial QNMs provide an additional source of information, since they are connected with the properties of spherical photon orbits around the BH~\cite{Teo:2020sey,Yang:2012he,Pappas:2018opz}. Those modes could also be modeled by a post-Kerr-like approximation~\cite{Glampedakis:2017dvb} in order to study spacetimes that deviate perturbatively from the Kerr BH. Notice however that, in both the metric-specific and agnostic approaches, additional parameters would be needed to model non-equatorial orbits.

The connection between metric and QNMs can be used to study compact objects beyond GR via test scalar perturbations~\cite{Pani:2026yzi}, and by combining techniques from frequency and time domain~\cite{DeSimone:2026mkz}. The same methodology can also be applied to other parametrized metrics~\cite{Johannsen:2013szh, Cardoso:2014rha}, which might be more suitable to include information from electromagnetic or gravitational perturbations. In Ref.~\cite{Suvorov:2021amy}, it was shown how specific theories can be constructed, such that they obey a given RZ metric as a solution, which then allows, in principle, for a consistent computation of QNMs from the gravitational perturbation equations. 

In addition to QNMs, it would be interesting to complement this study with information from GW inspiral~\cite{Shashank:2021giy}, for instance by modeling equatorial time-like orbits in modified spacetimes~\cite{Dai:2026ujj}, which are relevant for the study of Extreme-mass-ratio inspirals (EMRIs)~\cite{Amaro-Seoane:2007osp}. Similarly, other phenomena from the electromagnetic sector such as X-rays~\cite{Yu:2021xen, Ni:2016uik,Choudhury:2018zmf}, BH shadows~\cite{Johannsen:2013vgc,Psaltis:2015uza,Younsi:2016azx,Volkel:2020xlc,Lara:2021zth,Kocherlakota:2022jnz,EventHorizonTelescope:2022xqj} and BH jets~\cite{Pei:2016kka,Camilloni:2026irg} can also be implemented in this line of research to characterize other regions of the metric and set stronger constraints on the KRZ parameters. 

Finally, our results highlight that the inverse problem of BH spectroscopy can be formulated at different levels: one may seek to reconstruct a specific spacetime model or, alternatively, infer only the local geometric properties directly supported by the observations. The present work demonstrates that these two viewpoints are complementary, that multi-parameter tests are crucial for robust results beyond null tests, and should be pursued together in future analyses.

~\\
\acknowledgments
C.\,D.\,S. is grateful to the University of T\"ubingen for hospitality during the realization of this work. C.\,D.\,S. acknowledges the support of INFN, {\it sez. di Napoli}, {\it iniziativa specifica}  QGSKY. K.\,D.\,K. and S.\,H.\,V. acknowledge funding from the European Union’s Horizon MSCA-2022 research and innovation programme “EinsteinWaves” under grant agreement No. 101131233. The authors acknowledge the use of OpenAI's GPT-5.6 Sol and GPT-6 Astra to assist in providing feedback on the clarity, structure, and presentation of the manuscript.

\bibliography{literature}

\begin{thebibliography}{102}%
\makeatletter
\providecommand \@ifxundefined [1]{%
 \@ifx{#1\undefined}
}%
\providecommand \@ifnum [1]{%
 \ifnum #1\expandafter \@firstoftwo
 \else \expandafter \@secondoftwo
 \fi
}%
\providecommand \@ifx [1]{%
 \ifx #1\expandafter \@firstoftwo
 \else \expandafter \@secondoftwo
 \fi
}%
\providecommand \natexlab [1]{#1}%
\providecommand \enquote  [1]{``#1''}%
\providecommand \bibnamefont  [1]{#1}%
\providecommand \bibfnamefont [1]{#1}%
\providecommand \citenamefont [1]{#1}%
\providecommand \href@noop [0]{\@secondoftwo}%
\providecommand \href [0]{\begingroup \@sanitize@url \@href}%
\providecommand \@href[1]{\@@startlink{#1}\@@href}%
\providecommand \@@href[1]{\endgroup#1\@@endlink}%
\providecommand \@sanitize@url [0]{\catcode `\\12\catcode `\$12\catcode
  `\&12\catcode `\#12\catcode `\^12\catcode `\_12\catcode `\%12\relax}%
\providecommand \@@startlink[1]{}%
\providecommand \@@endlink[0]{}%
\providecommand \url  [0]{\begingroup\@sanitize@url \@url }%
\providecommand \@url [1]{\endgroup\@href {#1}{\urlprefix }}%
\providecommand \urlprefix  [0]{URL }%
\providecommand \Eprint [0]{\href }%
\providecommand \doibase [0]{https://doi.org/}%
\providecommand \selectlanguage [0]{\@gobble}%
\providecommand \bibinfo  [0]{\@secondoftwo}%
\providecommand \bibfield  [0]{\@secondoftwo}%
\providecommand \translation [1]{[#1]}%
\providecommand \BibitemOpen [0]{}%
\providecommand \bibitemStop [0]{}%
\providecommand \bibitemNoStop [0]{.\EOS\space}%
\providecommand \EOS [0]{\spacefactor3000\relax}%
\providecommand \BibitemShut  [1]{\csname bibitem#1\endcsname}%
\let\auto@bib@innerbib\@empty
\bibitem [{\citenamefont {Abbott}\ \emph {et~al.}(2016)\citenamefont {Abbott}
  \emph {et~al.}}]{LIGOScientific:2016aoc}%
  \BibitemOpen
  \bibfield  {author} {\bibinfo {author} {\bibfnamefont {B.~P.}\ \bibnamefont
  {Abbott}} \emph {et~al.} (\bibinfo {collaboration} {LIGO Scientific,
  Virgo}),\ }\bibfield  {title} {\bibinfo {title} {{Observation of
  Gravitational Waves from a Binary Black Hole Merger}},\ }\href
  {https://doi.org/10.1103/PhysRevLett.116.061102} {\bibfield  {journal}
  {\bibinfo  {journal} {Phys. Rev. Lett.}\ }\textbf {\bibinfo {volume} {116}},\
  \bibinfo {pages} {061102} (\bibinfo {year} {2016})},\ \Eprint
  {https://arxiv.org/abs/1602.03837} {arXiv:1602.03837 [gr-qc]} \BibitemShut
  {NoStop}%
\bibitem [{\citenamefont {Kerr}(1963)}]{Kerr:1963ud}%
  \BibitemOpen
  \bibfield  {author} {\bibinfo {author} {\bibfnamefont {R.~P.}\ \bibnamefont
  {Kerr}},\ }\bibfield  {title} {\bibinfo {title} {{Gravitational field of a
  spinning mass as an example of algebraically special metrics}},\ }\href
  {https://doi.org/10.1103/PhysRevLett.11.237} {\bibfield  {journal} {\bibinfo
  {journal} {Phys. Rev. Lett.}\ }\textbf {\bibinfo {volume} {11}},\ \bibinfo
  {pages} {237} (\bibinfo {year} {1963})}\BibitemShut {NoStop}%
\bibitem [{\citenamefont {Regge}\ and\ \citenamefont
  {Wheeler}(1957)}]{Regge:1957td}%
  \BibitemOpen
  \bibfield  {author} {\bibinfo {author} {\bibfnamefont {T.}~\bibnamefont
  {Regge}}\ and\ \bibinfo {author} {\bibfnamefont {J.~A.}\ \bibnamefont
  {Wheeler}},\ }\bibfield  {title} {\bibinfo {title} {{Stability of a
  Schwarzschild singularity}},\ }\href
  {https://doi.org/10.1103/PhysRev.108.1063} {\bibfield  {journal} {\bibinfo
  {journal} {Phys. Rev.}\ }\textbf {\bibinfo {volume} {108}},\ \bibinfo {pages}
  {1063} (\bibinfo {year} {1957})}\BibitemShut {NoStop}%
\bibitem [{\citenamefont {Zerilli}(1970)}]{Zerilli:1970se}%
  \BibitemOpen
  \bibfield  {author} {\bibinfo {author} {\bibfnamefont {F.~J.}\ \bibnamefont
  {Zerilli}},\ }\bibfield  {title} {\bibinfo {title} {{Effective potential for
  even parity Regge-Wheeler gravitational perturbation equations}},\ }\href
  {https://doi.org/10.1103/PhysRevLett.24.737} {\bibfield  {journal} {\bibinfo
  {journal} {Phys. Rev. Lett.}\ }\textbf {\bibinfo {volume} {24}},\ \bibinfo
  {pages} {737} (\bibinfo {year} {1970})}\BibitemShut {NoStop}%
\bibitem [{\citenamefont {Teukolsky}(1972)}]{Teukolsky1972}%
  \BibitemOpen
  \bibfield  {author} {\bibinfo {author} {\bibfnamefont {S.~A.}\ \bibnamefont
  {Teukolsky}},\ }\bibfield  {title} {\bibinfo {title} {{Rotating black holes -
  separable wave equations for gravitational and electromagnetic
  perturbations}},\ }\href {https://doi.org/10.1103/PhysRevLett.29.1114}
  {\bibfield  {journal} {\bibinfo  {journal} {Phys. Rev. Lett.}\ }\textbf
  {\bibinfo {volume} {29}},\ \bibinfo {pages} {1114} (\bibinfo {year}
  {1972})}\BibitemShut {NoStop}%
\bibitem [{\citenamefont {Vishveshwara}(1970)}]{Vishveshwara:1970zz}%
  \BibitemOpen
  \bibfield  {author} {\bibinfo {author} {\bibfnamefont {C.~V.}\ \bibnamefont
  {Vishveshwara}},\ }\bibfield  {title} {\bibinfo {title} {{Scattering of
  Gravitational Radiation by a Schwarzschild Black-hole}},\ }\href
  {https://doi.org/10.1038/227936a0} {\bibfield  {journal} {\bibinfo  {journal}
  {Nature}\ }\textbf {\bibinfo {volume} {227}},\ \bibinfo {pages} {936}
  (\bibinfo {year} {1970})}\BibitemShut {NoStop}%
\bibitem [{\citenamefont {Chandrasekhar}\ and\ \citenamefont
  {Detweiler}(1975)}]{Chandrasekhar:1975zza}%
  \BibitemOpen
  \bibfield  {author} {\bibinfo {author} {\bibfnamefont {S.}~\bibnamefont
  {Chandrasekhar}}\ and\ \bibinfo {author} {\bibfnamefont {S.~L.}\ \bibnamefont
  {Detweiler}},\ }\bibfield  {title} {\bibinfo {title} {{The quasi-normal modes
  of the Schwarzschild black hole}},\ }\href
  {https://doi.org/10.1098/rspa.1975.0112} {\bibfield  {journal} {\bibinfo
  {journal} {Proc. Roy. Soc. Lond. A}\ }\textbf {\bibinfo {volume} {344}},\
  \bibinfo {pages} {441} (\bibinfo {year} {1975})}\BibitemShut {NoStop}%
\bibitem [{\citenamefont {Leaver}(1985)}]{Leaver1985}%
  \BibitemOpen
  \bibfield  {author} {\bibinfo {author} {\bibfnamefont {E.~W.}\ \bibnamefont
  {Leaver}},\ }\bibfield  {title} {\bibinfo {title} {{An Analytic
  representation for the quasi normal modes of Kerr black holes}},\ }\href
  {https://doi.org/10.1098/rspa.1985.0119} {\bibfield  {journal} {\bibinfo
  {journal} {Proc. Roy. Soc. Lond. A}\ }\textbf {\bibinfo {volume} {402}},\
  \bibinfo {pages} {285} (\bibinfo {year} {1985})}\BibitemShut {NoStop}%
\bibitem [{\citenamefont {Chandrasekhar}(1985)}]{Chandrasekhar:1985kt}%
  \BibitemOpen
  \bibfield  {author} {\bibinfo {author} {\bibfnamefont {S.}~\bibnamefont
  {Chandrasekhar}},\ }\href@noop {} {\emph {\bibinfo {title} {{The mathematical
  theory of black holes}}}}\ (\bibinfo {year} {1985})\BibitemShut {NoStop}%
\bibitem [{\citenamefont {Kokkotas}\ and\ \citenamefont
  {Schmidt}(1999)}]{Kokkotas:1999bd}%
  \BibitemOpen
  \bibfield  {author} {\bibinfo {author} {\bibfnamefont {K.~D.}\ \bibnamefont
  {Kokkotas}}\ and\ \bibinfo {author} {\bibfnamefont {B.~G.}\ \bibnamefont
  {Schmidt}},\ }\bibfield  {title} {\bibinfo {title} {{Quasinormal modes of
  stars and black holes}},\ }\href {https://doi.org/10.12942/lrr-1999-2}
  {\bibfield  {journal} {\bibinfo  {journal} {Living Rev. Rel.}\ }\textbf
  {\bibinfo {volume} {2}},\ \bibinfo {pages} {2} (\bibinfo {year} {1999})},\
  \Eprint {https://arxiv.org/abs/gr-qc/9909058} {arXiv:gr-qc/9909058}
  \BibitemShut {NoStop}%
\bibitem [{\citenamefont {Berti}\ \emph {et~al.}(2009)\citenamefont {Berti},
  \citenamefont {Cardoso},\ and\ \citenamefont {Starinets}}]{Berti:2009kk}%
  \BibitemOpen
  \bibfield  {author} {\bibinfo {author} {\bibfnamefont {E.}~\bibnamefont
  {Berti}}, \bibinfo {author} {\bibfnamefont {V.}~\bibnamefont {Cardoso}},\
  and\ \bibinfo {author} {\bibfnamefont {A.~O.}\ \bibnamefont {Starinets}},\
  }\bibfield  {title} {\bibinfo {title} {{Quasinormal modes of black holes and
  black branes}},\ }\href {https://doi.org/10.1088/0264-9381/26/16/163001}
  {\bibfield  {journal} {\bibinfo  {journal} {Class. Quant. Grav.}\ }\textbf
  {\bibinfo {volume} {26}},\ \bibinfo {pages} {163001} (\bibinfo {year}
  {2009})},\ \Eprint {https://arxiv.org/abs/0905.2975} {arXiv:0905.2975
  [gr-qc]} \BibitemShut {NoStop}%
\bibitem [{\citenamefont {Konoplya}\ and\ \citenamefont
  {Zhidenko}(2011)}]{Konoplya:2011qq}%
  \BibitemOpen
  \bibfield  {author} {\bibinfo {author} {\bibfnamefont {R.~A.}\ \bibnamefont
  {Konoplya}}\ and\ \bibinfo {author} {\bibfnamefont {A.}~\bibnamefont
  {Zhidenko}},\ }\bibfield  {title} {\bibinfo {title} {{Quasinormal modes of
  black holes: From astrophysics to string theory}},\ }\href
  {https://doi.org/10.1103/RevModPhys.83.793} {\bibfield  {journal} {\bibinfo
  {journal} {Rev. Mod. Phys.}\ }\textbf {\bibinfo {volume} {83}},\ \bibinfo
  {pages} {793} (\bibinfo {year} {2011})},\ \Eprint
  {https://arxiv.org/abs/1102.4014} {arXiv:1102.4014 [gr-qc]} \BibitemShut
  {NoStop}%
\bibitem [{\citenamefont {Israel}(1967)}]{Israel:1967wq}%
  \BibitemOpen
  \bibfield  {author} {\bibinfo {author} {\bibfnamefont {W.}~\bibnamefont
  {Israel}},\ }\bibfield  {title} {\bibinfo {title} {{Event horizons in static
  vacuum space-times}},\ }\href {https://doi.org/10.1103/PhysRev.164.1776}
  {\bibfield  {journal} {\bibinfo  {journal} {Phys. Rev.}\ }\textbf {\bibinfo
  {volume} {164}},\ \bibinfo {pages} {1776} (\bibinfo {year}
  {1967})}\BibitemShut {NoStop}%
\bibitem [{\citenamefont {Carter}(1971)}]{Carter:1971zc}%
  \BibitemOpen
  \bibfield  {author} {\bibinfo {author} {\bibfnamefont {B.}~\bibnamefont
  {Carter}},\ }\bibfield  {title} {\bibinfo {title} {{Axisymmetric Black Hole
  Has Only Two Degrees of Freedom}},\ }\href
  {https://doi.org/10.1103/PhysRevLett.26.331} {\bibfield  {journal} {\bibinfo
  {journal} {Phys. Rev. Lett.}\ }\textbf {\bibinfo {volume} {26}},\ \bibinfo
  {pages} {331} (\bibinfo {year} {1971})}\BibitemShut {NoStop}%
\bibitem [{\citenamefont {Hawking}(1972)}]{Hawking:1971vc}%
  \BibitemOpen
  \bibfield  {author} {\bibinfo {author} {\bibfnamefont {S.~W.}\ \bibnamefont
  {Hawking}},\ }\bibfield  {title} {\bibinfo {title} {{Black holes in general
  relativity}},\ }\href {https://doi.org/10.1007/BF01877517} {\bibfield
  {journal} {\bibinfo  {journal} {Commun. Math. Phys.}\ }\textbf {\bibinfo
  {volume} {25}},\ \bibinfo {pages} {152} (\bibinfo {year} {1972})}\BibitemShut
  {NoStop}%
\bibitem [{\citenamefont {Robinson}(1975)}]{Robinson:1975bv}%
  \BibitemOpen
  \bibfield  {author} {\bibinfo {author} {\bibfnamefont {D.~C.}\ \bibnamefont
  {Robinson}},\ }\bibfield  {title} {\bibinfo {title} {{Uniqueness of the Kerr
  black hole}},\ }\href {https://doi.org/10.1103/PhysRevLett.34.905} {\bibfield
   {journal} {\bibinfo  {journal} {Phys. Rev. Lett.}\ }\textbf {\bibinfo
  {volume} {34}},\ \bibinfo {pages} {905} (\bibinfo {year} {1975})}\BibitemShut
  {NoStop}%
\bibitem [{\citenamefont {Detweiler}(1980)}]{Detweiler:1980gk}%
  \BibitemOpen
  \bibfield  {author} {\bibinfo {author} {\bibfnamefont {S.~L.}\ \bibnamefont
  {Detweiler}},\ }\bibfield  {title} {\bibinfo {title} {{Black Holes and
  Gravitational Waves. III. The Resonant Frequencies of Rotating Holes}},\
  }\href {https://doi.org/10.1086/158109} {\bibfield  {journal} {\bibinfo
  {journal} {Astrophys. J.}\ }\textbf {\bibinfo {volume} {239}},\ \bibinfo
  {pages} {292} (\bibinfo {year} {1980})}\BibitemShut {NoStop}%
\bibitem [{\citenamefont {Dreyer}\ \emph {et~al.}(2004)\citenamefont {Dreyer},
  \citenamefont {Kelly}, \citenamefont {Krishnan}, \citenamefont {Finn},
  \citenamefont {Garrison},\ and\ \citenamefont
  {Lopez-Aleman}}]{Dreyer:2003bv}%
  \BibitemOpen
  \bibfield  {author} {\bibinfo {author} {\bibfnamefont {O.}~\bibnamefont
  {Dreyer}}, \bibinfo {author} {\bibfnamefont {B.~J.}\ \bibnamefont {Kelly}},
  \bibinfo {author} {\bibfnamefont {B.}~\bibnamefont {Krishnan}}, \bibinfo
  {author} {\bibfnamefont {L.~S.}\ \bibnamefont {Finn}}, \bibinfo {author}
  {\bibfnamefont {D.}~\bibnamefont {Garrison}},\ and\ \bibinfo {author}
  {\bibfnamefont {R.}~\bibnamefont {Lopez-Aleman}},\ }\bibfield  {title}
  {\bibinfo {title} {{Black hole spectroscopy: Testing general relativity
  through gravitational wave observations}},\ }\href
  {https://doi.org/10.1088/0264-9381/21/4/003} {\bibfield  {journal} {\bibinfo
  {journal} {Class. Quant. Grav.}\ }\textbf {\bibinfo {volume} {21}},\ \bibinfo
  {pages} {787} (\bibinfo {year} {2004})},\ \Eprint
  {https://arxiv.org/abs/gr-qc/0309007} {arXiv:gr-qc/0309007} \BibitemShut
  {NoStop}%
\bibitem [{\citenamefont {Berti}\ \emph {et~al.}(2006)\citenamefont {Berti},
  \citenamefont {Cardoso},\ and\ \citenamefont {Will}}]{Berti:2005ys}%
  \BibitemOpen
  \bibfield  {author} {\bibinfo {author} {\bibfnamefont {E.}~\bibnamefont
  {Berti}}, \bibinfo {author} {\bibfnamefont {V.}~\bibnamefont {Cardoso}},\
  and\ \bibinfo {author} {\bibfnamefont {C.~M.}\ \bibnamefont {Will}},\
  }\bibfield  {title} {\bibinfo {title} {{On gravitational-wave spectroscopy of
  massive black holes with the space interferometer LISA}},\ }\href
  {https://doi.org/10.1103/PhysRevD.73.064030} {\bibfield  {journal} {\bibinfo
  {journal} {Phys. Rev. D}\ }\textbf {\bibinfo {volume} {73}},\ \bibinfo
  {pages} {064030} (\bibinfo {year} {2006})},\ \Eprint
  {https://arxiv.org/abs/gr-qc/0512160} {arXiv:gr-qc/0512160} \BibitemShut
  {NoStop}%
\bibitem [{\citenamefont {Cardoso}\ and\ \citenamefont
  {Gualtieri}(2016)}]{Cardoso:2016ryw}%
  \BibitemOpen
  \bibfield  {author} {\bibinfo {author} {\bibfnamefont {V.}~\bibnamefont
  {Cardoso}}\ and\ \bibinfo {author} {\bibfnamefont {L.}~\bibnamefont
  {Gualtieri}},\ }\bibfield  {title} {\bibinfo {title} {{Testing the black hole
  {\textquoteleft}no-hair{\textquoteright} hypothesis}},\ }\href
  {https://doi.org/10.1088/0264-9381/33/17/174001} {\bibfield  {journal}
  {\bibinfo  {journal} {Class. Quant. Grav.}\ }\textbf {\bibinfo {volume}
  {33}},\ \bibinfo {pages} {174001} (\bibinfo {year} {2016})},\ \Eprint
  {https://arxiv.org/abs/1607.03133} {arXiv:1607.03133 [gr-qc]} \BibitemShut
  {NoStop}%
\bibitem [{\citenamefont {Franchini}\ and\ \citenamefont
  {V{\"o}lkel}(2024)}]{Franchini:2023eda}%
  \BibitemOpen
  \bibfield  {author} {\bibinfo {author} {\bibfnamefont {N.}~\bibnamefont
  {Franchini}}\ and\ \bibinfo {author} {\bibfnamefont {S.~H.}\ \bibnamefont
  {V{\"o}lkel}},\ }\bibinfo {title} {{Testing General Relativity with Black
  Hole Quasi-normal Modes}}\ (\bibinfo {year} {2024})\ \Eprint
  {https://arxiv.org/abs/2305.01696} {arXiv:2305.01696 [gr-qc]} \BibitemShut
  {NoStop}%
\bibitem [{\citenamefont {Berti}\ \emph {et~al.}(2025)\citenamefont {Berti}
  \emph {et~al.}}]{Berti:2025hly}%
  \BibitemOpen
  \bibfield  {author} {\bibinfo {author} {\bibfnamefont {E.}~\bibnamefont
  {Berti}} \emph {et~al.},\ }\bibfield  {title} {\bibinfo {title} {{Black hole
  spectroscopy: from theory to experiment}},\ }\href@noop {} {\  (\bibinfo
  {year} {2025})},\ \Eprint {https://arxiv.org/abs/2505.23895}
  {arXiv:2505.23895 [gr-qc]} \BibitemShut {NoStop}%
\bibitem [{\citenamefont {{Goebel}}(1972)}]{1972ApJ...172L..95G}%
  \BibitemOpen
  \bibfield  {author} {\bibinfo {author} {\bibfnamefont {C.}~\bibnamefont
  {{Goebel}}},\ }\bibfield  {title} {\bibinfo {title} {{Comments on the
  ``vibrations'' of a Black Hole}},\ }\href {https://doi.org/10.1086/180898}
  {\bibfield  {journal} {\bibinfo  {journal} {Astrophysical Journal Letters}\
  }\textbf {\bibinfo {volume} {172}},\ \bibinfo {pages} {L95} (\bibinfo {year}
  {1972})}\BibitemShut {NoStop}%
\bibitem [{\citenamefont {Ferrari}\ and\ \citenamefont
  {Mashhoon}(1984)}]{Ferrari:1984zz}%
  \BibitemOpen
  \bibfield  {author} {\bibinfo {author} {\bibfnamefont {V.}~\bibnamefont
  {Ferrari}}\ and\ \bibinfo {author} {\bibfnamefont {B.}~\bibnamefont
  {Mashhoon}},\ }\bibfield  {title} {\bibinfo {title} {{New approach to the
  quasinormal modes of a black hole}},\ }\href
  {https://doi.org/10.1103/PhysRevD.30.295} {\bibfield  {journal} {\bibinfo
  {journal} {Phys. Rev. D}\ }\textbf {\bibinfo {volume} {30}},\ \bibinfo
  {pages} {295} (\bibinfo {year} {1984})}\BibitemShut {NoStop}%
\bibitem [{\citenamefont {Cardoso}\ \emph {et~al.}(2009)\citenamefont
  {Cardoso}, \citenamefont {Miranda}, \citenamefont {Berti}, \citenamefont
  {Witek},\ and\ \citenamefont {Zanchin}}]{Cardoso:2008bp}%
  \BibitemOpen
  \bibfield  {author} {\bibinfo {author} {\bibfnamefont {V.}~\bibnamefont
  {Cardoso}}, \bibinfo {author} {\bibfnamefont {A.~S.}\ \bibnamefont
  {Miranda}}, \bibinfo {author} {\bibfnamefont {E.}~\bibnamefont {Berti}},
  \bibinfo {author} {\bibfnamefont {H.}~\bibnamefont {Witek}},\ and\ \bibinfo
  {author} {\bibfnamefont {V.~T.}\ \bibnamefont {Zanchin}},\ }\bibfield
  {title} {\bibinfo {title} {{Geodesic stability, Lyapunov exponents and
  quasinormal modes}},\ }\href {https://doi.org/10.1103/PhysRevD.79.064016}
  {\bibfield  {journal} {\bibinfo  {journal} {Phys. Rev. D}\ }\textbf {\bibinfo
  {volume} {79}},\ \bibinfo {pages} {064016} (\bibinfo {year} {2009})},\
  \Eprint {https://arxiv.org/abs/0812.1806} {arXiv:0812.1806 [hep-th]}
  \BibitemShut {NoStop}%
\bibitem [{\citenamefont {Dolan}(2010)}]{Dolan:2010wr}%
  \BibitemOpen
  \bibfield  {author} {\bibinfo {author} {\bibfnamefont {S.~R.}\ \bibnamefont
  {Dolan}},\ }\bibfield  {title} {\bibinfo {title} {{The Quasinormal Mode
  Spectrum of a Kerr Black Hole in the Eikonal Limit}},\ }\href
  {https://doi.org/10.1103/PhysRevD.82.104003} {\bibfield  {journal} {\bibinfo
  {journal} {Phys. Rev. D}\ }\textbf {\bibinfo {volume} {82}},\ \bibinfo
  {pages} {104003} (\bibinfo {year} {2010})},\ \Eprint
  {https://arxiv.org/abs/1007.5097} {arXiv:1007.5097 [gr-qc]} \BibitemShut
  {NoStop}%
\bibitem [{\citenamefont {Yang}\ \emph {et~al.}(2012)\citenamefont {Yang},
  \citenamefont {Nichols}, \citenamefont {Zhang}, \citenamefont {Zimmerman},
  \citenamefont {Zhang},\ and\ \citenamefont {Chen}}]{Yang:2012he}%
  \BibitemOpen
  \bibfield  {author} {\bibinfo {author} {\bibfnamefont {H.}~\bibnamefont
  {Yang}}, \bibinfo {author} {\bibfnamefont {D.~A.}\ \bibnamefont {Nichols}},
  \bibinfo {author} {\bibfnamefont {F.}~\bibnamefont {Zhang}}, \bibinfo
  {author} {\bibfnamefont {A.}~\bibnamefont {Zimmerman}}, \bibinfo {author}
  {\bibfnamefont {Z.}~\bibnamefont {Zhang}},\ and\ \bibinfo {author}
  {\bibfnamefont {Y.}~\bibnamefont {Chen}},\ }\bibfield  {title} {\bibinfo
  {title} {{Quasinormal-mode spectrum of Kerr black holes and its geometric
  interpretation}},\ }\href {https://doi.org/10.1103/PhysRevD.86.104006}
  {\bibfield  {journal} {\bibinfo  {journal} {Phys. Rev. D}\ }\textbf {\bibinfo
  {volume} {86}},\ \bibinfo {pages} {104006} (\bibinfo {year} {2012})},\
  \Eprint {https://arxiv.org/abs/1207.4253} {arXiv:1207.4253 [gr-qc]}
  \BibitemShut {NoStop}%
\bibitem [{\citenamefont {Konoplya}\ and\ \citenamefont
  {Stuchl{\'\i}k}(2017)}]{Konoplya:2017wot}%
  \BibitemOpen
  \bibfield  {author} {\bibinfo {author} {\bibfnamefont {R.~A.}\ \bibnamefont
  {Konoplya}}\ and\ \bibinfo {author} {\bibfnamefont {Z.}~\bibnamefont
  {Stuchl{\'\i}k}},\ }\bibfield  {title} {\bibinfo {title} {{Are eikonal
  quasinormal modes linked to the unstable circular null geodesics?}},\ }\href
  {https://doi.org/10.1016/j.physletb.2017.06.015} {\bibfield  {journal}
  {\bibinfo  {journal} {Phys. Lett. B}\ }\textbf {\bibinfo {volume} {771}},\
  \bibinfo {pages} {597} (\bibinfo {year} {2017})},\ \Eprint
  {https://arxiv.org/abs/1705.05928} {arXiv:1705.05928 [gr-qc]} \BibitemShut
  {NoStop}%
\bibitem [{\citenamefont {Konoplya}(2023)}]{Konoplya:2022gjp}%
  \BibitemOpen
  \bibfield  {author} {\bibinfo {author} {\bibfnamefont {R.~A.}\ \bibnamefont
  {Konoplya}},\ }\bibfield  {title} {\bibinfo {title} {{Further clarification
  on quasinormal modes/circular null geodesics correspondence}},\ }\href
  {https://doi.org/10.1016/j.physletb.2023.137674} {\bibfield  {journal}
  {\bibinfo  {journal} {Phys. Lett. B}\ }\textbf {\bibinfo {volume} {838}},\
  \bibinfo {pages} {137674} (\bibinfo {year} {2023})},\ \Eprint
  {https://arxiv.org/abs/2210.08373} {arXiv:2210.08373 [gr-qc]} \BibitemShut
  {NoStop}%
\bibitem [{\citenamefont {Chen}\ \emph {et~al.}(2023)\citenamefont {Chen},
  \citenamefont {Chen}, \citenamefont {Ho},\ and\ \citenamefont
  {Tseng}}]{Chen:2022nlw}%
  \BibitemOpen
  \bibfield  {author} {\bibinfo {author} {\bibfnamefont {C.-Y.}\ \bibnamefont
  {Chen}}, \bibinfo {author} {\bibfnamefont {Y.-J.}\ \bibnamefont {Chen}},
  \bibinfo {author} {\bibfnamefont {M.-Y.}\ \bibnamefont {Ho}},\ and\ \bibinfo
  {author} {\bibfnamefont {Y.-H.}\ \bibnamefont {Tseng}},\ }\bibfield  {title}
  {\bibinfo {title} {{A novel test of gravity via black hole eikonal
  correspondence}},\ }\href {https://doi.org/10.1016/j.physletb.2023.138153}
  {\bibfield  {journal} {\bibinfo  {journal} {Phys. Lett. B}\ }\textbf
  {\bibinfo {volume} {845}},\ \bibinfo {pages} {138153} (\bibinfo {year}
  {2023})},\ \Eprint {https://arxiv.org/abs/2212.10028} {arXiv:2212.10028
  [gr-qc]} \BibitemShut {NoStop}%
\bibitem [{\citenamefont {Khanna}\ and\ \citenamefont
  {Price}(2017)}]{Khanna:2016yow}%
  \BibitemOpen
  \bibfield  {author} {\bibinfo {author} {\bibfnamefont {G.}~\bibnamefont
  {Khanna}}\ and\ \bibinfo {author} {\bibfnamefont {R.~H.}\ \bibnamefont
  {Price}},\ }\bibfield  {title} {\bibinfo {title} {{Black Hole Ringing,
  Quasinormal Modes, and Light Rings}},\ }\href
  {https://doi.org/10.1103/PhysRevD.95.081501} {\bibfield  {journal} {\bibinfo
  {journal} {Phys. Rev. D}\ }\textbf {\bibinfo {volume} {95}},\ \bibinfo
  {pages} {081501} (\bibinfo {year} {2017})},\ \Eprint
  {https://arxiv.org/abs/1609.00083} {arXiv:1609.00083 [gr-qc]} \BibitemShut
  {NoStop}%
\bibitem [{\citenamefont {Johannsen}(2013{\natexlab{a}})}]{Johannsen:2013szh}%
  \BibitemOpen
  \bibfield  {author} {\bibinfo {author} {\bibfnamefont {T.}~\bibnamefont
  {Johannsen}},\ }\bibfield  {title} {\bibinfo {title} {{Regular Black Hole
  Metric with Three Constants of Motion}},\ }\href
  {https://doi.org/10.1103/PhysRevD.88.044002} {\bibfield  {journal} {\bibinfo
  {journal} {Phys. Rev. D}\ }\textbf {\bibinfo {volume} {88}},\ \bibinfo
  {pages} {044002} (\bibinfo {year} {2013}{\natexlab{a}})},\ \Eprint
  {https://arxiv.org/abs/1501.02809} {arXiv:1501.02809 [gr-qc]} \BibitemShut
  {NoStop}%
\bibitem [{\citenamefont {Johannsen}\ and\ \citenamefont
  {Psaltis}(2011)}]{Johannsen:2011dh}%
  \BibitemOpen
  \bibfield  {author} {\bibinfo {author} {\bibfnamefont {T.}~\bibnamefont
  {Johannsen}}\ and\ \bibinfo {author} {\bibfnamefont {D.}~\bibnamefont
  {Psaltis}},\ }\bibfield  {title} {\bibinfo {title} {{A Metric for Rapidly
  Spinning Black Holes Suitable for Strong-Field Tests of the No-Hair
  Theorem}},\ }\href {https://doi.org/10.1103/PhysRevD.83.124015} {\bibfield
  {journal} {\bibinfo  {journal} {Phys. Rev. D}\ }\textbf {\bibinfo {volume}
  {83}},\ \bibinfo {pages} {124015} (\bibinfo {year} {2011})},\ \Eprint
  {https://arxiv.org/abs/1105.3191} {arXiv:1105.3191 [gr-qc]} \BibitemShut
  {NoStop}%
\bibitem [{\citenamefont {Cardoso}\ \emph {et~al.}(2014)\citenamefont
  {Cardoso}, \citenamefont {Pani},\ and\ \citenamefont
  {Rico}}]{Cardoso:2014rha}%
  \BibitemOpen
  \bibfield  {author} {\bibinfo {author} {\bibfnamefont {V.}~\bibnamefont
  {Cardoso}}, \bibinfo {author} {\bibfnamefont {P.}~\bibnamefont {Pani}},\ and\
  \bibinfo {author} {\bibfnamefont {J.}~\bibnamefont {Rico}},\ }\bibfield
  {title} {\bibinfo {title} {{On generic parametrizations of spinning
  black-hole geometries}},\ }\href {https://doi.org/10.1103/PhysRevD.89.064007}
  {\bibfield  {journal} {\bibinfo  {journal} {Phys. Rev. D}\ }\textbf {\bibinfo
  {volume} {89}},\ \bibinfo {pages} {064007} (\bibinfo {year} {2014})},\
  \Eprint {https://arxiv.org/abs/1401.0528} {arXiv:1401.0528 [gr-qc]}
  \BibitemShut {NoStop}%
\bibitem [{\citenamefont {Del~Piano}\ \emph
  {et~al.}(2024{\natexlab{a}})\citenamefont {Del~Piano}, \citenamefont
  {Hohenegger},\ and\ \citenamefont {Sannino}}]{DelPiano:2023fiw}%
  \BibitemOpen
  \bibfield  {author} {\bibinfo {author} {\bibfnamefont {M.}~\bibnamefont
  {Del~Piano}}, \bibinfo {author} {\bibfnamefont {S.}~\bibnamefont
  {Hohenegger}},\ and\ \bibinfo {author} {\bibfnamefont {F.}~\bibnamefont
  {Sannino}},\ }\bibfield  {title} {\bibinfo {title} {{Quantum black hole
  physics from the event horizon}},\ }\href
  {https://doi.org/10.1103/PhysRevD.109.024045} {\bibfield  {journal} {\bibinfo
   {journal} {Phys. Rev. D}\ }\textbf {\bibinfo {volume} {109}},\ \bibinfo
  {pages} {024045} (\bibinfo {year} {2024}{\natexlab{a}})},\ \Eprint
  {https://arxiv.org/abs/2307.13489} {arXiv:2307.13489 [gr-qc]} \BibitemShut
  {NoStop}%
\bibitem [{\citenamefont {Del~Piano}\ \emph
  {et~al.}(2024{\natexlab{b}})\citenamefont {Del~Piano}, \citenamefont
  {Hohenegger},\ and\ \citenamefont {Sannino}}]{DelPiano:2024gvw}%
  \BibitemOpen
  \bibfield  {author} {\bibinfo {author} {\bibfnamefont {M.}~\bibnamefont
  {Del~Piano}}, \bibinfo {author} {\bibfnamefont {S.}~\bibnamefont
  {Hohenegger}},\ and\ \bibinfo {author} {\bibfnamefont {F.}~\bibnamefont
  {Sannino}},\ }\bibfield  {title} {\bibinfo {title} {{Effective metric
  descriptions of quantum black holes}},\ }\href
  {https://doi.org/10.1140/epjc/s10052-024-13609-5} {\bibfield  {journal}
  {\bibinfo  {journal} {Eur. Phys. J. C}\ }\textbf {\bibinfo {volume} {84}},\
  \bibinfo {pages} {1273} (\bibinfo {year} {2024}{\natexlab{b}})},\ \Eprint
  {https://arxiv.org/abs/2403.12679} {arXiv:2403.12679 [gr-qc]} \BibitemShut
  {NoStop}%
\bibitem [{\citenamefont {Collins}\ and\ \citenamefont
  {Hughes}(2004)}]{Collins:2004ex}%
  \BibitemOpen
  \bibfield  {author} {\bibinfo {author} {\bibfnamefont {N.~A.}\ \bibnamefont
  {Collins}}\ and\ \bibinfo {author} {\bibfnamefont {S.~A.}\ \bibnamefont
  {Hughes}},\ }\bibfield  {title} {\bibinfo {title} {{Towards a formalism for
  mapping the space-times of massive compact objects: Bumpy black holes and
  their orbits}},\ }\href {https://doi.org/10.1103/PhysRevD.69.124022}
  {\bibfield  {journal} {\bibinfo  {journal} {Phys. Rev. D}\ }\textbf {\bibinfo
  {volume} {69}},\ \bibinfo {pages} {124022} (\bibinfo {year} {2004})},\
  \Eprint {https://arxiv.org/abs/gr-qc/0402063} {arXiv:gr-qc/0402063}
  \BibitemShut {NoStop}%
\bibitem [{\citenamefont {Vigeland}\ and\ \citenamefont
  {Hughes}(2010)}]{Vigeland:2009pr}%
  \BibitemOpen
  \bibfield  {author} {\bibinfo {author} {\bibfnamefont {S.~J.}\ \bibnamefont
  {Vigeland}}\ and\ \bibinfo {author} {\bibfnamefont {S.~A.}\ \bibnamefont
  {Hughes}},\ }\bibfield  {title} {\bibinfo {title} {{Spacetime and orbits of
  bumpy black holes}},\ }\href {https://doi.org/10.1103/PhysRevD.81.024030}
  {\bibfield  {journal} {\bibinfo  {journal} {Phys. Rev. D}\ }\textbf {\bibinfo
  {volume} {81}},\ \bibinfo {pages} {024030} (\bibinfo {year} {2010})},\
  \Eprint {https://arxiv.org/abs/0911.1756} {arXiv:0911.1756 [gr-qc]}
  \BibitemShut {NoStop}%
\bibitem [{\citenamefont {Rezzolla}\ and\ \citenamefont
  {Zhidenko}(2014)}]{Rezzolla:2014mua}%
  \BibitemOpen
  \bibfield  {author} {\bibinfo {author} {\bibfnamefont {L.}~\bibnamefont
  {Rezzolla}}\ and\ \bibinfo {author} {\bibfnamefont {A.}~\bibnamefont
  {Zhidenko}},\ }\bibfield  {title} {\bibinfo {title} {{New parametrization for
  spherically symmetric black holes in metric theories of gravity}},\ }\href
  {https://doi.org/10.1103/PhysRevD.90.084009} {\bibfield  {journal} {\bibinfo
  {journal} {Phys. Rev. D}\ }\textbf {\bibinfo {volume} {90}},\ \bibinfo
  {pages} {084009} (\bibinfo {year} {2014})},\ \Eprint
  {https://arxiv.org/abs/1407.3086} {arXiv:1407.3086 [gr-qc]} \BibitemShut
  {NoStop}%
\bibitem [{\citenamefont {Konoplya}\ \emph {et~al.}(2016)\citenamefont
  {Konoplya}, \citenamefont {Rezzolla},\ and\ \citenamefont
  {Zhidenko}}]{Konoplya:2016jvv}%
  \BibitemOpen
  \bibfield  {author} {\bibinfo {author} {\bibfnamefont {R.}~\bibnamefont
  {Konoplya}}, \bibinfo {author} {\bibfnamefont {L.}~\bibnamefont {Rezzolla}},\
  and\ \bibinfo {author} {\bibfnamefont {A.}~\bibnamefont {Zhidenko}},\
  }\bibfield  {title} {\bibinfo {title} {{General parametrization of
  axisymmetric black holes in metric theories of gravity}},\ }\href
  {https://doi.org/10.1103/PhysRevD.93.064015} {\bibfield  {journal} {\bibinfo
  {journal} {Phys. Rev. D}\ }\textbf {\bibinfo {volume} {93}},\ \bibinfo
  {pages} {064015} (\bibinfo {year} {2016})},\ \Eprint
  {https://arxiv.org/abs/1602.02378} {arXiv:1602.02378 [gr-qc]} \BibitemShut
  {NoStop}%
\bibitem [{\citenamefont {Konoplya}\ \emph {et~al.}(2018)\citenamefont
  {Konoplya}, \citenamefont {Stuchl{\'\i}k},\ and\ \citenamefont
  {Zhidenko}}]{Konoplya:2018arm}%
  \BibitemOpen
  \bibfield  {author} {\bibinfo {author} {\bibfnamefont {R.~A.}\ \bibnamefont
  {Konoplya}}, \bibinfo {author} {\bibfnamefont {Z.}~\bibnamefont
  {Stuchl{\'\i}k}},\ and\ \bibinfo {author} {\bibfnamefont {A.}~\bibnamefont
  {Zhidenko}},\ }\bibfield  {title} {\bibinfo {title} {{Axisymmetric black
  holes allowing for separation of variables in the Klein-Gordon and
  Hamilton-Jacobi equations}},\ }\href
  {https://doi.org/10.1103/PhysRevD.97.084044} {\bibfield  {journal} {\bibinfo
  {journal} {Phys. Rev. D}\ }\textbf {\bibinfo {volume} {97}},\ \bibinfo
  {pages} {084044} (\bibinfo {year} {2018})},\ \Eprint
  {https://arxiv.org/abs/1801.07195} {arXiv:1801.07195 [gr-qc]} \BibitemShut
  {NoStop}%
\bibitem [{\citenamefont {Mukazhanov}\ \emph {et~al.}(2024)\citenamefont
  {Mukazhanov}, \citenamefont {Roy}, \citenamefont {Mirzaev},\ and\
  \citenamefont {Bambi}}]{Mukazhanov:2024rka}%
  \BibitemOpen
  \bibfield  {author} {\bibinfo {author} {\bibfnamefont {O.}~\bibnamefont
  {Mukazhanov}}, \bibinfo {author} {\bibfnamefont {R.}~\bibnamefont {Roy}},
  \bibinfo {author} {\bibfnamefont {T.}~\bibnamefont {Mirzaev}},\ and\ \bibinfo
  {author} {\bibfnamefont {C.}~\bibnamefont {Bambi}},\ }\bibfield  {title}
  {\bibinfo {title} {{Numerical parametrization of stationary axisymmetric
  black holes in a theory agnostic framework}},\ }\href
  {https://doi.org/10.1103/PhysRevD.110.024060} {\bibfield  {journal} {\bibinfo
   {journal} {Phys. Rev. D}\ }\textbf {\bibinfo {volume} {110}},\ \bibinfo
  {pages} {024060} (\bibinfo {year} {2024})},\ \Eprint
  {https://arxiv.org/abs/2404.17055} {arXiv:2404.17055 [gr-qc]} \BibitemShut
  {NoStop}%
\bibitem [{\citenamefont {Konoplya}\ and\ \citenamefont
  {Zhidenko}(2020)}]{Konoplya:2020hyk}%
  \BibitemOpen
  \bibfield  {author} {\bibinfo {author} {\bibfnamefont {R.~A.}\ \bibnamefont
  {Konoplya}}\ and\ \bibinfo {author} {\bibfnamefont {A.}~\bibnamefont
  {Zhidenko}},\ }\bibfield  {title} {\bibinfo {title} {{General parametrization
  of black holes: The only parameters that matter}},\ }\href
  {https://doi.org/10.1103/PhysRevD.101.124004} {\bibfield  {journal} {\bibinfo
   {journal} {Phys. Rev. D}\ }\textbf {\bibinfo {volume} {101}},\ \bibinfo
  {pages} {124004} (\bibinfo {year} {2020})},\ \Eprint
  {https://arxiv.org/abs/2001.06100} {arXiv:2001.06100 [gr-qc]} \BibitemShut
  {NoStop}%
\bibitem [{\citenamefont {Konoplya}\ and\ \citenamefont
  {Zhidenko}(2022)}]{Konoplya:2022tvv}%
  \BibitemOpen
  \bibfield  {author} {\bibinfo {author} {\bibfnamefont {R.~A.}\ \bibnamefont
  {Konoplya}}\ and\ \bibinfo {author} {\bibfnamefont {A.}~\bibnamefont
  {Zhidenko}},\ }\bibfield  {title} {\bibinfo {title} {{Quasinormal ringing of
  general spherically symmetric parametrized black holes}},\ }\href
  {https://doi.org/10.1103/PhysRevD.105.104032} {\bibfield  {journal} {\bibinfo
   {journal} {Phys. Rev. D}\ }\textbf {\bibinfo {volume} {105}},\ \bibinfo
  {pages} {104032} (\bibinfo {year} {2022})},\ \Eprint
  {https://arxiv.org/abs/2201.12897} {arXiv:2201.12897 [gr-qc]} \BibitemShut
  {NoStop}%
\bibitem [{\citenamefont {V{\"o}lkel}\ and\ \citenamefont
  {Barausse}(2020)}]{Volkel:2020daa}%
  \BibitemOpen
  \bibfield  {author} {\bibinfo {author} {\bibfnamefont {S.~H.}\ \bibnamefont
  {V{\"o}lkel}}\ and\ \bibinfo {author} {\bibfnamefont {E.}~\bibnamefont
  {Barausse}},\ }\bibfield  {title} {\bibinfo {title} {{Bayesian Metric
  Reconstruction with Gravitational Wave Observations}},\ }\href
  {https://doi.org/10.1103/PhysRevD.102.084025} {\bibfield  {journal} {\bibinfo
   {journal} {Phys. Rev. D}\ }\textbf {\bibinfo {volume} {102}},\ \bibinfo
  {pages} {084025} (\bibinfo {year} {2020})},\ \Eprint
  {https://arxiv.org/abs/2007.02986} {arXiv:2007.02986 [gr-qc]} \BibitemShut
  {NoStop}%
\bibitem [{\citenamefont {Del~Piano}\ \emph {et~al.}(2026)\citenamefont
  {Del~Piano}, \citenamefont {De~Simone}, \citenamefont {Damia~Paciarini},
  \citenamefont {De~Falco}, \citenamefont {Myszkowski}, \citenamefont
  {Sannino},\ and\ \citenamefont {Vellucci}}]{DelPiano:2025ykr}%
  \BibitemOpen
  \bibfield  {author} {\bibinfo {author} {\bibfnamefont {M.}~\bibnamefont
  {Del~Piano}}, \bibinfo {author} {\bibfnamefont {C.}~\bibnamefont
  {De~Simone}}, \bibinfo {author} {\bibfnamefont {M.}~\bibnamefont
  {Damia~Paciarini}}, \bibinfo {author} {\bibfnamefont {V.}~\bibnamefont
  {De~Falco}}, \bibinfo {author} {\bibfnamefont {M.}~\bibnamefont
  {Myszkowski}}, \bibinfo {author} {\bibfnamefont {F.}~\bibnamefont
  {Sannino}},\ and\ \bibinfo {author} {\bibfnamefont {V.}~\bibnamefont
  {Vellucci}},\ }\bibfield  {title} {\bibinfo {title} {{Toward a unified view
  of agnostic parametrizations for deformed black holes}},\ }\href
  {https://doi.org/10.1103/x9fw-lrv7} {\bibfield  {journal} {\bibinfo
  {journal} {Phys. Rev. D}\ }\textbf {\bibinfo {volume} {113}},\ \bibinfo
  {pages} {064010} (\bibinfo {year} {2026})},\ \Eprint
  {https://arxiv.org/abs/2510.14707} {arXiv:2510.14707 [gr-qc]} \BibitemShut
  {NoStop}%
\bibitem [{\citenamefont {Shashank}\ and\ \citenamefont
  {Bambi}(2022)}]{Shashank:2021giy}%
  \BibitemOpen
  \bibfield  {author} {\bibinfo {author} {\bibfnamefont {S.}~\bibnamefont
  {Shashank}}\ and\ \bibinfo {author} {\bibfnamefont {C.}~\bibnamefont
  {Bambi}},\ }\bibfield  {title} {\bibinfo {title} {{Constraining the
  Konoplya-Rezzolla-Zhidenko deformation parameters III: Limits from
  stellar-mass black holes using gravitational-wave observations}},\ }\href
  {https://doi.org/10.1103/PhysRevD.105.104004} {\bibfield  {journal} {\bibinfo
   {journal} {Phys. Rev. D}\ }\textbf {\bibinfo {volume} {105}},\ \bibinfo
  {pages} {104004} (\bibinfo {year} {2022})},\ \Eprint
  {https://arxiv.org/abs/2112.05388} {arXiv:2112.05388 [gr-qc]} \BibitemShut
  {NoStop}%
\bibitem [{\citenamefont {Dey}\ \emph {et~al.}(2023)\citenamefont {Dey},
  \citenamefont {Barausse},\ and\ \citenamefont {Basak}}]{Dey:2022pmv}%
  \BibitemOpen
  \bibfield  {author} {\bibinfo {author} {\bibfnamefont {K.}~\bibnamefont
  {Dey}}, \bibinfo {author} {\bibfnamefont {E.}~\bibnamefont {Barausse}},\ and\
  \bibinfo {author} {\bibfnamefont {S.}~\bibnamefont {Basak}},\ }\bibfield
  {title} {\bibinfo {title} {{Measuring deviations from the Kerr geometry with
  black hole ringdown}},\ }\href {https://doi.org/10.1103/PhysRevD.108.024064}
  {\bibfield  {journal} {\bibinfo  {journal} {Phys. Rev. D}\ }\textbf {\bibinfo
  {volume} {108}},\ \bibinfo {pages} {024064} (\bibinfo {year} {2023})},\
  \Eprint {https://arxiv.org/abs/2212.10725} {arXiv:2212.10725 [gr-qc]}
  \BibitemShut {NoStop}%
\bibitem [{\citenamefont {Cardoso}\ \emph {et~al.}(2019)\citenamefont
  {Cardoso}, \citenamefont {Kimura}, \citenamefont {Maselli}, \citenamefont
  {Berti}, \citenamefont {Macedo},\ and\ \citenamefont
  {McManus}}]{Cardoso:2019mqo}%
  \BibitemOpen
  \bibfield  {author} {\bibinfo {author} {\bibfnamefont {V.}~\bibnamefont
  {Cardoso}}, \bibinfo {author} {\bibfnamefont {M.}~\bibnamefont {Kimura}},
  \bibinfo {author} {\bibfnamefont {A.}~\bibnamefont {Maselli}}, \bibinfo
  {author} {\bibfnamefont {E.}~\bibnamefont {Berti}}, \bibinfo {author}
  {\bibfnamefont {C.~F.~B.}\ \bibnamefont {Macedo}},\ and\ \bibinfo {author}
  {\bibfnamefont {R.}~\bibnamefont {McManus}},\ }\bibfield  {title} {\bibinfo
  {title} {{Parametrized black hole quasinormal ringdown: Decoupled equations
  for nonrotating black holes}},\ }\href
  {https://doi.org/10.1103/PhysRevD.99.104077} {\bibfield  {journal} {\bibinfo
  {journal} {Phys. Rev. D}\ }\textbf {\bibinfo {volume} {99}},\ \bibinfo
  {pages} {104077} (\bibinfo {year} {2019})},\ \Eprint
  {https://arxiv.org/abs/1901.01265} {arXiv:1901.01265 [gr-qc]} \BibitemShut
  {NoStop}%
\bibitem [{\citenamefont {McManus}\ \emph {et~al.}(2019)\citenamefont
  {McManus}, \citenamefont {Berti}, \citenamefont {Macedo}, \citenamefont
  {Kimura}, \citenamefont {Maselli},\ and\ \citenamefont
  {Cardoso}}]{McManus:2019ulj}%
  \BibitemOpen
  \bibfield  {author} {\bibinfo {author} {\bibfnamefont {R.}~\bibnamefont
  {McManus}}, \bibinfo {author} {\bibfnamefont {E.}~\bibnamefont {Berti}},
  \bibinfo {author} {\bibfnamefont {C.~F.~B.}\ \bibnamefont {Macedo}}, \bibinfo
  {author} {\bibfnamefont {M.}~\bibnamefont {Kimura}}, \bibinfo {author}
  {\bibfnamefont {A.}~\bibnamefont {Maselli}},\ and\ \bibinfo {author}
  {\bibfnamefont {V.}~\bibnamefont {Cardoso}},\ }\bibfield  {title} {\bibinfo
  {title} {{Parametrized black hole quasinormal ringdown. II. Coupled equations
  and quadratic corrections for nonrotating black holes}},\ }\href
  {https://doi.org/10.1103/PhysRevD.100.044061} {\bibfield  {journal} {\bibinfo
   {journal} {Phys. Rev. D}\ }\textbf {\bibinfo {volume} {100}},\ \bibinfo
  {pages} {044061} (\bibinfo {year} {2019})},\ \Eprint
  {https://arxiv.org/abs/1906.05155} {arXiv:1906.05155 [gr-qc]} \BibitemShut
  {NoStop}%
\bibitem [{\citenamefont {Kimura}(2020)}]{Kimura:2020mrh}%
  \BibitemOpen
  \bibfield  {author} {\bibinfo {author} {\bibfnamefont {M.}~\bibnamefont
  {Kimura}},\ }\bibfield  {title} {\bibinfo {title} {{Note on the parametrized
  black hole quasinormal ringdown formalism}},\ }\href
  {https://doi.org/10.1103/PhysRevD.101.064031} {\bibfield  {journal} {\bibinfo
   {journal} {Phys. Rev. D}\ }\textbf {\bibinfo {volume} {101}},\ \bibinfo
  {pages} {064031} (\bibinfo {year} {2020})},\ \Eprint
  {https://arxiv.org/abs/2001.09613} {arXiv:2001.09613 [gr-qc]} \BibitemShut
  {NoStop}%
\bibitem [{\citenamefont {V{\"o}lkel}\ \emph
  {et~al.}(2022{\natexlab{a}})\citenamefont {V{\"o}lkel}, \citenamefont
  {Franchini},\ and\ \citenamefont {Barausse}}]{Volkel:2022aca}%
  \BibitemOpen
  \bibfield  {author} {\bibinfo {author} {\bibfnamefont {S.~H.}\ \bibnamefont
  {V{\"o}lkel}}, \bibinfo {author} {\bibfnamefont {N.}~\bibnamefont
  {Franchini}},\ and\ \bibinfo {author} {\bibfnamefont {E.}~\bibnamefont
  {Barausse}},\ }\bibfield  {title} {\bibinfo {title} {{Theory-agnostic
  reconstruction of potential and couplings from quasinormal modes}},\ }\href
  {https://doi.org/10.1103/PhysRevD.105.084046} {\bibfield  {journal} {\bibinfo
   {journal} {Phys. Rev. D}\ }\textbf {\bibinfo {volume} {105}},\ \bibinfo
  {pages} {084046} (\bibinfo {year} {2022}{\natexlab{a}})},\ \Eprint
  {https://arxiv.org/abs/2202.08655} {arXiv:2202.08655 [gr-qc]} \BibitemShut
  {NoStop}%
\bibitem [{\citenamefont {V{\"o}lkel}\ \emph
  {et~al.}(2022{\natexlab{b}})\citenamefont {V{\"o}lkel}, \citenamefont
  {Franchini}, \citenamefont {Barausse},\ and\ \citenamefont
  {Berti}}]{Volkel:2022khh}%
  \BibitemOpen
  \bibfield  {author} {\bibinfo {author} {\bibfnamefont {S.~H.}\ \bibnamefont
  {V{\"o}lkel}}, \bibinfo {author} {\bibfnamefont {N.}~\bibnamefont
  {Franchini}}, \bibinfo {author} {\bibfnamefont {E.}~\bibnamefont
  {Barausse}},\ and\ \bibinfo {author} {\bibfnamefont {E.}~\bibnamefont
  {Berti}},\ }\bibfield  {title} {\bibinfo {title} {{Constraining modifications
  of black hole perturbation potentials near the light ring with quasinormal
  modes}},\ }\href {https://doi.org/10.1103/PhysRevD.106.124036} {\bibfield
  {journal} {\bibinfo  {journal} {Phys. Rev. D}\ }\textbf {\bibinfo {volume}
  {106}},\ \bibinfo {pages} {124036} (\bibinfo {year} {2022}{\natexlab{b}})},\
  \Eprint {https://arxiv.org/abs/2209.10564} {arXiv:2209.10564 [gr-qc]}
  \BibitemShut {NoStop}%
\bibitem [{\citenamefont {Franchini}\ and\ \citenamefont
  {V{\"o}lkel}(2023)}]{Franchini:2022axs}%
  \BibitemOpen
  \bibfield  {author} {\bibinfo {author} {\bibfnamefont {N.}~\bibnamefont
  {Franchini}}\ and\ \bibinfo {author} {\bibfnamefont {S.~H.}\ \bibnamefont
  {V{\"o}lkel}},\ }\bibfield  {title} {\bibinfo {title} {{Parametrized
  quasinormal mode framework for non-Schwarzschild metrics}},\ }\href
  {https://doi.org/10.1103/PhysRevD.107.124063} {\bibfield  {journal} {\bibinfo
   {journal} {Phys. Rev. D}\ }\textbf {\bibinfo {volume} {107}},\ \bibinfo
  {pages} {124063} (\bibinfo {year} {2023})},\ \Eprint
  {https://arxiv.org/abs/2210.14020} {arXiv:2210.14020 [gr-qc]} \BibitemShut
  {NoStop}%
\bibitem [{\citenamefont {Thomopoulos}\ \emph {et~al.}(2025)\citenamefont
  {Thomopoulos}, \citenamefont {V{\"o}lkel},\ and\ \citenamefont
  {Pfeiffer}}]{Thomopoulos:2025nuf}%
  \BibitemOpen
  \bibfield  {author} {\bibinfo {author} {\bibfnamefont {S.}~\bibnamefont
  {Thomopoulos}}, \bibinfo {author} {\bibfnamefont {S.~H.}\ \bibnamefont
  {V{\"o}lkel}},\ and\ \bibinfo {author} {\bibfnamefont {H.~P.}\ \bibnamefont
  {Pfeiffer}},\ }\bibfield  {title} {\bibinfo {title} {{Ringdown spectroscopy
  of phenomenologically modified black holes}},\ }\href
  {https://doi.org/10.1103/xtzl-lyn6} {\bibfield  {journal} {\bibinfo
  {journal} {Phys. Rev. D}\ }\textbf {\bibinfo {volume} {112}},\ \bibinfo
  {pages} {064054} (\bibinfo {year} {2025})},\ \Eprint
  {https://arxiv.org/abs/2504.17848} {arXiv:2504.17848 [gr-qc]} \BibitemShut
  {NoStop}%
\bibitem [{\citenamefont {Cano}\ \emph {et~al.}(2024)\citenamefont {Cano},
  \citenamefont {Capuano}, \citenamefont {Franchini}, \citenamefont {Maenaut},\
  and\ \citenamefont {V{\"o}lkel}}]{Cano:2024jkd}%
  \BibitemOpen
  \bibfield  {author} {\bibinfo {author} {\bibfnamefont {P.~A.}\ \bibnamefont
  {Cano}}, \bibinfo {author} {\bibfnamefont {L.}~\bibnamefont {Capuano}},
  \bibinfo {author} {\bibfnamefont {N.}~\bibnamefont {Franchini}}, \bibinfo
  {author} {\bibfnamefont {S.}~\bibnamefont {Maenaut}},\ and\ \bibinfo {author}
  {\bibfnamefont {S.~H.}\ \bibnamefont {V{\"o}lkel}},\ }\bibfield  {title}
  {\bibinfo {title} {{Parametrized quasinormal mode framework for modified
  Teukolsky equations}},\ }\href {https://doi.org/10.1103/PhysRevD.110.104007}
  {\bibfield  {journal} {\bibinfo  {journal} {Phys. Rev. D}\ }\textbf {\bibinfo
  {volume} {110}},\ \bibinfo {pages} {104007} (\bibinfo {year} {2024})},\
  \bibinfo {note} {[Erratum: Phys.Rev.D 113, 069902 (2026)]},\ \Eprint
  {https://arxiv.org/abs/2407.15947} {arXiv:2407.15947 [gr-qc]} \BibitemShut
  {NoStop}%
\bibitem [{\citenamefont {De~Simone}\ \emph
  {et~al.}(2026{\natexlab{a}})\citenamefont {De~Simone}, \citenamefont
  {V{\"o}lkel}, \citenamefont {Kokkotas},\ and\ \citenamefont
  {Capozziello}}]{DeSimone:2026waz}%
  \BibitemOpen
  \bibfield  {author} {\bibinfo {author} {\bibfnamefont {C.}~\bibnamefont
  {De~Simone}}, \bibinfo {author} {\bibfnamefont {S.~H.}\ \bibnamefont
  {V{\"o}lkel}}, \bibinfo {author} {\bibfnamefont {K.~D.}\ \bibnamefont
  {Kokkotas}},\ and\ \bibinfo {author} {\bibfnamefont {S.}~\bibnamefont
  {Capozziello}},\ }\bibfield  {title} {\bibinfo {title} {{Parametrized
  beyond-Teukolsky framework in the time domain}},\ }\href
  {https://doi.org/10.1103/93dw-jlcg} {\bibfield  {journal} {\bibinfo
  {journal} {Phys. Rev. D}\ }\textbf {\bibinfo {volume} {114}},\ \bibinfo
  {pages} {064060} (\bibinfo {year} {2026}{\natexlab{a}})},\ \Eprint
  {https://arxiv.org/abs/2607.25311} {arXiv:2607.25311 [gr-qc]} \BibitemShut
  {NoStop}%
\bibitem [{\citenamefont {Li}\ \emph {et~al.}(2023)\citenamefont {Li},
  \citenamefont {Wagle}, \citenamefont {Chen},\ and\ \citenamefont
  {Yunes}}]{Li:2022pcy}%
  \BibitemOpen
  \bibfield  {author} {\bibinfo {author} {\bibfnamefont {D.}~\bibnamefont
  {Li}}, \bibinfo {author} {\bibfnamefont {P.}~\bibnamefont {Wagle}}, \bibinfo
  {author} {\bibfnamefont {Y.}~\bibnamefont {Chen}},\ and\ \bibinfo {author}
  {\bibfnamefont {N.}~\bibnamefont {Yunes}},\ }\bibfield  {title} {\bibinfo
  {title} {{Perturbations of Spinning Black Holes beyond General Relativity:
  Modified Teukolsky Equation}},\ }\href
  {https://doi.org/10.1103/PhysRevX.13.021029} {\bibfield  {journal} {\bibinfo
  {journal} {Phys. Rev. X}\ }\textbf {\bibinfo {volume} {13}},\ \bibinfo
  {pages} {021029} (\bibinfo {year} {2023})},\ \Eprint
  {https://arxiv.org/abs/2206.10652} {arXiv:2206.10652 [gr-qc]} \BibitemShut
  {NoStop}%
\bibitem [{\citenamefont {Hussain}\ and\ \citenamefont
  {Zimmerman}(2022)}]{Hussain:2022ins}%
  \BibitemOpen
  \bibfield  {author} {\bibinfo {author} {\bibfnamefont {A.}~\bibnamefont
  {Hussain}}\ and\ \bibinfo {author} {\bibfnamefont {A.}~\bibnamefont
  {Zimmerman}},\ }\bibfield  {title} {\bibinfo {title} {{Approach to computing
  spectral shifts for black holes beyond Kerr}},\ }\href
  {https://doi.org/10.1103/PhysRevD.106.104018} {\bibfield  {journal} {\bibinfo
   {journal} {Phys. Rev. D}\ }\textbf {\bibinfo {volume} {106}},\ \bibinfo
  {pages} {104018} (\bibinfo {year} {2022})},\ \Eprint
  {https://arxiv.org/abs/2206.10653} {arXiv:2206.10653 [gr-qc]} \BibitemShut
  {NoStop}%
\bibitem [{\citenamefont {Cano}\ \emph {et~al.}(2023)\citenamefont {Cano},
  \citenamefont {Fransen}, \citenamefont {Hertog},\ and\ \citenamefont
  {Maenaut}}]{Cano:2023tmv}%
  \BibitemOpen
  \bibfield  {author} {\bibinfo {author} {\bibfnamefont {P.~A.}\ \bibnamefont
  {Cano}}, \bibinfo {author} {\bibfnamefont {K.}~\bibnamefont {Fransen}},
  \bibinfo {author} {\bibfnamefont {T.}~\bibnamefont {Hertog}},\ and\ \bibinfo
  {author} {\bibfnamefont {S.}~\bibnamefont {Maenaut}},\ }\bibfield  {title}
  {\bibinfo {title} {{Universal Teukolsky equations and black hole
  perturbations in higher-derivative gravity}},\ }\href
  {https://doi.org/10.1103/PhysRevD.108.024040} {\bibfield  {journal} {\bibinfo
   {journal} {Phys. Rev. D}\ }\textbf {\bibinfo {volume} {108}},\ \bibinfo
  {pages} {024040} (\bibinfo {year} {2023})},\ \Eprint
  {https://arxiv.org/abs/2304.02663} {arXiv:2304.02663 [gr-qc]} \BibitemShut
  {NoStop}%
\bibitem [{\citenamefont {Tang}\ \emph {et~al.}(2026)\citenamefont {Tang},
  \citenamefont {Franchini}, \citenamefont {V{\"o}lkel},\ and\ \citenamefont
  {Berti}}]{Tang:2025qaq}%
  \BibitemOpen
  \bibfield  {author} {\bibinfo {author} {\bibfnamefont {R.}~\bibnamefont
  {Tang}}, \bibinfo {author} {\bibfnamefont {N.}~\bibnamefont {Franchini}},
  \bibinfo {author} {\bibfnamefont {S.~H.}\ \bibnamefont {V{\"o}lkel}},\ and\
  \bibinfo {author} {\bibfnamefont {E.}~\bibnamefont {Berti}},\ }\bibfield
  {title} {\bibinfo {title} {{Quasinormal modes of rotating black holes beyond
  general relativity in the WKB approximation}},\ }\href
  {https://doi.org/10.1103/dvv2-sw54} {\bibfield  {journal} {\bibinfo
  {journal} {Phys. Rev. D}\ }\textbf {\bibinfo {volume} {113}},\ \bibinfo
  {pages} {104052} (\bibinfo {year} {2026})},\ \Eprint
  {https://arxiv.org/abs/2512.17786} {arXiv:2512.17786 [gr-qc]} \BibitemShut
  {NoStop}%
\bibitem [{\citenamefont {Abac}\ \emph {et~al.}(2025)\citenamefont {Abac} \emph
  {et~al.}}]{LIGOScientific:2025rid}%
  \BibitemOpen
  \bibfield  {author} {\bibinfo {author} {\bibfnamefont {A.~G.}\ \bibnamefont
  {Abac}} \emph {et~al.} (\bibinfo {collaboration} {LIGO Scientific, Virgo,
  KAGRA}),\ }\bibfield  {title} {\bibinfo {title} {{GW250114: Testing
  Hawking{\textquoteright}s Area Law and the Kerr Nature of Black Holes}},\
  }\href {https://doi.org/10.1103/kw5g-d732} {\bibfield  {journal} {\bibinfo
  {journal} {Phys. Rev. Lett.}\ }\textbf {\bibinfo {volume} {135}},\ \bibinfo
  {pages} {111403} (\bibinfo {year} {2025})},\ \Eprint
  {https://arxiv.org/abs/2509.08054} {arXiv:2509.08054 [gr-qc]} \BibitemShut
  {NoStop}%
\bibitem [{\citenamefont {Abac}\ \emph
  {et~al.}(2026{\natexlab{a}})\citenamefont {Abac} \emph
  {et~al.}}]{LIGOScientific:2025wao}%
  \BibitemOpen
  \bibfield  {author} {\bibinfo {author} {\bibfnamefont {A.~G.}\ \bibnamefont
  {Abac}} \emph {et~al.} (\bibinfo {collaboration} {LIGO Scientific, Virgo,
  KAGRA}),\ }\bibfield  {title} {\bibinfo {title} {{Black Hole Spectroscopy and
  Tests of General Relativity with GW250114}},\ }\href
  {https://doi.org/10.1103/6c61-fm1n} {\bibfield  {journal} {\bibinfo
  {journal} {Phys. Rev. Lett.}\ }\textbf {\bibinfo {volume} {136}},\ \bibinfo
  {pages} {041403} (\bibinfo {year} {2026}{\natexlab{a}})},\ \Eprint
  {https://arxiv.org/abs/2509.08099} {arXiv:2509.08099 [gr-qc]} \BibitemShut
  {NoStop}%
\bibitem [{\citenamefont {V{\"o}lkel}\ and\ \citenamefont
  {Franchini}(2026)}]{Volkel:2026qqz}%
  \BibitemOpen
  \bibfield  {author} {\bibinfo {author} {\bibfnamefont {S.~H.}\ \bibnamefont
  {V{\"o}lkel}}\ and\ \bibinfo {author} {\bibfnamefont {N.}~\bibnamefont
  {Franchini}},\ }\bibfield  {title} {\bibinfo {title} {{Constraining
  deviations from the Teukolsky equation with GW250114}},\ }\href@noop {} {\
  (\bibinfo {year} {2026})},\ \Eprint {https://arxiv.org/abs/2607.26561}
  {arXiv:2607.26561 [gr-qc]} \BibitemShut {NoStop}%
\bibitem [{\citenamefont {Albuquerque}\ and\ \citenamefont
  {V{\"o}lkel}(2025)}]{Albuquerque:2025eny}%
  \BibitemOpen
  \bibfield  {author} {\bibinfo {author} {\bibfnamefont {S.}~\bibnamefont
  {Albuquerque}}\ and\ \bibinfo {author} {\bibfnamefont {S.~H.}\ \bibnamefont
  {V{\"o}lkel}},\ }\bibfield  {title} {\bibinfo {title} {{Bayesian analysis of
  analog gravity systems with the Rezzolla-Zhidenko metric}},\ }\href
  {https://doi.org/10.1103/kwrg-rs71} {\bibfield  {journal} {\bibinfo
  {journal} {Phys. Rev. D}\ }\textbf {\bibinfo {volume} {111}},\ \bibinfo
  {pages} {124020} (\bibinfo {year} {2025})},\ \Eprint
  {https://arxiv.org/abs/2501.09000} {arXiv:2501.09000 [gr-qc]} \BibitemShut
  {NoStop}%
\bibitem [{\citenamefont {Albuquerque}\ and\ \citenamefont
  {V{\"o}lkel}(2026)}]{Albuquerque:2026bgc}%
  \BibitemOpen
  \bibfield  {author} {\bibinfo {author} {\bibfnamefont {S.}~\bibnamefont
  {Albuquerque}}\ and\ \bibinfo {author} {\bibfnamefont {S.~H.}\ \bibnamefont
  {V{\"o}lkel}},\ }\bibfield  {title} {\bibinfo {title} {{Optimal frequency
  scales for probing black-hole geometries}},\ }\href@noop {} {\  (\bibinfo
  {year} {2026})},\ \Eprint {https://arxiv.org/abs/2608.01061}
  {arXiv:2608.01061 [gr-qc]} \BibitemShut {NoStop}%
\bibitem [{\citenamefont {Pedrotti}\ and\ \citenamefont
  {Vagnozzi}(2024)}]{Pedrotti:2024znu}%
  \BibitemOpen
  \bibfield  {author} {\bibinfo {author} {\bibfnamefont {D.}~\bibnamefont
  {Pedrotti}}\ and\ \bibinfo {author} {\bibfnamefont {S.}~\bibnamefont
  {Vagnozzi}},\ }\bibfield  {title} {\bibinfo {title} {{Quasinormal
  modes-shadow correspondence for rotating regular black holes}},\ }\href
  {https://doi.org/10.1103/PhysRevD.110.084075} {\bibfield  {journal} {\bibinfo
   {journal} {Phys. Rev. D}\ }\textbf {\bibinfo {volume} {110}},\ \bibinfo
  {pages} {084075} (\bibinfo {year} {2024})},\ \Eprint
  {https://arxiv.org/abs/2404.07589} {arXiv:2404.07589 [gr-qc]} \BibitemShut
  {NoStop}%
\bibitem [{\citenamefont {Glampedakis}\ \emph {et~al.}(2017)\citenamefont
  {Glampedakis}, \citenamefont {Pappas}, \citenamefont {Silva},\ and\
  \citenamefont {Berti}}]{Glampedakis:2017dvb}%
  \BibitemOpen
  \bibfield  {author} {\bibinfo {author} {\bibfnamefont {K.}~\bibnamefont
  {Glampedakis}}, \bibinfo {author} {\bibfnamefont {G.}~\bibnamefont {Pappas}},
  \bibinfo {author} {\bibfnamefont {H.~O.}\ \bibnamefont {Silva}},\ and\
  \bibinfo {author} {\bibfnamefont {E.}~\bibnamefont {Berti}},\ }\bibfield
  {title} {\bibinfo {title} {{Post-Kerr black hole spectroscopy}},\ }\href
  {https://doi.org/10.1103/PhysRevD.96.064054} {\bibfield  {journal} {\bibinfo
  {journal} {Phys. Rev. D}\ }\textbf {\bibinfo {volume} {96}},\ \bibinfo
  {pages} {064054} (\bibinfo {year} {2017})},\ \Eprint
  {https://arxiv.org/abs/1706.07658} {arXiv:1706.07658 [gr-qc]} \BibitemShut
  {NoStop}%
\bibitem [{\citenamefont {Berti}\ \emph {et~al.}(2016)\citenamefont {Berti},
  \citenamefont {Sesana}, \citenamefont {Barausse}, \citenamefont {Cardoso},\
  and\ \citenamefont {Belczynski}}]{Berti:2016lat}%
  \BibitemOpen
  \bibfield  {author} {\bibinfo {author} {\bibfnamefont {E.}~\bibnamefont
  {Berti}}, \bibinfo {author} {\bibfnamefont {A.}~\bibnamefont {Sesana}},
  \bibinfo {author} {\bibfnamefont {E.}~\bibnamefont {Barausse}}, \bibinfo
  {author} {\bibfnamefont {V.}~\bibnamefont {Cardoso}},\ and\ \bibinfo {author}
  {\bibfnamefont {K.}~\bibnamefont {Belczynski}},\ }\bibfield  {title}
  {\bibinfo {title} {{Spectroscopy of Kerr black holes with Earth- and
  space-based interferometers}},\ }\href
  {https://doi.org/10.1103/PhysRevLett.117.101102} {\bibfield  {journal}
  {\bibinfo  {journal} {Phys. Rev. Lett.}\ }\textbf {\bibinfo {volume} {117}},\
  \bibinfo {pages} {101102} (\bibinfo {year} {2016})},\ \Eprint
  {https://arxiv.org/abs/1605.09286} {arXiv:1605.09286 [gr-qc]} \BibitemShut
  {NoStop}%
\bibitem [{\citenamefont {Barausse}\ \emph {et~al.}(2020)\citenamefont
  {Barausse} \emph {et~al.}}]{Barausse:2020rsu}%
  \BibitemOpen
  \bibfield  {author} {\bibinfo {author} {\bibfnamefont {E.}~\bibnamefont
  {Barausse}} \emph {et~al.},\ }\bibfield  {title} {\bibinfo {title}
  {{Prospects for Fundamental Physics with LISA}},\ }\href
  {https://doi.org/10.1007/s10714-020-02691-1} {\bibfield  {journal} {\bibinfo
  {journal} {Gen. Rel. Grav.}\ }\textbf {\bibinfo {volume} {52}},\ \bibinfo
  {pages} {81} (\bibinfo {year} {2020})},\ \Eprint
  {https://arxiv.org/abs/2001.09793} {arXiv:2001.09793 [gr-qc]} \BibitemShut
  {NoStop}%
\bibitem [{\citenamefont {Seoane}\ \emph {et~al.}(2023)\citenamefont {Seoane}
  \emph {et~al.}}]{LISA:2022yao}%
  \BibitemOpen
  \bibfield  {author} {\bibinfo {author} {\bibfnamefont {P.~A.}\ \bibnamefont
  {Seoane}} \emph {et~al.} (\bibinfo {collaboration} {LISA}),\ }\bibfield
  {title} {\bibinfo {title} {{Astrophysics with the Laser Interferometer Space
  Antenna}},\ }\href {https://doi.org/10.1007/s41114-022-00041-y} {\bibfield
  {journal} {\bibinfo  {journal} {Living Rev. Rel.}\ }\textbf {\bibinfo
  {volume} {26}},\ \bibinfo {pages} {2} (\bibinfo {year} {2023})},\ \Eprint
  {https://arxiv.org/abs/2203.06016} {arXiv:2203.06016 [gr-qc]} \BibitemShut
  {NoStop}%
\bibitem [{\citenamefont {Colpi}\ \emph {et~al.}(2024)\citenamefont {Colpi}
  \emph {et~al.}}]{LISA:2024hlh}%
  \BibitemOpen
  \bibfield  {author} {\bibinfo {author} {\bibfnamefont {M.}~\bibnamefont
  {Colpi}} \emph {et~al.} (\bibinfo {collaboration} {LISA}),\ }\bibfield
  {title} {\bibinfo {title} {{LISA Definition Study Report}},\ }\href@noop {}
  {\  (\bibinfo {year} {2024})},\ \Eprint {https://arxiv.org/abs/2402.07571}
  {arXiv:2402.07571 [astro-ph.CO]} \BibitemShut {NoStop}%
\bibitem [{\citenamefont {Pani}\ and\ \citenamefont
  {Sanna}(2026)}]{Pani:2026yzi}%
  \BibitemOpen
  \bibfield  {author} {\bibinfo {author} {\bibfnamefont {P.}~\bibnamefont
  {Pani}}\ and\ \bibinfo {author} {\bibfnamefont {A.~P.}\ \bibnamefont
  {Sanna}},\ }\bibfield  {title} {\bibinfo {title} {{Scalar shortcut to
  beyond-Kerr ringdown tests and their complementarity with black-hole shadow
  observations}},\ }\href@noop {} {\  (\bibinfo {year} {2026})},\ \Eprint
  {https://arxiv.org/abs/2603.08782} {arXiv:2603.08782 [gr-qc]} \BibitemShut
  {NoStop}%
\bibitem [{\citenamefont {De~Simone}\ \emph
  {et~al.}(2026{\natexlab{b}})\citenamefont {De~Simone}, \citenamefont
  {V{\"o}lkel}, \citenamefont {Kokkotas}, \citenamefont {De~Falco},\ and\
  \citenamefont {Capozziello}}]{DeSimone:2026mkz}%
  \BibitemOpen
  \bibfield  {author} {\bibinfo {author} {\bibfnamefont {C.}~\bibnamefont
  {De~Simone}}, \bibinfo {author} {\bibfnamefont {S.~H.}\ \bibnamefont
  {V{\"o}lkel}}, \bibinfo {author} {\bibfnamefont {K.~D.}\ \bibnamefont
  {Kokkotas}}, \bibinfo {author} {\bibfnamefont {V.}~\bibnamefont {De~Falco}},\
  and\ \bibinfo {author} {\bibfnamefont {S.}~\bibnamefont {Capozziello}},\
  }\bibfield  {title} {\bibinfo {title} {{Confronting eikonal and post-Kerr
  methods with numerical evolution of scalar field perturbations in spacetimes
  beyond Kerr}},\ }\href {https://doi.org/10.1103/ym27-xbz2} {\bibfield
  {journal} {\bibinfo  {journal} {Phys. Rev. D}\ }\textbf {\bibinfo {volume}
  {113}},\ \bibinfo {pages} {104004} (\bibinfo {year} {2026}{\natexlab{b}})},\
  \Eprint {https://arxiv.org/abs/2601.09607} {arXiv:2601.09607 [gr-qc]}
  \BibitemShut {NoStop}%
\bibitem [{\citenamefont {Will}(2014)}]{Will:2014kxa}%
  \BibitemOpen
  \bibfield  {author} {\bibinfo {author} {\bibfnamefont {C.~M.}\ \bibnamefont
  {Will}},\ }\bibfield  {title} {\bibinfo {title} {{The Confrontation between
  General Relativity and Experiment}},\ }\href
  {https://doi.org/10.12942/lrr-2014-4} {\bibfield  {journal} {\bibinfo
  {journal} {Living Rev. Rel.}\ }\textbf {\bibinfo {volume} {17}},\ \bibinfo
  {pages} {4} (\bibinfo {year} {2014})},\ \Eprint
  {https://arxiv.org/abs/1403.7377} {arXiv:1403.7377 [gr-qc]} \BibitemShut
  {NoStop}%
\bibitem [{\citenamefont {Abdikamalov}\ \emph {et~al.}(2021)\citenamefont
  {Abdikamalov}, \citenamefont {Ayzenberg}, \citenamefont {Bambi},
  \citenamefont {Nampalliwar},\ and\ \citenamefont
  {Tripathi}}]{Abdikamalov:2021zwv}%
  \BibitemOpen
  \bibfield  {author} {\bibinfo {author} {\bibfnamefont {A.~B.}\ \bibnamefont
  {Abdikamalov}}, \bibinfo {author} {\bibfnamefont {D.}~\bibnamefont
  {Ayzenberg}}, \bibinfo {author} {\bibfnamefont {C.}~\bibnamefont {Bambi}},
  \bibinfo {author} {\bibfnamefont {S.}~\bibnamefont {Nampalliwar}},\ and\
  \bibinfo {author} {\bibfnamefont {A.}~\bibnamefont {Tripathi}},\ }\bibfield
  {title} {\bibinfo {title} {{Constraining the Konoplya-Rezzolla-Zhidenko
  deformation parameters: Limits from supermassive black hole x-ray data}},\
  }\href {https://doi.org/10.1103/PhysRevD.104.024058} {\bibfield  {journal}
  {\bibinfo  {journal} {Phys. Rev. D}\ }\textbf {\bibinfo {volume} {104}},\
  \bibinfo {pages} {024058} (\bibinfo {year} {2021})},\ \Eprint
  {https://arxiv.org/abs/2104.04183} {arXiv:2104.04183 [astro-ph.HE]}
  \BibitemShut {NoStop}%
\bibitem [{\citenamefont {Foreman-Mackey}\ \emph {et~al.}(2013)\citenamefont
  {Foreman-Mackey}, \citenamefont {Hogg}, \citenamefont {Lang},\ and\
  \citenamefont {Goodman}}]{Foreman-Mackey:2012any}%
  \BibitemOpen
  \bibfield  {author} {\bibinfo {author} {\bibfnamefont {D.}~\bibnamefont
  {Foreman-Mackey}}, \bibinfo {author} {\bibfnamefont {D.~W.}\ \bibnamefont
  {Hogg}}, \bibinfo {author} {\bibfnamefont {D.}~\bibnamefont {Lang}},\ and\
  \bibinfo {author} {\bibfnamefont {J.}~\bibnamefont {Goodman}},\ }\bibfield
  {title} {\bibinfo {title} {{emcee: The MCMC Hammer}},\ }\href
  {https://doi.org/10.1086/670067} {\bibfield  {journal} {\bibinfo  {journal}
  {Publ. Astron. Soc. Pac.}\ }\textbf {\bibinfo {volume} {125}},\ \bibinfo
  {pages} {306} (\bibinfo {year} {2013})},\ \Eprint
  {https://arxiv.org/abs/1202.3665} {arXiv:1202.3665 [astro-ph.IM]}
  \BibitemShut {NoStop}%
\bibitem [{\citenamefont {{Goodman}}\ and\ \citenamefont
  {{Weare}}(2010)}]{2010CAMCS...5...65G}%
  \BibitemOpen
  \bibfield  {author} {\bibinfo {author} {\bibfnamefont {J.}~\bibnamefont
  {{Goodman}}}\ and\ \bibinfo {author} {\bibfnamefont {J.}~\bibnamefont
  {{Weare}}},\ }\bibfield  {title} {\bibinfo {title} {{Ensemble samplers with
  affine invariance}},\ }\href {https://doi.org/10.2140/camcos.2010.5.65}
  {\bibfield  {journal} {\bibinfo  {journal} {Communications in Applied
  Mathematics and Computational Science}\ }\textbf {\bibinfo {volume} {5}},\
  \bibinfo {pages} {65} (\bibinfo {year} {2010})}\BibitemShut {NoStop}%
\bibitem [{\citenamefont {Kocherlakota}\ and\ \citenamefont
  {Rezzolla}(2022{\natexlab{a}})}]{Kocherlakota:2022mro}%
  \BibitemOpen
  \bibfield  {author} {\bibinfo {author} {\bibfnamefont {P.}~\bibnamefont
  {Kocherlakota}}\ and\ \bibinfo {author} {\bibfnamefont {L.}~\bibnamefont
  {Rezzolla}},\ }\bibfield  {title} {\bibinfo {title} {{Comment on the
  Analytical Bounds in the Rezzolla-Zhidenko Parametrization}},\ }\href@noop {}
  {\  (\bibinfo {year} {2022}{\natexlab{a}})},\ \Eprint
  {https://arxiv.org/abs/2206.03146} {arXiv:2206.03146 [gr-qc]} \BibitemShut
  {NoStop}%
\bibitem [{\citenamefont {Lim}\ \emph {et~al.}(2019)\citenamefont {Lim},
  \citenamefont {Khanna}, \citenamefont {Apte},\ and\ \citenamefont
  {Hughes}}]{Lim:2019xrb}%
  \BibitemOpen
  \bibfield  {author} {\bibinfo {author} {\bibfnamefont {H.}~\bibnamefont
  {Lim}}, \bibinfo {author} {\bibfnamefont {G.}~\bibnamefont {Khanna}},
  \bibinfo {author} {\bibfnamefont {A.}~\bibnamefont {Apte}},\ and\ \bibinfo
  {author} {\bibfnamefont {S.~A.}\ \bibnamefont {Hughes}},\ }\bibfield  {title}
  {\bibinfo {title} {{Exciting black hole modes via misaligned coalescences:
  II. The mode content of late-time coalescence waveforms}},\ }\href
  {https://doi.org/10.1103/PhysRevD.100.084032} {\bibfield  {journal} {\bibinfo
   {journal} {Phys. Rev. D}\ }\textbf {\bibinfo {volume} {100}},\ \bibinfo
  {pages} {084032} (\bibinfo {year} {2019})},\ \Eprint
  {https://arxiv.org/abs/1901.05902} {arXiv:1901.05902 [gr-qc]} \BibitemShut
  {NoStop}%
\bibitem [{\citenamefont {Abac}\ \emph
  {et~al.}(2026{\natexlab{b}})\citenamefont {Abac} \emph
  {et~al.}}]{ET:2025xjr}%
  \BibitemOpen
  \bibfield  {author} {\bibinfo {author} {\bibfnamefont {A.}~\bibnamefont
  {Abac}} \emph {et~al.} (\bibinfo {collaboration} {ET}),\ }\bibfield  {title}
  {\bibinfo {title} {{The Science of the Einstein Telescope}},\ }\href
  {https://doi.org/10.1088/1475-7516/2026/03/081} {\bibfield  {journal}
  {\bibinfo  {journal} {JCAP}\ }\textbf {\bibinfo {volume} {03}},\ \bibinfo
  {pages} {081}},\ \Eprint {https://arxiv.org/abs/2503.12263} {arXiv:2503.12263
  [gr-qc]} \BibitemShut {NoStop}%
\bibitem [{\citenamefont {Teo}(2021)}]{Teo:2020sey}%
  \BibitemOpen
  \bibfield  {author} {\bibinfo {author} {\bibfnamefont {E.}~\bibnamefont
  {Teo}},\ }\bibfield  {title} {\bibinfo {title} {{Spherical orbits around a
  Kerr black hole}},\ }\href {https://doi.org/10.1007/s10714-020-02782-z}
  {\bibfield  {journal} {\bibinfo  {journal} {Gen. Rel. Grav.}\ }\textbf
  {\bibinfo {volume} {53}},\ \bibinfo {pages} {10} (\bibinfo {year} {2021})},\
  \Eprint {https://arxiv.org/abs/2007.04022} {arXiv:2007.04022 [gr-qc]}
  \BibitemShut {NoStop}%
\bibitem [{\citenamefont {Pappas}\ and\ \citenamefont
  {Glampedakis}(2018)}]{Pappas:2018opz}%
  \BibitemOpen
  \bibfield  {author} {\bibinfo {author} {\bibfnamefont {G.}~\bibnamefont
  {Pappas}}\ and\ \bibinfo {author} {\bibfnamefont {K.}~\bibnamefont
  {Glampedakis}},\ }\bibfield  {title} {\bibinfo {title} {{On the connection of
  spacetime separability and spherical photon orbits}},\ }\href@noop {} {\
  (\bibinfo {year} {2018})},\ \Eprint {https://arxiv.org/abs/1806.04091}
  {arXiv:1806.04091 [gr-qc]} \BibitemShut {NoStop}%
\bibitem [{\citenamefont {Suvorov}\ and\ \citenamefont
  {V{\"o}lkel}(2021)}]{Suvorov:2021amy}%
  \BibitemOpen
  \bibfield  {author} {\bibinfo {author} {\bibfnamefont {A.~G.}\ \bibnamefont
  {Suvorov}}\ and\ \bibinfo {author} {\bibfnamefont {S.~H.}\ \bibnamefont
  {V{\"o}lkel}},\ }\bibfield  {title} {\bibinfo {title} {{Exact theory for the
  Rezzolla-Zhidenko metric and self-consistent calculation of quasinormal
  modes}},\ }\href {https://doi.org/10.1103/PhysRevD.103.044027} {\bibfield
  {journal} {\bibinfo  {journal} {Phys. Rev. D}\ }\textbf {\bibinfo {volume}
  {103}},\ \bibinfo {pages} {044027} (\bibinfo {year} {2021})},\ \Eprint
  {https://arxiv.org/abs/2101.09697} {arXiv:2101.09697 [gr-qc]} \BibitemShut
  {NoStop}%
\bibitem [{\citenamefont {Dai}\ \emph {et~al.}(2026)\citenamefont {Dai},
  \citenamefont {Fang}, \citenamefont {Kuang},\ and\ \citenamefont
  {Jing}}]{Dai:2026ujj}%
  \BibitemOpen
  \bibfield  {author} {\bibinfo {author} {\bibfnamefont {Q.}~\bibnamefont
  {Dai}}, \bibinfo {author} {\bibfnamefont {X.}~\bibnamefont {Fang}}, \bibinfo
  {author} {\bibfnamefont {X.-M.}\ \bibnamefont {Kuang}},\ and\ \bibinfo
  {author} {\bibfnamefont {J.}~\bibnamefont {Jing}},\ }\bibfield  {title}
  {\bibinfo {title} {{Resonant bound orbits and kludge waveforms in rotating
  Konoplya-Zhidenko black hole spacetime}},\ }\href@noop {} {\  (\bibinfo
  {year} {2026})},\ \Eprint {https://arxiv.org/abs/2608.29234}
  {arXiv:2608.29234 [gr-qc]} \BibitemShut {NoStop}%
\bibitem [{\citenamefont {Amaro-Seoane}\ \emph {et~al.}(2007)\citenamefont
  {Amaro-Seoane}, \citenamefont {Gair}, \citenamefont {Freitag}, \citenamefont
  {Coleman~Miller}, \citenamefont {Mandel}, \citenamefont {Cutler},\ and\
  \citenamefont {Babak}}]{Amaro-Seoane:2007osp}%
  \BibitemOpen
  \bibfield  {author} {\bibinfo {author} {\bibfnamefont {P.}~\bibnamefont
  {Amaro-Seoane}}, \bibinfo {author} {\bibfnamefont {J.~R.}\ \bibnamefont
  {Gair}}, \bibinfo {author} {\bibfnamefont {M.}~\bibnamefont {Freitag}},
  \bibinfo {author} {\bibfnamefont {M.}~\bibnamefont {Coleman~Miller}},
  \bibinfo {author} {\bibfnamefont {I.}~\bibnamefont {Mandel}}, \bibinfo
  {author} {\bibfnamefont {C.~J.}\ \bibnamefont {Cutler}},\ and\ \bibinfo
  {author} {\bibfnamefont {S.}~\bibnamefont {Babak}},\ }\bibfield  {title}
  {\bibinfo {title} {{Astrophysics, detection and science applications of
  intermediate- and extreme mass-ratio inspirals}},\ }\href
  {https://doi.org/10.1088/0264-9381/24/17/R01} {\bibfield  {journal} {\bibinfo
   {journal} {Class. Quant. Grav.}\ }\textbf {\bibinfo {volume} {24}},\
  \bibinfo {pages} {R113} (\bibinfo {year} {2007})},\ \Eprint
  {https://arxiv.org/abs/astro-ph/0703495} {arXiv:astro-ph/0703495}
  \BibitemShut {NoStop}%
\bibitem [{\citenamefont {Yu}\ \emph {et~al.}(2021)\citenamefont {Yu},
  \citenamefont {Jiang}, \citenamefont {Abdikamalov}, \citenamefont
  {Ayzenberg}, \citenamefont {Bambi}, \citenamefont {Liu}, \citenamefont
  {Nampalliwar},\ and\ \citenamefont {Tripathi}}]{Yu:2021xen}%
  \BibitemOpen
  \bibfield  {author} {\bibinfo {author} {\bibfnamefont {Z.}~\bibnamefont
  {Yu}}, \bibinfo {author} {\bibfnamefont {Q.}~\bibnamefont {Jiang}}, \bibinfo
  {author} {\bibfnamefont {A.~B.}\ \bibnamefont {Abdikamalov}}, \bibinfo
  {author} {\bibfnamefont {D.}~\bibnamefont {Ayzenberg}}, \bibinfo {author}
  {\bibfnamefont {C.}~\bibnamefont {Bambi}}, \bibinfo {author} {\bibfnamefont
  {H.}~\bibnamefont {Liu}}, \bibinfo {author} {\bibfnamefont {S.}~\bibnamefont
  {Nampalliwar}},\ and\ \bibinfo {author} {\bibfnamefont {A.}~\bibnamefont
  {Tripathi}},\ }\bibfield  {title} {\bibinfo {title} {{Constraining the
  Konoplya-Rezzolla-Zhidenko deformation parameters. II. Limits from
  stellar-mass black hole x-ray data}},\ }\href
  {https://doi.org/10.1103/PhysRevD.104.084035} {\bibfield  {journal} {\bibinfo
   {journal} {Phys. Rev. D}\ }\textbf {\bibinfo {volume} {104}},\ \bibinfo
  {pages} {084035} (\bibinfo {year} {2021})},\ \Eprint
  {https://arxiv.org/abs/2106.11658} {arXiv:2106.11658 [astro-ph.HE]}
  \BibitemShut {NoStop}%
\bibitem [{\citenamefont {Ni}\ \emph {et~al.}(2016)\citenamefont {Ni},
  \citenamefont {Jiang},\ and\ \citenamefont {Bambi}}]{Ni:2016uik}%
  \BibitemOpen
  \bibfield  {author} {\bibinfo {author} {\bibfnamefont {Y.}~\bibnamefont
  {Ni}}, \bibinfo {author} {\bibfnamefont {J.}~\bibnamefont {Jiang}},\ and\
  \bibinfo {author} {\bibfnamefont {C.}~\bibnamefont {Bambi}},\ }\bibfield
  {title} {\bibinfo {title} {{Testing the Kerr metric with the iron line and
  the KRZ parametrization}},\ }\href
  {https://doi.org/10.1088/1475-7516/2016/09/014} {\bibfield  {journal}
  {\bibinfo  {journal} {JCAP}\ }\textbf {\bibinfo {volume} {09}},\ \bibinfo
  {pages} {014}},\ \Eprint {https://arxiv.org/abs/1607.04893} {arXiv:1607.04893
  [gr-qc]} \BibitemShut {NoStop}%
\bibitem [{\citenamefont {Choudhury}\ \emph {et~al.}(2019)\citenamefont
  {Choudhury}, \citenamefont {Nampalliwar}, \citenamefont {Abdikamalov},
  \citenamefont {Ayzenberg}, \citenamefont {Bambi}, \citenamefont {Dauser},\
  and\ \citenamefont {Garcia}}]{Choudhury:2018zmf}%
  \BibitemOpen
  \bibfield  {author} {\bibinfo {author} {\bibfnamefont {K.}~\bibnamefont
  {Choudhury}}, \bibinfo {author} {\bibfnamefont {S.}~\bibnamefont
  {Nampalliwar}}, \bibinfo {author} {\bibfnamefont {A.~B.}\ \bibnamefont
  {Abdikamalov}}, \bibinfo {author} {\bibfnamefont {D.}~\bibnamefont
  {Ayzenberg}}, \bibinfo {author} {\bibfnamefont {C.}~\bibnamefont {Bambi}},
  \bibinfo {author} {\bibfnamefont {T.}~\bibnamefont {Dauser}},\ and\ \bibinfo
  {author} {\bibfnamefont {J.~A.}\ \bibnamefont {Garcia}},\ }\bibfield  {title}
  {\bibinfo {title} {{Testing the Kerr metric with X-ray Reflection
  Spectroscopy of Mrk 335 Suzaku data}},\ }\href
  {https://doi.org/10.3847/1538-4357/ab24d6} {\bibfield  {journal} {\bibinfo
  {journal} {Astrophys. J.}\ }\textbf {\bibinfo {volume} {879}},\ \bibinfo
  {pages} {80} (\bibinfo {year} {2019})},\ \Eprint
  {https://arxiv.org/abs/1809.06669} {arXiv:1809.06669 [gr-qc]} \BibitemShut
  {NoStop}%
\bibitem [{\citenamefont {Johannsen}(2013{\natexlab{b}})}]{Johannsen:2013vgc}%
  \BibitemOpen
  \bibfield  {author} {\bibinfo {author} {\bibfnamefont {T.}~\bibnamefont
  {Johannsen}},\ }\bibfield  {title} {\bibinfo {title} {{Photon Rings around
  Kerr and Kerr-like Black Holes}},\ }\href
  {https://doi.org/10.1088/0004-637X/777/2/170} {\bibfield  {journal} {\bibinfo
   {journal} {Astrophys. J.}\ }\textbf {\bibinfo {volume} {777}},\ \bibinfo
  {pages} {170} (\bibinfo {year} {2013}{\natexlab{b}})},\ \Eprint
  {https://arxiv.org/abs/1501.02814} {arXiv:1501.02814 [astro-ph.HE]}
  \BibitemShut {NoStop}%
\bibitem [{\citenamefont {Psaltis}\ \emph {et~al.}(2016)\citenamefont
  {Psaltis}, \citenamefont {Wex},\ and\ \citenamefont
  {Kramer}}]{Psaltis:2015uza}%
  \BibitemOpen
  \bibfield  {author} {\bibinfo {author} {\bibfnamefont {D.}~\bibnamefont
  {Psaltis}}, \bibinfo {author} {\bibfnamefont {N.}~\bibnamefont {Wex}},\ and\
  \bibinfo {author} {\bibfnamefont {M.}~\bibnamefont {Kramer}},\ }\bibfield
  {title} {\bibinfo {title} {{A Quantitative Test of the No-Hair Theorem with
  Sgr A* using stars, pulsars, and the Event Horizon Telescope}},\ }\href
  {https://doi.org/10.3847/0004-637X/818/2/121} {\bibfield  {journal} {\bibinfo
   {journal} {Astrophys. J.}\ }\textbf {\bibinfo {volume} {818}},\ \bibinfo
  {pages} {121} (\bibinfo {year} {2016})},\ \Eprint
  {https://arxiv.org/abs/1510.00394} {arXiv:1510.00394 [astro-ph.HE]}
  \BibitemShut {NoStop}%
\bibitem [{\citenamefont {Younsi}\ \emph {et~al.}(2016)\citenamefont {Younsi},
  \citenamefont {Zhidenko}, \citenamefont {Rezzolla}, \citenamefont
  {Konoplya},\ and\ \citenamefont {Mizuno}}]{Younsi:2016azx}%
  \BibitemOpen
  \bibfield  {author} {\bibinfo {author} {\bibfnamefont {Z.}~\bibnamefont
  {Younsi}}, \bibinfo {author} {\bibfnamefont {A.}~\bibnamefont {Zhidenko}},
  \bibinfo {author} {\bibfnamefont {L.}~\bibnamefont {Rezzolla}}, \bibinfo
  {author} {\bibfnamefont {R.}~\bibnamefont {Konoplya}},\ and\ \bibinfo
  {author} {\bibfnamefont {Y.}~\bibnamefont {Mizuno}},\ }\bibfield  {title}
  {\bibinfo {title} {{New method for shadow calculations: Application to
  parametrized axisymmetric black holes}},\ }\href
  {https://doi.org/10.1103/PhysRevD.94.084025} {\bibfield  {journal} {\bibinfo
  {journal} {Phys. Rev. D}\ }\textbf {\bibinfo {volume} {94}},\ \bibinfo
  {pages} {084025} (\bibinfo {year} {2016})},\ \Eprint
  {https://arxiv.org/abs/1607.05767} {arXiv:1607.05767 [gr-qc]} \BibitemShut
  {NoStop}%
\bibitem [{\citenamefont {V{\"o}lkel}\ \emph {et~al.}(2021)\citenamefont
  {V{\"o}lkel}, \citenamefont {Barausse}, \citenamefont {Franchini},\ and\
  \citenamefont {Broderick}}]{Volkel:2020xlc}%
  \BibitemOpen
  \bibfield  {author} {\bibinfo {author} {\bibfnamefont {S.~H.}\ \bibnamefont
  {V{\"o}lkel}}, \bibinfo {author} {\bibfnamefont {E.}~\bibnamefont
  {Barausse}}, \bibinfo {author} {\bibfnamefont {N.}~\bibnamefont
  {Franchini}},\ and\ \bibinfo {author} {\bibfnamefont {A.~E.}\ \bibnamefont
  {Broderick}},\ }\bibfield  {title} {\bibinfo {title} {{EHT tests of the
  strong-field regime of general relativity}},\ }\href
  {https://doi.org/10.1088/1361-6382/ac27ed} {\bibfield  {journal} {\bibinfo
  {journal} {Class. Quant. Grav.}\ }\textbf {\bibinfo {volume} {38}},\ \bibinfo
  {pages} {21LT01} (\bibinfo {year} {2021})},\ \Eprint
  {https://arxiv.org/abs/2011.06812} {arXiv:2011.06812 [gr-qc]} \BibitemShut
  {NoStop}%
\bibitem [{\citenamefont {Lara}\ \emph {et~al.}(2021)\citenamefont {Lara},
  \citenamefont {V{\"o}lkel},\ and\ \citenamefont {Barausse}}]{Lara:2021zth}%
  \BibitemOpen
  \bibfield  {author} {\bibinfo {author} {\bibfnamefont {G.}~\bibnamefont
  {Lara}}, \bibinfo {author} {\bibfnamefont {S.~H.}\ \bibnamefont
  {V{\"o}lkel}},\ and\ \bibinfo {author} {\bibfnamefont {E.}~\bibnamefont
  {Barausse}},\ }\bibfield  {title} {\bibinfo {title} {{Separating astrophysics
  and geometry in black hole images}},\ }\href
  {https://doi.org/10.1103/PhysRevD.104.124041} {\bibfield  {journal} {\bibinfo
   {journal} {Phys. Rev. D}\ }\textbf {\bibinfo {volume} {104}},\ \bibinfo
  {pages} {124041} (\bibinfo {year} {2021})},\ \Eprint
  {https://arxiv.org/abs/2110.00026} {arXiv:2110.00026 [gr-qc]} \BibitemShut
  {NoStop}%
\bibitem [{\citenamefont {Kocherlakota}\ and\ \citenamefont
  {Rezzolla}(2022{\natexlab{b}})}]{Kocherlakota:2022jnz}%
  \BibitemOpen
  \bibfield  {author} {\bibinfo {author} {\bibfnamefont {P.}~\bibnamefont
  {Kocherlakota}}\ and\ \bibinfo {author} {\bibfnamefont {L.}~\bibnamefont
  {Rezzolla}},\ }\bibfield  {title} {\bibinfo {title} {{Distinguishing
  gravitational and emission physics in black hole imaging: spherical
  symmetry}},\ }\href {https://doi.org/10.1093/mnras/stac891} {\bibfield
  {journal} {\bibinfo  {journal} {Mon. Not. Roy. Astron. Soc.}\ }\textbf
  {\bibinfo {volume} {513}},\ \bibinfo {pages} {1229} (\bibinfo {year}
  {2022}{\natexlab{b}})},\ \Eprint {https://arxiv.org/abs/2201.05641}
  {arXiv:2201.05641 [gr-qc]} \BibitemShut {NoStop}%
\bibitem [{\citenamefont {Akiyama}\ \emph {et~al.}(2022)\citenamefont {Akiyama}
  \emph {et~al.}}]{EventHorizonTelescope:2022xqj}%
  \BibitemOpen
  \bibfield  {author} {\bibinfo {author} {\bibfnamefont {K.}~\bibnamefont
  {Akiyama}} \emph {et~al.} (\bibinfo {collaboration} {Event Horizon
  Telescope}),\ }\bibfield  {title} {\bibinfo {title} {{First Sagittarius A*
  Event Horizon Telescope Results. VI. Testing the Black Hole Metric}},\ }\href
  {https://doi.org/10.3847/2041-8213/ac6756} {\bibfield  {journal} {\bibinfo
  {journal} {Astrophys. J. Lett.}\ }\textbf {\bibinfo {volume} {930}},\
  \bibinfo {pages} {L17} (\bibinfo {year} {2022})},\ \Eprint
  {https://arxiv.org/abs/2311.09484} {arXiv:2311.09484 [astro-ph.HE]}
  \BibitemShut {NoStop}%
\bibitem [{\citenamefont {Pei}\ \emph {et~al.}(2016)\citenamefont {Pei},
  \citenamefont {Nampalliwar}, \citenamefont {Bambi},\ and\ \citenamefont
  {Middleton}}]{Pei:2016kka}%
  \BibitemOpen
  \bibfield  {author} {\bibinfo {author} {\bibfnamefont {G.}~\bibnamefont
  {Pei}}, \bibinfo {author} {\bibfnamefont {S.}~\bibnamefont {Nampalliwar}},
  \bibinfo {author} {\bibfnamefont {C.}~\bibnamefont {Bambi}},\ and\ \bibinfo
  {author} {\bibfnamefont {M.~J.}\ \bibnamefont {Middleton}},\ }\bibfield
  {title} {\bibinfo {title} {{Blandford-Znajek mechanism in black holes in
  alternative theories of gravity}},\ }\href
  {https://doi.org/10.1140/epjc/s10052-016-4387-z} {\bibfield  {journal}
  {\bibinfo  {journal} {Eur. Phys. J. C}\ }\textbf {\bibinfo {volume} {76}},\
  \bibinfo {pages} {534} (\bibinfo {year} {2016})},\ \Eprint
  {https://arxiv.org/abs/1606.04643} {arXiv:1606.04643 [gr-qc]} \BibitemShut
  {NoStop}%
\bibitem [{\citenamefont {Camilloni}\ and\ \citenamefont
  {Rezzolla}(2026)}]{Camilloni:2026irg}%
  \BibitemOpen
  \bibfield  {author} {\bibinfo {author} {\bibfnamefont {F.}~\bibnamefont
  {Camilloni}}\ and\ \bibinfo {author} {\bibfnamefont {L.}~\bibnamefont
  {Rezzolla}},\ }\bibfield  {title} {\bibinfo {title} {{Universality of the
  Blandford-Znajek emission in stationary and axisymmetric spacetimes}},\
  }\href@noop {} {\  (\bibinfo {year} {2026})},\ \Eprint
  {https://arxiv.org/abs/2602.23417} {arXiv:2602.23417 [gr-qc]} \BibitemShut
  {NoStop}%
\bibitem [{\citenamefont {Li}\ \emph {et~al.}(2022)\citenamefont {Li},
  \citenamefont {Sun}, \citenamefont {Lo}, \citenamefont {Payne},\ and\
  \citenamefont {Chen}}]{Li:2021wgz}%
  \BibitemOpen
  \bibfield  {author} {\bibinfo {author} {\bibfnamefont {X.}~\bibnamefont
  {Li}}, \bibinfo {author} {\bibfnamefont {L.}~\bibnamefont {Sun}}, \bibinfo
  {author} {\bibfnamefont {R.~K.~L.}\ \bibnamefont {Lo}}, \bibinfo {author}
  {\bibfnamefont {E.}~\bibnamefont {Payne}},\ and\ \bibinfo {author}
  {\bibfnamefont {Y.}~\bibnamefont {Chen}},\ }\bibfield  {title} {\bibinfo
  {title} {{Angular emission patterns of remnant black holes}},\ }\href
  {https://doi.org/10.1103/PhysRevD.105.024016} {\bibfield  {journal} {\bibinfo
   {journal} {Phys. Rev. D}\ }\textbf {\bibinfo {volume} {105}},\ \bibinfo
  {pages} {024016} (\bibinfo {year} {2022})},\ \Eprint
  {https://arxiv.org/abs/2110.03116} {arXiv:2110.03116 [gr-qc]} \BibitemShut
  {NoStop}%
\bibitem [{\citenamefont {Ma}\ \emph {et~al.}(2022)\citenamefont {Ma},
  \citenamefont {Mitman}, \citenamefont {Sun}, \citenamefont {Deppe},
  \citenamefont {H{\'e}bert}, \citenamefont {Kidder}, \citenamefont {Moxon},
  \citenamefont {Throwe}, \citenamefont {Vu},\ and\ \citenamefont
  {Chen}}]{Ma:2022wpv}%
  \BibitemOpen
  \bibfield  {author} {\bibinfo {author} {\bibfnamefont {S.}~\bibnamefont
  {Ma}}, \bibinfo {author} {\bibfnamefont {K.}~\bibnamefont {Mitman}}, \bibinfo
  {author} {\bibfnamefont {L.}~\bibnamefont {Sun}}, \bibinfo {author}
  {\bibfnamefont {N.}~\bibnamefont {Deppe}}, \bibinfo {author} {\bibfnamefont
  {F.}~\bibnamefont {H{\'e}bert}}, \bibinfo {author} {\bibfnamefont {L.~E.}\
  \bibnamefont {Kidder}}, \bibinfo {author} {\bibfnamefont {J.}~\bibnamefont
  {Moxon}}, \bibinfo {author} {\bibfnamefont {W.}~\bibnamefont {Throwe}},
  \bibinfo {author} {\bibfnamefont {N.~L.}\ \bibnamefont {Vu}},\ and\ \bibinfo
  {author} {\bibfnamefont {Y.}~\bibnamefont {Chen}},\ }\bibfield  {title}
  {\bibinfo {title} {{Quasinormal-mode filters: A new approach to analyze the
  gravitational-wave ringdown of binary black-hole mergers}},\ }\href
  {https://doi.org/10.1103/PhysRevD.106.084036} {\bibfield  {journal} {\bibinfo
   {journal} {Phys. Rev. D}\ }\textbf {\bibinfo {volume} {106}},\ \bibinfo
  {pages} {084036} (\bibinfo {year} {2022})},\ \Eprint
  {https://arxiv.org/abs/2207.10870} {arXiv:2207.10870 [gr-qc]} \BibitemShut
  {NoStop}%
\bibitem [{\citenamefont {Dhani}(2021)}]{Dhani:2020nik}%
  \BibitemOpen
  \bibfield  {author} {\bibinfo {author} {\bibfnamefont {A.}~\bibnamefont
  {Dhani}},\ }\bibfield  {title} {\bibinfo {title} {{Importance of mirror modes
  in binary black hole ringdown waveform}},\ }\href
  {https://doi.org/10.1103/PhysRevD.103.104048} {\bibfield  {journal} {\bibinfo
   {journal} {Phys. Rev. D}\ }\textbf {\bibinfo {volume} {103}},\ \bibinfo
  {pages} {104048} (\bibinfo {year} {2021})},\ \Eprint
  {https://arxiv.org/abs/2010.08602} {arXiv:2010.08602 [gr-qc]} \BibitemShut
  {NoStop}%
\bibitem [{\citenamefont {Jim{\'e}nez~Forteza}\ \emph
  {et~al.}(2020)\citenamefont {Jim{\'e}nez~Forteza}, \citenamefont {Bhagwat},
  \citenamefont {Pani},\ and\ \citenamefont
  {Ferrari}}]{JimenezForteza:2020cve}%
  \BibitemOpen
  \bibfield  {author} {\bibinfo {author} {\bibfnamefont {X.}~\bibnamefont
  {Jim{\'e}nez~Forteza}}, \bibinfo {author} {\bibfnamefont {S.}~\bibnamefont
  {Bhagwat}}, \bibinfo {author} {\bibfnamefont {P.}~\bibnamefont {Pani}},\ and\
  \bibinfo {author} {\bibfnamefont {V.}~\bibnamefont {Ferrari}},\ }\bibfield
  {title} {\bibinfo {title} {{Spectroscopy of binary black hole ringdown using
  overtones and angular modes}},\ }\href
  {https://doi.org/10.1103/PhysRevD.102.044053} {\bibfield  {journal} {\bibinfo
   {journal} {Phys. Rev. D}\ }\textbf {\bibinfo {volume} {102}},\ \bibinfo
  {pages} {044053} (\bibinfo {year} {2020})},\ \Eprint
  {https://arxiv.org/abs/2005.03260} {arXiv:2005.03260 [gr-qc]} \BibitemShut
  {NoStop}%
\end{thebibliography}%

\appendix

\section{Retrograde modes}
\label{appendix:retro}

\begin{figure*}[!t]
\centering
\includegraphics[width=1.0\linewidth]{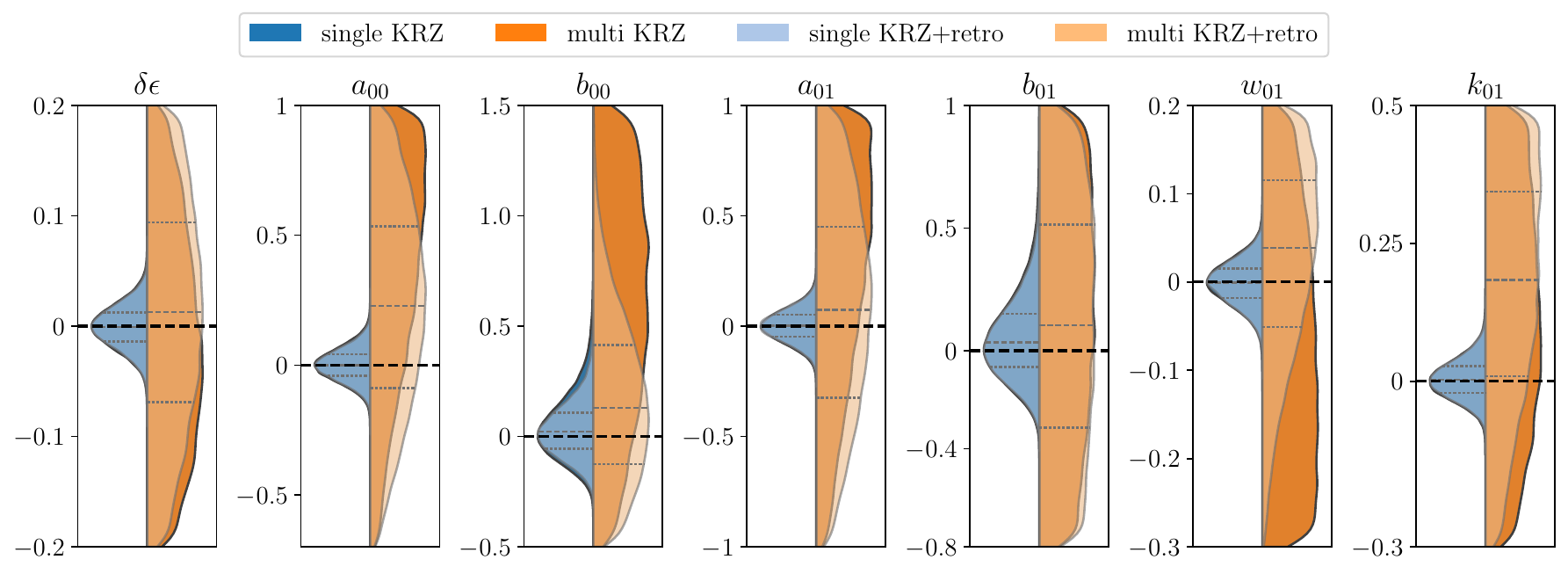}
\caption{Each panel shows the posterior distribution of a KRZ parameter for the single-parameter (left violins) and multi-parameter (right violins) KRZ analyses of the GR injection. The distributions are obtained using either the prograde QNM alone (single KRZ, multi KRZ) or both the prograde and retrograde QNMs (single KRZ+retro, multi KRZ+retro). The black dashed lines represent the corresponding GR values, while the black dotted lines identify the $1/4$, $1/2$, and $3/4$ quantiles.}
\label{GR_retro_KRZ}
\end{figure*}

\begin{figure*}[!t]
\centering
\includegraphics[width=1.0\linewidth]{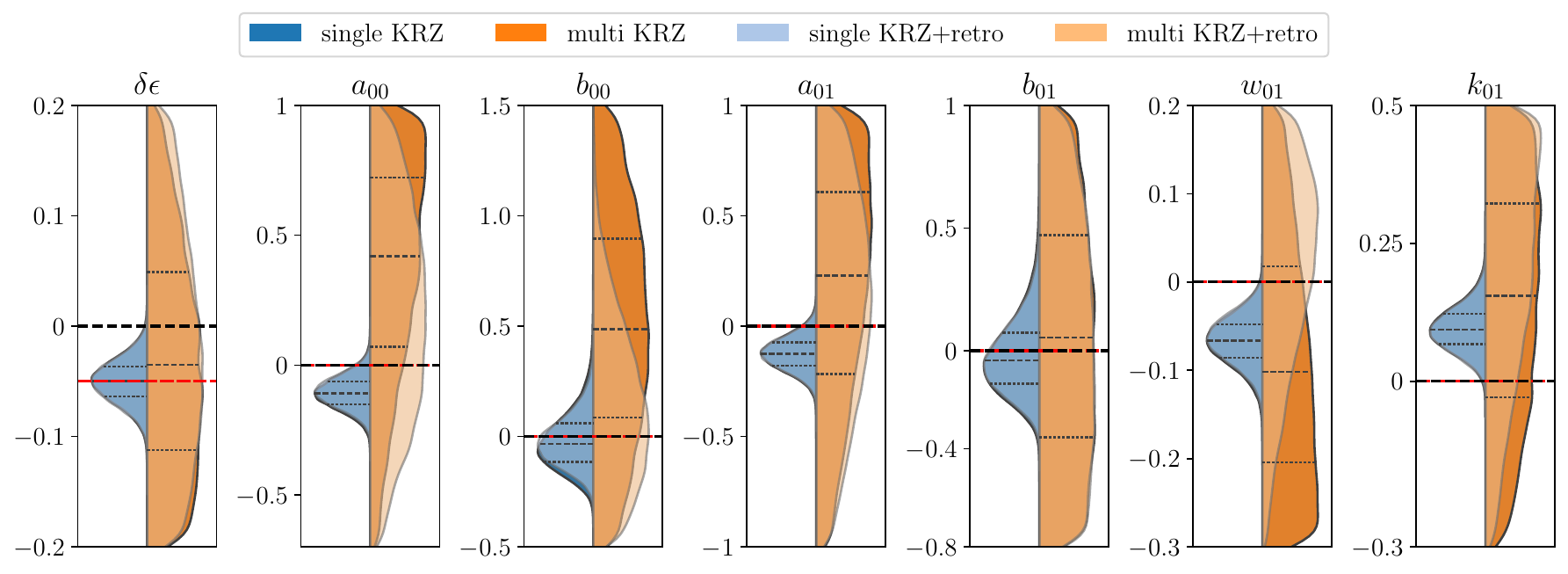}
\caption{Each panel shows the posterior distribution of a KRZ parameter for the single-parameter (left violins) and multi-parameter (right violins) KRZ analyses of the non-GR injection. The distributions are obtained using either the prograde QNM alone (single KRZ, multi KRZ) or both the prograde and retrograde QNMs (single KRZ+retro, multi KRZ+retro). The red dashed lines represent the injected non-GR values (only non-zero for $\delta \epsilon$), the black dashed lines represent the corresponding GR values, and the black dotted lines identify the $1/4$, $1/2$, and $3/4$ quantiles.}
\label{inj_retro_KRZ}
\end{figure*}

In this appendix, we discuss how the posteriors on the KRZ parameters are affected by adding the information on a retrograde mode in the GR and non-GR case. Those modes can be excited in binary BH mergers depending on the properties of the progenitors~\cite{Li:2021wgz,Ma:2022wpv} and may be relevant in ringdown analyses~\cite{Dhani:2020nik,JimenezForteza:2020cve}.

In Fig.~\ref{GR_retro_KRZ}, we compare the posteriors on the KRZ parameters when taking into account also the information on the retrograde mode $\ell=2,\,m=-2$, assuming the same uncertainty as the prograde one. The retrograde mode has been included only in the metric-specific case, where the KRZ parametrization determines the metric functions over the entire spacetime. In the metric-agnostic case, instead, adding a retrograde mode implies doubling the number of free parameters, since the eikonal formula depends on the metric functions evaluated at the retrograde photon orbit.

It turns out that including the additional information on the retrograde modes does not lead to any significant improvement in the metric shifts near the prograde photon sphere. Similarly, the posteriors for the KRZ parameters in the one at a time case are largely unaffected by the retrograde mode, with a small difference only in the $b_{00}$ posterior. This result is partially expected, since in the considered range of the KRZ parameters, the QNM shifts in the retrograde modes are typically subdominant compared to the prograde ones. However, we have checked that, by reducing the uncertainty $\sigma$ on the retrograde QNM shifts by a factor of ten, the resulting posteriors on the KRZ parameters are significantly more informative and entirely compatible with GR. 

Finally, in Fig.~\ref{inj_retro_KRZ} we report the results for the non-GR case. The results are coherent with the GR analysis since retrograde modes do not lead to any significant improvement on the one-at-a-time KRZ posteriors or in the case where all the parameters are allowed to vary. In this case reducing the uncertainty on the QNM shifts leads to narrower posteriors but the injection is still not recovered very accurately.

\end{document}